\documentclass[12pt]{iopart}
\usepackage{setspace,graphicx,enumerate,amssymb,sidecap}

\usepackage[left=1.54cm,top=1.54cm,bottom=1.54cm,right=1.54cm]{geometry}
\renewcommand{\theequation}{\arabic{equation}}  %this removes section numbers from equations
\date{\today}
\usepackage[usenames]{color}

\begin{document}
\newcommand{\bm}[1]{\mbox{\boldmath{$#1$}}}
%\captionsetup[subfigure]{style=default, margin=0pt, parskip=0pt, hangindent=0pt,indention=0pt, singlelincheck=false}

\title[Cold and Ultracold Molecules]{\begin{center}Cold and Ultracold Molecules:\\ Science, Technology, and Applications \end{center}}
\author{Lincoln D. Carr}
\address{Department of Physics, Colorado School of Mines, Golden, Colorado 80401, USA}
\author{David DeMille}
\address{{Department of Physics,} Yale University, PO Box 208120, New Haven, CT 06520, USA}
\author{Roman V. Krems}
\address{Department of Chemistry, University of British Columbia, Vancouver, B.C. V6T 1Z1, Canada}
\author{Jun Ye}
\address{JILA, National Institute of Standards and Technology and University of Colorado\\
Department of Physics, University of Colorado, Boulder, Colorado 80309-0440, USA}

\begin{abstract}
This article presents a review of the current state of the art in the research field of
cold and ultracold molecules. It serves as an introduction to the Special Issue of the New Journal of Physics on Cold and Ultracold Molecules and describes new prospects for fundamental research and technological development. Cold and ultracold molecules may revolutionize physical chemistry and few body physics, provide techniques for probing new states of quantum matter, allow for precision measurements of both fundamental and applied interest, and enable quantum simulations of condensed-matter phenomena. Ultracold molecules offer promising applications such as new platforms for quantum computing, precise control of molecular dynamics, nanolithography, and Bose-enhanced chemistry. The discussion is based on recent experimental and theoretical work and concludes with a summary of anticipated future directions and open questions in this rapidly expanding research field.

\end{abstract}

\pagebreak[4]

\tableofcontents
\pagebreak[4]

\section{Introduction}
\label{sec:introduction}

\begin{figure}[t]
 \begin{center}
\vspace*{2mm}
\includegraphics[scale=0.45]{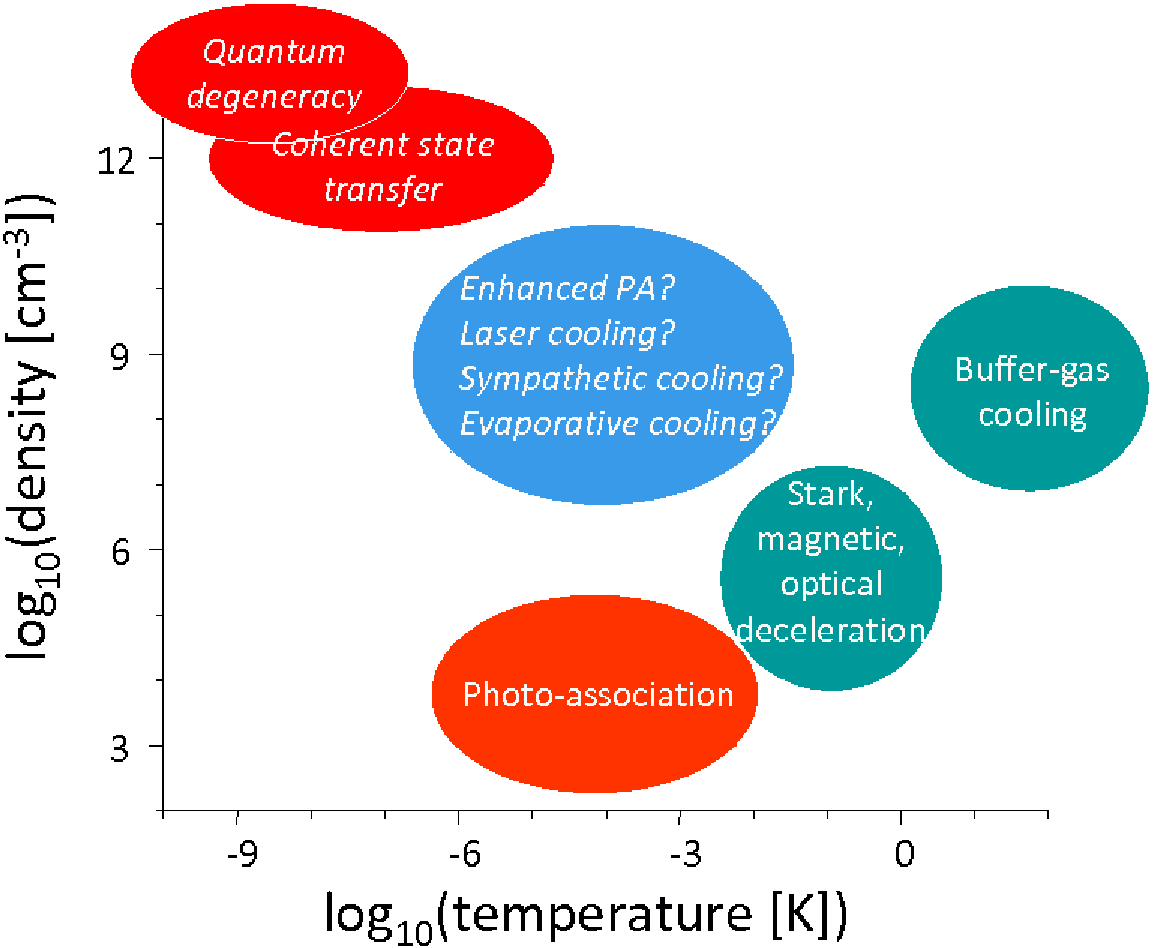}
\hspace*{12mm}
\includegraphics[scale=0.45]{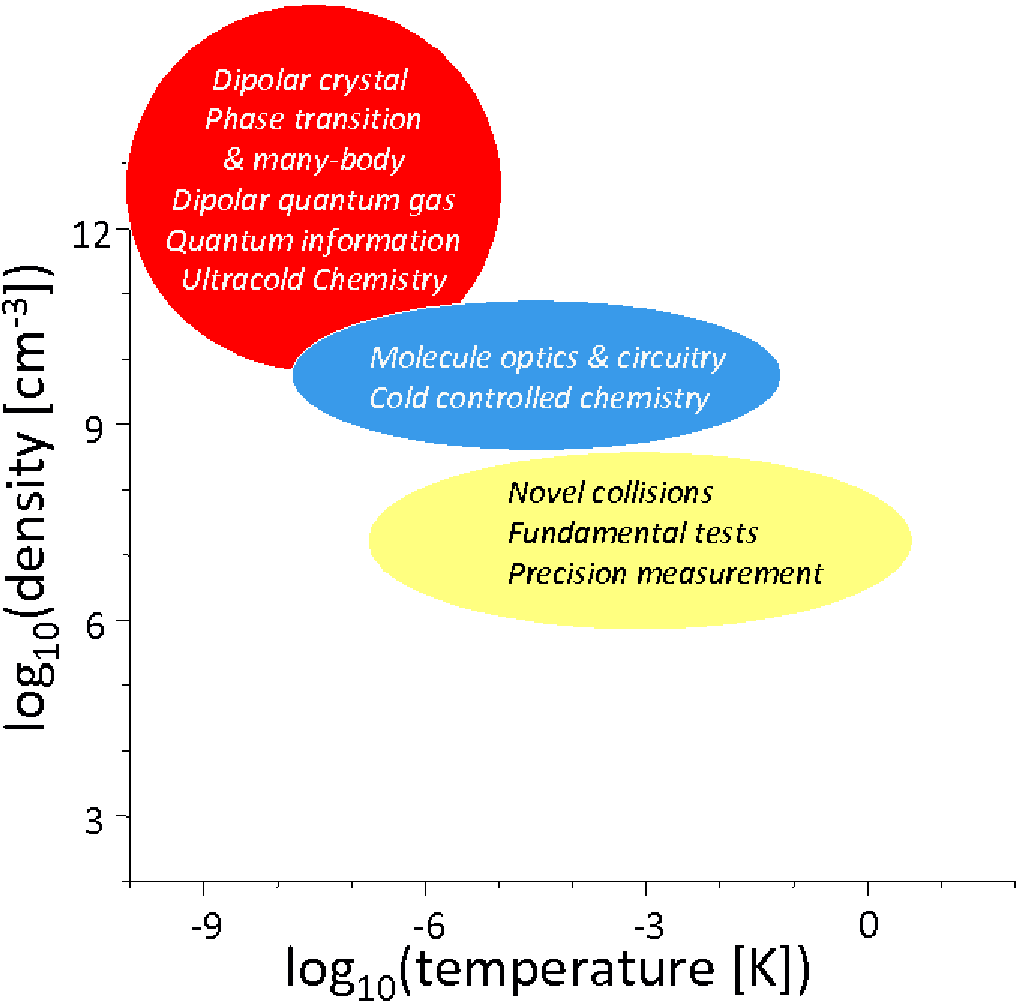}
\caption{(a) The production of cold and ultracold molecules in different regions of spatial density ($n$) and temperature ($T$). Some technical approaches that are yet to be demonstrated in experiments can potentially address the important region of $n\sim10^7 - 10^{10}$ cm$^{-3}$ and $T\sim1$ mK -- 1 $\mu$K (the panel in the middle).  (b) Applications of cold and ultracold molecules to various scientific explorations are shown with the required values of $n$ and $T$. The various bounds shown here are not meant to be strictly applied, but rather they serve as general guidelines for the technical requirements necessary for specific scientific topics.}
\label{fig:schematic}
 \end{center}
\end{figure}
\begin{table}[t]
\begin{center}
\begin{tabular}{|c|c|c|c|}
\hline Phase Space Density & Number Density  & Temperature & Scientific Goal\\
\hline  $10^{-17} - 10^{-14}$ & $10^6 - 10^9$  cm$^{-3}$ & $< 1$ K &Tests of fundamental forces of nature \\
\hline  $10^{-14}$ & $> 10^9$  cm$^{-3}$ & $< 1$ K & Electric dipole interactions \\
\hline  $10^{-13} - 10^{-10}$ & $> 10^{10}$  cm$^{-3}$ & $< 1$ K & Cold controlled chemistry \\
\hline  $10^{-5}$ & $> 10^{9}$  cm$^{-3}$ & $< 1$ $\mu$K & Ultracold chemistry \\
\hline  $1$ & $> 10^{13}$  cm$^{-3}$ & $ 100$ nK & Quantum degeneracy with molecules \\
\hline  $1$ & $> 10^{13}$  cm$^{-3}$ & $ 100$ nK & Optical lattices of molecules \\
\hline  10 & $> 10^{14}$  cm$^{-3}$ & $< 100$ nK & Novel quantum phase transitions  \\
\hline  100 & $> 10^{14}$  cm$^{-3}$ & $< 30$ nK & Dipolar crystals \\
\hline
\end{tabular}
  \caption{Orders of magnitude estimate of the phase space density and temperature of molecular ensembles required to achieve the main scientific goals that stimulate the development of ultracold molecule research. For dipolar physics, we assume a permanent dipole moment of 1 Debye. The molecular mass is assumed to be 100 amu.}
  \label{tab:phaseSpace}
\end{center}
\end{table}
Experimental work with molecular gases cooled to ultralow temperatures offers new insights into many body physics, quantum dynamics of complex systems, quantum chemistry, and fundamental forces in nature.  From magnetic and electric decelerators to magneto- and photo-association to buffer gas cooling, a variety of newly developed experimental techniques provide new routes to physical discoveries based on cold and ultracold molecules.  The research field of cold molecules brings together two of three main thrusts of modern atomic, molecular and optical (AMO) physics: the ultracold and the  ultraprecise.  It also brings together researchers from a variety of fields, including AMO physics, chemistry, quantum information science and quantum simulations, condensed matter physics, nuclear physics, and astrophysics.  In order to explore exciting applications of cold matter experiments, it is necessary to produce large ensembles of molecules at temperatures below 1 Kelvin.  We distinguish \emph{cold} ($1$ mK - 1 K) and \emph{ultracold} ($< 1$ mK) temperature regimes.  In this review article we give a broad overview of the state of the art in the research field of cold and ultracold molecules, describe the experimental methods developed to create molecular ensembles at extremely low temperatures, and discuss potential applications of cold and ultracold molecules anticipated in the near future.  Our discussion focuses on the specific scientific goals that motivate most of the current experiments.  These scientific goals are summarized in Table~\ref{tab:phaseSpace} and schematically depicted in Fig.~\ref{fig:schematic}.

Starting from a few research groups in the late 1990s working on the study of cold and ultracold molecules, the field has grown to encompass hundreds of researchers.  This review is a part of and serves as an introduction to the Special Issue of the New Journal of Physics on Cold and Ultracold Molecules.  The special issue contains 40 contributions, of which 38 are original research papers
\cite{Kajita09,Ghosal09,Klawunn09,GonzalezFerez09,Raitzsch09,Lu09,Avdeenkov09,Sogo09,Patterson09,Xu09,Kim09,Tscherbul09,Brickman09,Meek09,Salzburger09,Hojbjerre09,Wall09,Kuznetsova09,Barletta09,Motsch09,Parazzoli09,Metz09,Takase09,Deiglmayr09,Inouye09,Danzl09,Sofikitis09,Tokunaga09,Cavagnero09,Bohn09,Roschilde09,Haimberger09,Kotochigova09,Thalhammer09,Ortner09,Narevicius09,Pellegrini09,Kuma09},
which give a snapshot of part of the current experimental and theoretical work in this research field, and two are reviews, including this article and an additional short review dedicated to time variation of fundamental constants~\cite{Chin09}.

The production of ultracold ensembles of atoms has revolutionized the field of AMO physics and generated much interest among researchers in other, traditionally disjoint fields. The impact of creating ultracold molecules is expected to be as large and profound as that made by the work performed with ultracold atoms. Molecules offer microscopic degrees of freedom absent in atomic gases. This gives ultracold molecular gases unique properties that may allow for the study of new physical phenomena and lead to discoveries, reaching far beyond the focus of traditional molecular science~\cite{Doyle04}.
For example, a Bose-Einstein condensate (BEC) of polar molecules would represent a quantum fluid of strongly and anisotropically interacting particles and thereby greatly enhance the scope for study and applications of collective quantum phenomena.  It could be used to elucidate the link between BECs in dilute gases and in dense liquids.  The study of ultracold fermionic molecules is also of great interest: the electric dipole-dipole interaction may give rise to a molecular superfluid via Bardeen-Cooper-Schrieffer (BCS) pairing. The dipole-dipole interaction is both long range and anisotropic and it leads to fundamentally new condensed-matter phases and new complex quantum dynamics.  Ensembles of trapped polar molecules dressed by microwave fields may form topologically ordered states supporting excitations with anyonic statistics~\cite{Micheli06,Buchler07}.  The association of ultracold molecules into chains might provide a novel system to model rheological phenomena extending the study of elasticity to materials with non-classical mechanical behavior~\cite{Wang06}.

Ultracold polar molecules trapped on optical lattices, or near mesoscopic electrical circuits, provide a promising platform for quantum information processing.  They offer a significant advantage over neutral atoms because they have additional tunable experimental parameters: an electric dipole moment can be induced in ultracold polar molecules by a static DC electric field, and transitions between internal rotational states can be driven with resonant microwave fields.  The presence of rotationally excited states allows for the possibility to dynamically tailor the dipole-dipole interactions to be effectively short or long range~\cite{Pupillo08}.  This is in addition to the long coherence times, extreme purity of the sample, and ability to hold and manipulate molecules with off-resonant AC trapping fields, e.g., confinement in an optical lattice, all features which they share with neutral atoms.   The choice of molecule can also lead to fine or hyperfine structure which can be addressed together with the rotational states or $\Lambda$-doublet states~\cite{Lev06}.  These additional structures lead to a large internal Hilbert space, which, when combined with vibrational states in an external trap, is analogous to the system of trapped ions~\cite{Haeffner08}, successfully used for a manifold of fundamental applications.  Quantum computing schemes based on ultracold molecules have been explored theoretically for some years now.  Thus there is a well-developed literature to build from, encompassing everything from molecular vapors~\cite{Zadoyan01} to artificial lattice systems~\cite{Demille02,Wall09} to hybrid mesoscopic devices~\cite{Andre06,Rabl06}.

Similarly to how ultracold atoms have revolutionized AMO physics, cold and ultracold molecules have the potential to dramatically influence precision tests of fundamental physics, physical chemistry, and few body physics. Indeed, ultracold molecules represent an exciting new frontier in this endeavor. The additional degrees of freedom available in molecules, while presenting complexities and challenges for experimental control, offer a rich playground for precision measurement and unique opportunities for quantum control. For the first time we can now envision that ultracold molecules are prepared in single quantum states for both their internal and external degrees of freedom. The control of the molecular internal structure and its external motions will be intricately related and mutually influential. Precision spectroscopy will be connected to high resolution quantum control~\cite{Peer07,Shapiro08}. The presence of electronic, rovibrational, $\Lambda$- or $\Omega$-doublet, and hyperfine structure allow more accurate spectroscopy schemes to facilitate tests of fundamental laws of nature. Indeed, ultracold molecules will open important scientific directions such as precise control of chemical reactions~\cite{Hudson06a,Krems05, Krems08}, study of novel dynamics in low-energy collisions~\cite{Avdeenkov03}, long-range collective quantum effects and quantum phase transitions~\cite{Yi00,Goral02,Santos03,Micheli06,Buchler07}, and tests of fundamental symmetries such as parity and time-reversal~\cite{Hudson2002} and time variation of fundamental constants~\cite{Hudson06b,Chin09}. Simply put, molecular interactions control everything from chemical reactions for making new materials to generation of energy.

Chemical reactions of molecules have been predicted to occur rapidly at ultralow temperatures; the creation of dense ensembles of ultracold molecules will allow the study of ultracold chemistry~\cite{Krems08}.  The large de Broglie wavelength of ultracold molecules entirely changes the nature of reaction dynamics. At such low temperatures, even collisions of large molecules exhibit significant quantum effects. Energy barriers on the potential energy surface play a different role because, in this extremely quantum regime, tunneling  becomes the dominant reaction pathway. Barrierless reactions are accelerated by threshold phenomena. When the de Broglie wavelength increases, many-body interactions become significant and the outcome of a chemical reaction may depend on the spatial confinement of molecules.  This suggests the possibility of studying chemical processes in restricted geometries, which may lead to the development of new methods for manipulating chemical reactions with laser fields and help to elucidate the mechanisms of complex chemical reactions~\cite{Li08}.

Since early experiments on chemical reaction dynamics~\cite{Taylor55}, the efforts of many researchers have been to achieve external control over chemical reactions.  Many ground-breaking experiments demonstrated the possibility of controlling uni-molecular reactions, such as molecular dissociation and selective bond breaking, by external laser fields~\cite{Shapiro03}. However, external field control of bi-molecular chemical reactions remains an unrealized goal. It is complicated by thermal motion of molecules that randomizes molecular encounters and decreases the effects of external fields on molecular collisions. The thermal motion of molecules becomes insignificant when the temperature of molecular ensembles is reduced to below 1 $\mu$K. Cooling molecules to low temperatures may therefore allow for external field control of molecular collisions. In particular, the production of ultracold molecules will make possible coherent control of bi-molecular reactions~\cite{Krems07}. Coherent control of molecular dynamics is based on quantum interference effects, and has been applied with great success to the study of uni-molecular reactions such as photodissociation~\cite{Shapiro03}. However, such control of molecular collisions is complicated by the need to entangle the internal degrees of freedom of the colliding molecules with the wave function for the relative motion and the center-of-mass motion of the collision pairs. At ultralow temperatures, the center-of-mass motion will either be insignificant or controlled and the experiments with ultracold molecules may finally provide a test bed for the theory of coherent control of  molecular encounters.

\emph{Phase space density} is a primary organizing theme of this article, as indicated in Table~\ref{tab:phaseSpace} and Fig.~\ref{fig:schematic}.  The phase space density $\Omega$ is defined in free space as $\Omega = n \lambda_{\mathrm{dB}}^3$, where $n$ is the density of molecules, $\lambda_{\mathrm{dB}}=h/\sqrt{2\pi m k_B T}$ is the thermal de Broglie wavelength, $m$ is the molecular mass, and $T$ is the temperature. Although large phase space density is required for achieving quantum degenerate gases of molecules, low particle number density ensures the absence of destructive collisions. This may be desirable for precise spectroscopic experiments such as tests of fundamental symmetries.  Temperatures in the cold (as opposed to ultracold) regime may allow for collision processes involving excited rotational states of the collision complex (non-zero partial waves)~\cite{Gilijamse06,Sawyer08a}, leading to interesting quantum interference phenomena and differential scattering. We will therefore argue that different scientific goals require different experimental techniques for cooling molecules, rather than espousing the notion that all cooling experiments must aim at producing molecular ensembles with the highest possible phase space density. While in general most experiments benefit from a high phase space density molecular sample, other considerations such as chemical diversity, experimental complexity, and specific scientific goals need to be taken into account. The development of distinct experimental techniques for cooling is necessary to ensure the progress of the research field of cold and ultracold molecules. For example, we note that while substantial recent progress has been made for the production of ground-state polar molecules in the cold~\cite{Weinstein98,Bethlem99,Bochinski03,Merakker08}, ultracold~\cite{Sage05,Deiglmayr08}, and nearly quantum degenerate regime~\cite{Ni08}, general and versatile experimental techniques are yet to be realized to produce a large quantity of ultracold molecules, as shown in the middle panel of Fig. 1(a).

This article is organized as follows.  Section~\ref{sec:fundamental} discusses the fundamental research enabled by the creation of ultracold molecules.   Section~\ref{sec:current} describes the latest experimental techniques for the production of cold and ultracold ensembles of molecules. Section~\ref{sec:applications} gives an overview of practical applications of cold and ultracold molecules and outlines possible future directions of this research field. Section~\ref{sec:conclusions} presents the conclusions, including a list of fundamental open questions that need to be addressed.

\section{Fundamental Science}
\label{sec:fundamental}

As shown in Fig.~\ref{fig:schematic}(b), the fundamental science based on ensembles of cold and ultracold molecules encompasses three major research directions: few body physics and chemistry; precision spectroscopy and probes of fundamental symmetries and constants; and many body physics.  In the following, we discuss each of these areas separately, beginning with a review of molecular structure.  Experts in molecular physics can skip Sec.~\ref{ssec:structure}.

\subsection{Highlighted Features of Molecular Structure}
\label{ssec:structure}

\begin{SCfigure}
  \centering
%\vspace*{4mm}
\includegraphics[scale=1.0]{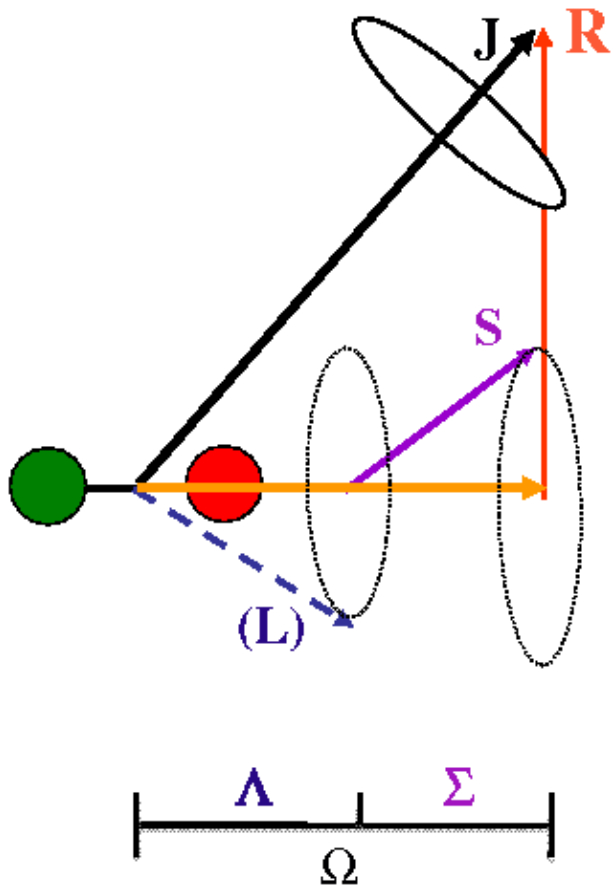}
\caption{Molecular electronic state labeling under the Hund's case (a) coupling. Heteronuclear diatomic molecules possess only axial symmetry. The projection of the electronic orbital angular momentum ($\mathbf{L}$) along the internuclear axis $\hat{e}_n$ is $\Lambda$. The projection of the electronic spin angular momentum $\mathbf{S}$ along $\hat{e}_n$ is $\Sigma$. $\hbox{\boldmath{$\mathcal{R}$}}$ is the mechanical rotational angular momentum of the nuclei about the center of mass of the molecule.  $\mathbf{J} = \hbox{\boldmath{$\mathcal{R}$}} + \mathbf{L} + \mathbf{S}$ is the total angular momentum. The electronic potentials are labeled as $^{2\Sigma+1}\Lambda_\Omega$, where $\Sigma$, $\Pi$, $\Delta$, ... states stand for $\Lambda$ = 1, 2, 3 ..., etc.  The good quantum numbers are $\Lambda$, $\Sigma$, $\Omega = \Lambda +\Sigma$, $J$, and the projection of $\mathbf{J}$ on the laboratory-fixed quantization axis, $M_J$.}
\label{fig:structure}
\end{SCfigure}
%

%hierarchy of energy scales

Molecules, as compared to atoms, have additional internal degrees of freedom that give rise to a far more complex energy level structure.  The interesting new science made possible with ultracold molecules is, in one way or another, associated with these novel features. Hence an understanding of their nature is important for sharing the excitement associated with this rapidly growing field.  However, the complexity of molecular structure can be daunting.  In this section we describe the essential features of diatomic molecules needed for a non-expert to grasp the physics discussed in this review.  A simple presentation of many of these features can be found in Ref.~\cite{Budker08}; thorough discussions of molecular structure are given e.g. in Refs.~\cite{Herzberg50,Field04,Brown03}.

We begin with a discussion of the relevant energy scales in molecular structure.  The highest-energy excitations in molecules correspond to changes in electronic orbitals.  The typical energy $E_{\mathrm{el}}$ associated with electronic excitation is determined parametrically by the atomic unit of energy: $E_{\mathrm{el}} \sim e^2/a_0$, where $e$ is the electron charge and $a_0$ is the Bohr radius. In practice, typical values of $E_{\mathrm{el}}$ are a few eV.  For each electronic state, the energy of the molecule depends on the separation between the constituent nuclei, $R$.  Hence each electronic orbital configuration has an associated potential energy curve $V(R)$.  In the limit of large separations ($R \rightarrow \infty$), these curves approach the energy of the isolated constituent atoms in their individual electronic states.  The exact form of $V(R)$ is largely determined by the electrostatic binding energies of the valence electrons to the two atomic cores, which change with $R$ as the orbitals of the individual atoms hybridize into new molecular orbitals.  The potential energy $V(R)$ has a minimum at an equilibrium distance $R_e$, with typical values of a few $a_0$ and binding energy of a few eV;  it is usually harmonic near $R=R_e$; and it has a long tail for large $R$ (often scaling as $V(R) \propto R^{-6}$ in this region due to van der Waals interactions) and a steep repulsive wall at small $R$ (where the electron cores of the two atoms begin to overlap).

The relative motion of the atoms in the potential $V(R)$ represents a qualitative new feature of the internal structure of molecules, as compared to isolated atoms.  The typical spacing  $E_{\mathrm{vib}}$ between energy levels of this vibrational motion, for deeply-bound levels in such potentials, is given parametrically by $E_{\mathrm{vib}} \sim \sqrt{\frac{m_e}{Am_p}} \frac{e^2}{a_0}$, where the reduced mass of the atoms is $\mu \equiv Am_p$. Because of the large value of $m_p/m_e$,
$E_{\mathrm{vib}} \sim 10^{-2} E_{\mathrm{el}}$.  Near the dissociation limit of any potential $V(R)$, the splitting between vibrational energy levels becomes systematically smaller: since the slope of the potential curve becomes small near its asymptotic limit, classically the motion in this potential has a longer period for weakly bound levels.

Finally, the framework of the molecule in any given vibrational state can rotate about its center of mass.
$\hbox{\boldmath{$\mathcal{R}$}}$ is the angular momentum associated with this mechanical rotation of the nuclei.
The total angular momentum of the diatomic system, $\mathbf{J}$, is defined as the vector sum of $\hbox{\boldmath{$\mathcal{R}$}}$ with other internal angular momenta (associated with electron spin and orbital motion). For a simple ``rigid rotor'' molecule with no internal structure, $\mathbf{J} = \hbox{\boldmath{$\mathcal{R}$}}$ and the energy of rotation $E_{\mathrm{rot}}$ is given by $E_{\mathrm{rot}} = \hbar^2 \frac{J(J+1)}{2I}$, where $I = \mu \left\langle R \right\rangle^2$ is the moment of inertia, and $\left\langle R \right\rangle$ is the expectation value of $R$.  Parametrically, $E_{\mathrm{rot}} \sim \frac{m_e}{Am_p} \frac{e^2}{a_0}$; typically $E_{\mathrm{rot}} \sim 10^{-4} E_{\mathrm{el}}$.  This corresponds to rotational splittings equivalent to temperatures $T \sim 1$ K for deeply bound vibrational levels where $\left\langle R \right\rangle \approx R_e$.  The rotational splittings also become smaller for weakly-bound vibrational levels: here $\left\langle R \right\rangle$ is increased, corresponding to the fact that a particle undergoing classical motion in $V(R)$ spends most of its time in the large-$R$ region where the potential is shallow.

One can write the total wavefunction of the molecule as a product of electronic, vibrational, and rotational parts: $\psi^{\gamma vJM_J}(\mathbf{r}_e; R, \theta, \phi) = \psi^\gamma_{el}(\mathbf{r}_e, R)\psi^v_{\mathrm{vib}}(R)\psi^{JM_J}_{\mathrm{rot}}(\theta, \phi)$.  Here $\gamma$ is a set of quantum numbers describing the electronic state and the associated form of $V(R)$; $v$ is the quantum number of the vibrational motion in $V(R)$; and the angles $\theta$ and $\phi$ describe the direction of the intermolecular axis $\hat{e}_n$ in the laboratory frame.  This factorized form of the wave function is based on the Born-Oppenheimer approximation, assuming that the motion of the nuclei in $V(R)$ is adiabatic with respect to the electronic state energy.  While this assumption makes the discussion of molecular structure simpler, we note below that some important phenomena arise when it is not valid.

The set of quantum numbers $\gamma$ associated with a given electronic state is determined by the cylindrical symmetry of the molecule, and depends on the hierarchy of couplings of the internal angular momenta of the molecule to $\hat{e}_n$ and to each other. These couplings can be quite complicated for molecules with unpaired electrons and non-zero nuclear spins. Different limiting cases of the coupling hierarchy are described by the so-called ``Hund's cases''.  For example, many molecules can be described as Hund's case (a), where spin-orbit effects are small compared to $V(R)$ (see Fig. \ref{fig:structure}).  In this case the total orbital angular momentum $\mathbf{L}$ of the electrons is coupled strongly to $\hat{e}_n$, such that $\mathbf{L}$ precesses rapidly about $\hat{e}_n$ and only the projection $\Lambda \equiv \mathbf{L}\cdot\hat{e}_n$ is well defined.  Due to spin-orbit effects, the electron spin $\mathbf{S}$ precesses about $\hbox{\boldmath{$\Lambda$}}=\Lambda \hat{e}_n$ and hence about the molecular axis; unlike $L$, however, $S$ remains a good quantum number.  The overall spin-orbit energy is determined by the average value of $\mathbf{L}\cdot\mathbf{S}$, which depends on $\Lambda$ and on $\Sigma \equiv \mathbf{S}\cdot\hat{e}_n$.  Hence the good quantum numbers are $\Lambda, S, \Sigma$, and (redundantly) $\Omega \equiv \Lambda + \Sigma$.  Such a state is denoted with the molecular term $^{2S+1}\Lambda_{\Omega}$.  The example of a Hund's case (a) molecular state with a $^1\Sigma_0$ term is perhaps the simplest type of molecule: it acts as a ``rigid rotor'' in which there is no net internal angular momentum, and hence $\mathbf{J} = \hbox{\boldmath{$\mathcal{R}$}}$ as described above.

Electromagnetic fields can be used to modify the internal and motional states of molecules. The effects of such fields (energy level shifts or transition rates) are determined by the relevant matrix elements.  Zeeman shifts can occur at $1^{\mathrm{st}}$ order in perturbation theory.  However, since the electric dipole operator connects states of opposite parity, the effect of electric fields always occurs at $2^{\mathrm{nd}}$ order.  Hence, the smaller energy scales associated with vibration and especially rotation in molecules can lead to a dramatic increase in response to electric fields, compared to atoms.  Indeed, this qualitatively new feature is at the heart of many of the new types of control envisioned for ultracold molecules.

As an example, we consider the effect of a DC electric field $\hbox{\boldmath{$\mathcal{E}$}}$ on a simple polar molecule (e.g. one in a $^1\Sigma_0$ state with no internal spin structure).  The largest effect of the field comes from mixing of the closely adjacent rotational states.  The relevant electric dipole Hamiltonian is ${H}_{E1} = -{\mathbf{D}}\cdot\hbox{\boldmath{$\mathcal{E}$}}$, where ${\mathbf{D}} = {D}\hat{e}_n$ is the ``permanent'' electric dipole moment associated with the molecule, i.e., the dipole moment one would observe in a frame where the molecular framework is fixed in space.  We note in passing that essentially \textit{only} homonuclear molecules are non-polar; for nearly every heteronuclear species $D$ lies in the range $D \sim 0.01-1~ e\,a_0$.  The molecular electric dipole moment is often measured in units of Debye, where 1 Debye $\cong$ 3.336$\times 10^{-30}$ C$\,$m = 0.393 $e\,a_0$.  The rotational wavefunction is given simply by $\psi^{JM_J}_{\mathrm{rot}}(\theta, \phi) = Y^J_{M_J}(\theta,\phi)$, where $Y^J_{M_J}(\theta,\phi)$ is a spherical harmonic.  One can show that ${H}_{E1}$ mixes states with $\Delta J = \pm 1$.  This mixing leads to an induced electric dipole moment $\langle {\mathbf{D}} \rangle \parallel \hbox{\boldmath{$\mathcal{E}$}}$.  For small fields, simple perturbation theory shows that for low-lying levels $\langle {D} \rangle \equiv |\langle {\mathbf{D}} \rangle | \sim D \hbox{{$\mathcal{E}$}}/B$, where $B \equiv \hbar^2/(2I)$ is the rotational constant.

This expression reveals some key and possibly unfamiliar facts.  For example, even for molecules with a ``permanent'' dipole moment, $\langle {{D}} \rangle = 0$ when the applied field $\mathcal{E}=0$.  Moreover, by noting that $\langle {D} \rangle = -\partial \left\langle {H}_{E1} \right\rangle / \partial \mathcal{E}$,  it is easy to see that $\langle {{D}} \rangle > 0$ for the lowest ($J=0$) rotational state, while excited states (which can repel from lower states due to the effect of ${H}_{E1}$) can have $\langle {{D}} \rangle < 0$.  Hence, in the presence of the field $\hbox{\boldmath{$\mathcal{E}$}}$ different energy levels have different values and even signs of $\langle {{D}} \rangle$.  For sufficiently large fields (such that $\mathcal{E} \gtrsim \mathcal{E}_c \equiv B/D$), perturbation theory breaks down and $\langle {{D}} \rangle$ approaches its maximal value such that $( \langle {{D}} \rangle\cdot \mathbf{\mathcal{E}})_{\mathrm{max}} \sim D |\mathcal{E}|$ for states associated with low values of $J$.  For typical molecules, $\mathcal{E}_c \sim 1-100$ kV/cm, field values that are routinely achieved in the lab.  Hence nearly full electrical polarization of molecules is easily achieved, exactly because of the small energy scale associated with rotation. Moreover, for ultralow temperatures the associated Stark shifts can easily be the dominant energy scale in the system, since in the fully polarized regime $\left\langle {H}_{E1} \right\rangle \sim B \gg k_BT$.

Although the response of molecules to DC electric fields is dominated by the rotational structure (for polar species), for oscillating electric fields the situation is more complicated.  The complex AC polarizability function $\alpha^{qq'}_{\gamma v J}(\omega)$ of a diatomic molecule in the state $\psi^{\gamma vJ}$ determines the complex energy shifts $\Delta E$ due to an applied AC field $\hbox{\boldmath{$\mathcal{E}$}}(\omega)$ ~\cite{Braun74,Friedrich95,Kotochigova06}.  In particular, $\Delta E = -\frac{1}{2}\sum_{q,q'}\alpha^{qq'}_{\gamma v J}(\omega)\left\langle \mathcal{E}_q \mathcal{E}^*_{q'}\right\rangle$, where $q,q'$ are spatial vector components and $\left\langle \right\rangle$ indicates a time average~\cite{Bonin97}.  In general $\alpha^{qq'}_{\gamma v J}(\omega)$ is a reducible rank-two tensor, with irreducible tensor, vector, and scalar parts.  The tensor (vector) part leads to splittings between $M_J$ sublevels in linearly (circularly) polarized fields, when $J > 1/2$. For simplicity, however, we suppress the spatial indices $qq'$ and (unless otherwise indicated) ignore the distinction between the irreducible tensor components of $\alpha_{\gamma v J}(\omega)$ in the following discussion.

Typically $\alpha_{\gamma v J}(\omega)$ has resonant features at many frequencies $\omega=\omega_{i0}$ in the optical frequency range, corresponding to electronic excitations.  For polar species, there are additional resonances in the infrared and the microwave ranges, corresponding to vibrational and rotational excitations, respectively. As in atoms, near-resonant fields can lead to population transfer between levels, described by the imaginary part of $f$, while the energy shifts, described by the real part of $f$, associated with far off-resonant fields can be used to apply mechanical forces for trapping and slowing molecules.  The width of the $i^{\mathrm{th}}$ resonant feature in $\alpha_{\gamma v J}(\omega)$ is determined by the spontaneous emission rate $\gamma_i$ from the $i^{\mathrm{th}}$ excited state (labeled by quantum numbers $\gamma' v' J'$) at energy $E_{\gamma' v' J'} = E_{\gamma v J} + \hbar\omega_{i0}$.  The height of the feature is determined by $|D_{\gamma v J,\gamma' v' J'}|^2$, where $D_{\gamma v J, \gamma' v' J'}$ is the electric dipole matrix element connecting the two states.  The width $\gamma_i$ is determined by the sum of partial widths $\gamma_{ij}$ for decay of the $i^{\mathrm{th}}$ state ($\gamma', v', J'$) to the $j^{\mathrm{th}}$ state (labeled by $\gamma'', v'', J''$); quite generally, $\gamma_{ij} \propto D_{ij}^2 \omega_{ij}^3$.  Note that rotational and, to a lesser extent, rovibrational sublevels of the lowest electronic state always have extremely small natural width and correspondingly long lifetimes, due simply to the fact that the transition frequency $\omega_{ij}$ is small for these low-lying levels.  For typical molecules, the natural lifetime $\tau = \gamma^{-1}$ is within a few orders of magnitude of $\tau \sim 10^5 $ s for rotational sublevels of the lowest vibronic level X$(v=0)$; and $\tau \sim 0.1 $ s for vibrationally excited states of the lowest electronic level X.  Hence, the natural lifetime of rotational levels for molecules in the X$(v=0)$ state is effectively infinite for most experiments.  By contrast, as in atoms, the lifetime of electronically excited states is typically very short, $\tau \sim 10$ ns (unless a strong selection rule suppresses all the matrix elements $d_{ij}$ leading to its decay, see below).  In cases of interest here, the width of individual rovibronic levels is typically much less than the rotational splittings ($\gamma \ll B$), so individual levels can be addressed.

The electric dipole matrix elements $D_{ij}$ are governed by certain exact and approximate selection rules, analogous to those in atoms.  For example, $D_{ij} \neq 0$ only when $\Delta J = \pm 1, 0$, so that within any vibronic (meaning electronic + vibrational) manifold of levels, at most three of the rotational levels contribute to $\alpha_{\gamma v J}(\omega)$.  Similarly, $D_{ij}$ vanishes for coupling between idealized electronic states with different total valence electron spin $S$. However, there are \textit{no} selection rules restricting the change in vibrational states.  Rather, for transitions between $\psi^{\gamma vJ}$ governed by the potential $V(R)$ and $\psi^{\gamma' v'J'}$ governed by a different potential $V'(R)$, the matrix element $d_{\gamma v J,\gamma', v', J'}$ is proportional to the Franck-Condon factor $F_{vv'} = \left| \int\!dR\,{\psi^v_{\mathrm{vib}}(R)\psi'^{\,v'}_{\mathrm{vib}}(R)} \right|^2$ which describes the overlap of the initial and final vibrational wavefunctions. (This is valid in the Born-Oppenheimer approximation.)  A simple rule of thumb is that Franck-Condon factors will be large if the states $\psi^v_{\mathrm{vib}}(R)$ and $\psi'^{v'}_{\mathrm{vib}}(R)$ share a common classical turning point; this can be seen from the correspondence principle, since the wavefunctions will be peaked near the turning points.  Because the potentials $V(R)$ and $V'(R)$ have no simple relationship, the Franck-Condon factors for any given $v$ are often substantial over a wide range of $v'$ .  However, they can also be extraordinarily small ($F_{vv'} \ll 10^{-10}$) for states with poor overlap.  The additional difficulty of controlling molecules with lasers, as compared to atoms, is in many ways due to the complex structure of Franck-Condon factors, which often require ingenious measures to work around.

In this discussion we have made a number of assumptions that, while often valid, can lead to interesting new features if not valid.  As one example, we have implicitly assumed that the molecule has no net internal angular momentum.  However, in many cases of interest the molecular states involve unpaired electronic orbital and spin angular momenta, as well as nuclear spins.  These give rise to spin-orbit, spin-rotation, and hyperfine substructure that is roughly analogous to that in atoms (see e.g.~\cite{Lev06}). However, a few features that are peculiar to molecules are notable.  For example, consider a molecule with internal electronic angular momentum $\mathbf{J}_e \neq 0$, in a frame where the molecular axis $\hat{e}_n$ is fixed.  Here states can be defined by their projection quantum number $\Omega \equiv \mathbf{J}_e \cdot \hat{e}_n$. (In states with small spin-orbit couplings including singlet states, $\Lambda$ rather than $\Omega$ is the relevant quantity.) However, for any possible value $M \neq 0$, the two states with $\Omega$ (or $\Lambda$) $= \pm M$ are mirror images of each other, and hence nominally degenerate in this frame.  In the laboratory frame (where the molecule can freely rotate), this degeneracy can be split by Coriolis-type effects coupling $\mathbf{J}_e$ with the rotational angular momentum $\hbox{\boldmath{$\mathcal{R}$}}$. These effects are typically small, so that the actual energy eigenstates consist of closely-spaced doublets of opposite parity, corresponding to the symmetric and antisymmetric superpositions of states with $\Omega$ (or $\Lambda$) $ = \pm M$.  These so-called ``$\Omega$ (or $\Lambda$)-doublets'' further enhance the DC polarizability of molecules. A good example is the hydroxyl radical (OH)~\cite{Bochinski04}.

Another interesting case involves the breakdown of the Born-Oppenheimer approximation.  A prototypical example arises because of fine or hyperfine substructure.  Note that for sufficiently large values of $R$, fine and even hyperfine energies of the separated constituent atoms can dominate over $V(R)$.  Here the constituent angular momenta of each atom are tightly coupled internally.  However, for internuclear separations $R \approx R_e$, $V(R)$ is almost always much larger than these effects.  Since $V(R)$ is determined to first order by electrostatic effects, there is no reason why the internal angular momenta  should couple in the same way at short range where $V(R)$ is substantial.  In general this leads to multiple potential curves $V(R)$, which undergo avoided crossings in the regime where the electrostatic binding energy is comparable to the atomic substructure~\cite{Field04}.  In the crossing region, these potentials are mixed as the internal spins rearrange from atomic-type coupling to molecular-type coupling.  In semiclassical terms, a particle moving in these coupled potentials can experience non-adiabatic transitions between the electronic states, as it moves through the crossing region.  This causes the true vibrational wavefunctions of the molecule to be a mixture of the idealized wavefunctions associated with either one of the diabatic potentials.  As might be expected from perturbation theory, the mixing is particularly large when the energies of two idealized levels are nearly degenerate.  In this case the true vibrational wavefunction can exhibit peaks at multiple pairs of classical turning points, each associated with one diabatic potential.  Because of the high density of states in molecules, accidental near-degeneracies of this type are not infrequent; in addition, they can be engineered in some cases by tuning one potential relative to another with applied fields.  In the special case where one of the nearly-degenerate levels is actually the zero-energy continuum level (just above the dissociation limit of one potential), this type of degeneracy is known as a Fano-Feshbach resonance~\cite{Kohler06,Chin09b}. Both accidental and engineered near-degeneracies are proving increasingly useful as a means of manipulating the dynamics of molecules.
For example, Gonz\'alez-F\'erez and Schmelcher \cite{GonzalezFerez09} discuss in this special issue the effects of strong DC electric fields on the dynamics of
LiCs molecules in highly excited vibrational levels. Their calculation shows that electric fields can be used to induce avoided crossings between different ro-vibrational states near the dissociation limit, leading to interesting dynamics of ro-vibrational motion.

\subsection{Chemistry and Few Body Physics}
\label{ssec:chemistry}

In this section we describe recent fundamental studies of chemistry and few-body collision physics of molecules in the cold and ultracold temperature regimes.

When molecules react, the rotational motion of the collision complex gives rise to a centrifugal force that may suppress collisions at low temperatures. The total scattering wave function of the collision complex can be decomposed into contributions from different angular momenta of the rotational motion called partial waves. The ultracold temperature regime is often loosely defined as $T < 1$ mK. A more rigorous concept defines ``ultracold'' as the temperature regime where the collision dynamics of particles in a gas is dominated by single partial-wave scattering. To distinguish the physics of single partial-wave and multiple partial-wave scattering,  we will use the latter definition in this section and assume that collisions at ultralow temperatures are entirely determined by single partial-wave scattering: $s$-wave scattering for  collisions of bosons or distinct particles and $p$-wave scattering for collisions of identical fermions, except in cases where these partial waves are intentionally suppressed through external manipulations. The boundary between the cold and ultracold temperature regimes therefore depends on the mass of the molecules and the magnitude of the long-range interactions between the colliding particles. For polar molecules in external electric fields, the onset of ultracold scattering occurs at extremely low temperatures. Collisions in the cold temperature regime are characterized by contributions of several partial waves, which may give rise to multiple scattering resonances and partial wave interference phenomena.  The cold and ultracold temperature regimes thus represent different scattering regimes, each providing new unique possibilities to study few-body collision physics and chemical processes from an entirely new perspective.

Much of the research on cold molecular collisions is based on tuning the long-range dipole-dipole interaction between molecules with external fields. The dipole-dipole interaction is often expressed in the molecule-fixed coordinate frame as

\begin{eqnarray}
{V}_{\mathrm{dd}}(\mathbf{r}) = \left ( \frac{1}{r^3} \right ) \left \{ {\mathbf{D}}_\mathrm{A} \cdot {\mathbf{D}}_\mathrm{B} - 3 ({\mathbf{D}}_\mathrm{A} \cdot \hat{e}_r ) (\hat{e}_{\rm r} \cdot {\mathbf{D}}_\mathrm{B}) \right \},
\label{eqn:dipole}\,,
\end{eqnarray}
where ${\mathbf{D}}_\mathrm{A}$ and ${\mathbf{D}}_\mathrm{B}$ are the dipole moment operators of the two interacting molecules A and B.  In Eq.~(\ref{eqn:dipole}) ${r}$ is the center-of-mass separation between the molecules, and $\hat{e}_r$ denotes a unit vector in the direction of $\mathbf{r}$. While this expression is simple, it does not show how the dipole-dipole interaction can be modified with electric fields. In order to analyze the effects of external fields on the intermolecular interaction, it is necessary to rewrite Eq.~(\ref{eqn:dipole}) in the space-fixed coordinate frame defined by the direction of an applied field. This leads to a spherical tensor representation of the dipole-dipole interaction~\cite{Zare88}
\begin{eqnarray}
{V}_{\mathrm{dd}} = -\frac{\sqrt{4\pi}}{\sqrt{5}}\frac{\sqrt{6}}{r^3} \sum_{q} (-1)^q{Y}^2_{-q}(\hat{e}_r) \left [ {\mathbf{D}}_{\rm A} \otimes {\mathbf{D}}_{\rm B} \right ]^{(2)}_q\,,
\end{eqnarray}
where ${Y}^2_{-q}(\hat{e}_r)$  is a spherical harmonic describing the orientation of the intermolecular axis in the space-fixed coordinate frame and $\left [ {\mathrm{D_{\rm A}}} \otimes \mathrm{{D_{\rm B}}} \right ]^{(2)}_q$ is a tensor product of two rank-1 tensors representing the dipole moment operators for molecules A and B. We emphasize that Eq. (2) is the same as Eq. (1) written in a different coordinate system.

It is most convenient to think about the dipole-dipole interaction between $^1\Sigma$ molecules as a matrix expressed in the basis of direct products $| J M_J \rangle_{\rm A} | J M_J \rangle_{\rm B} | l m_l \rangle$, where $| J M_J \rangle$ are the rotational wave functions of the isolated molecules and $| l m_l \rangle$ describes the rotational motion of the two-molecule complex as a whole~\cite{Krems04}. The wave functions $|lm_l \rangle$ represent the partial wave scattering states of two colliding molecules. The tensor ${Y}^2_{-q}(\hat{e}_r)$  acts only on the wave functions $| l m_l \rangle$ in these products and the integrals $\langle l m_l | {Y}^2_q | l m_l \rangle$  will average to zero, if $l=0$.  The reader must be familiar with the shape of the $d$ electron orbitals in the hydrogen atom, also described by the spherical harmonics of rank 2. The $s$-wave functions $|l m_l \rangle$ are independent of the spherical polar angles so the matrix elements $\langle l m_l | {Y}^2_q | l m_l \rangle$ are equivalent to the integrals of the spherical harmonics $Y^2_q(\theta, \phi)$ over the polar angles, which vanish.  That is why the expectation value of the dipole-dipole interaction between two ultracold molecules in an $s$-wave collision is zero. This has significant consequences for the collision dynamics of ultracold polar molecules. The energy dependence of elastic scattering is determined in the limit of small collision energies by Wigner's threshold laws~\cite{Wigner48}, which can be modified by long-range interactions such as the dipole-dipole interaction. That the dipole-dipole interaction in the $s$-wave collision channel averages to zero indicates that the threshold laws derived for systems with van der Waals interactions will also apply to $s$-wave collisions of polar molecules. That said, the dipole-dipole interactions may modify the dynamics of ultracold $s$-wave scattering by inducing couplings between $s$-wave and $d$-wave collision channels.

The dipole-dipole interaction will also vanish if the molecules are prepared in states with a particular angular momentum $J$.
The expectation value of the dipole-dipole interaction operator as defined in Eq. (2) can be written as
\begin{eqnarray}
\langle  {V}_{\mathrm{dd}}   \rangle  \propto \sum_{q} \langle l m_l | Y^2_q | l m_l \rangle \times \langle J M_J | {\mathbf{D}}_{\rm A} | J M_J \rangle_{\rm A}
\times \langle J M_J | {\mathbf{D}}_{\rm B} | J M_J \rangle_{\rm B}.
\end{eqnarray}
As explained in Sec.~\ref{ssec:structure}, the integrals $\langle J M_J | {\mathbf{D}}_{\rm A} | J M_J \rangle$ and $\langle J M_J | {\mathbf{D}}_{\rm B} | J M_J \rangle$ must vanish, because the integrands are the products of an even and an odd function. In the presence of electric fields, the molecular states correspond to linear combinations of the rotational states of different parity
and the expectation value of the dipole-dipole interaction operator becomes
\begin{eqnarray}
\langle  {V}_{\mathrm{dd}}   \rangle  \propto \sum_{q} \langle l m_l | {Y}^2_q | l m_l \rangle \times \left|\langle  {{D}}_{\rm A}  \rangle \right| \times \left|\langle  {{D}}_{\rm B}  \rangle\right|\,,
\end{eqnarray}
so it is non-zero, providing $l \ge 1$.
The collision physics of molecular dipoles at low and ultralow temperatures is very well described in this special issue in the contributions of Cavagnero and Newell~\cite{Cavagnero09} and Bohn, Cavagnero and Ticknor~\cite{Bohn09}. These papers provide detailed expressions for the matrix elements of the dipole-dipole interaction operator in the basis of molecular rotational states
and describe the scattering dynamics of molecular dipoles at low and ultralow temperatures.
In particular, this work  shows that collisions dynamics of molecular dipoles in the cold temperature regime exhibit a universal behavior well described by a semiclassical approximation.

The dipole-dipole interaction will also be non-zero if the molecules are confined by laser fields in one (1D) or two dimensions (2D) and subjected to an electric field. If the molecules are confined to move in low dimensions, the operator $Y^2_{-q}(\hat{e}_r)$ becomes a simple function of one angle (in 2D) or a multiplicative  factor (in 1D) and the integration involving this operator is restricted. The matrix elements of this operator will therefore be non-zero, even if the molecules are in the ultracold partial wave regime.

\subsubsection{Ultracold Temperature Regime.} Molecules at ultralow temperatures are characterized by unique interaction properties.  As mentioned above, a single unit of angular momentum in the rotational motion of the collision complex suppresses ultracold scattering. At the same time, inelastic scattering and chemical reaction processes at ultralow temperatures are dramatically enhanced by quantum threshold phenomena~\cite{Balakrishnan98,Balakrishnan01,Balakrishnan03,Hutson07, Bodo02, Flasher02, Florian04,Tilford04,Bodo04,Weck04,Weck05,Cvitas05,Cvitas05b,Ticknor05a,Mack06,Yang06,Lara06,Lara07,Gonzalez-Sanchez07,Gonzalez-Sanchez08}.
Several calculations and measurements have shown that the rates of chemical reactions in the limit of absolute zero may be extremely large.  For example, the work of Sold'{a}n {\it et al.}~\cite{Soldan02} showed that  the rate of the atom exchange reaction leading to vibrational relaxation in  Na + Na$_2$ collisions is about $10^{-10}$ cm$^3$/sec. The work of Balakrishnan and Dalgarno~\cite{Balakrishnan01} showed that the chemical reaction F + H$_2$ $\rightarrow$ HF + F proceeds at ultralow temperatures at the rate $\sim 10^{-12}$ cm$^3$/sec.
Mukaiyama and coworkers \cite{Mukaiyama04} observed that highly excited ultracold Na$_2$ molecules undergo inelastic relaxation in collisions with Na atoms or Na$_2$ molecules at the rates   $5 \times 10^{-11}$ cm$^3$/sec.
The measurements of Staanum {\it et al.}~\cite{Staanum06} and Zahzam {\it et al.}~\cite{Zahzam06} demonstrated that Cs + Cs$_2$ inelastic collisions in ultracold mixtures of atoms and molecules occur at the rates greater than $10^{-11}$ cm$^3$/sec. In the work of Zirbel \textit{et al}.~\cite{Zirbel08a,Zirbel08b}, atom-molecule inelastic collisions were studied where the molecule is a composite fermion. The inelastic collision rates can be tuned by varying the strength of the atom-molecule interactions. The differences are obvious when either atomic bosons or fermions are allowed to collide with the fermionic molecules in weakly bound vibrational states, namely enhanced loss rates when the atoms are bosons and suppressed loss rates when the atoms are fermions \cite{Petrov04}. Because the inelastic collisions and chemical reactions in ultracold gases are so efficient, ultracold molecules may offer unique opportunities to study chemistry in a completely new, previously unaccessible,  temperature regime, which may potentially lead to many fundamental discoveries.
For example, interactions of molecules in a Bose-Einstein condensate are determined by many-body dynamics and quantum statistics effects. Coherence of matter waves and interference of single partial-wave scattering states can be exploited to develop new schemes for quantum control of intermolecular interactions~\cite{Shapiro03}.
Moore and Vardi showed that the branching ratios for photodissociation of polyatomic molecules in a Bose-Einstein condensate may be dramatically modified due to Bose stimulation of collective many-body processes. This study may give rise to Bose-enhanced chemistry~\cite{Moore02}.  The effects of electromagnetic fields on ultracold molecules normally exceed the energy of their translational motion, even in a weak field limit. Collisions of ultracold molecules can therefore be controlled by static and laser fields through a variety of mechanisms. This can be exploited to study {\it controlled chemistry}.
The novelty and applications of cold controlled chemistry have recently been described in detail by one of us~\cite{Krems08}.

An illustrative example of how ultracold collisions can be manipulated by external fields can be found in the work of Volpi and Bohn~\cite{Volpi02}. Volpi and Bohn discovered that angular momentum transfer such as Zeeman or Stark relaxation in collisions of ultracold molecules is extremely sensitive to the magnitude of an external magnetic or electric field ~\cite{Volpi02}. Due to conservation of the total angular momentum projection on the external field axis, collisionally-induced Zeeman or Stark relaxation must be accompanied by changes of the orbital angular momentum associated with the rotation of the collision complex. This gives rise to long-range centrifugal barriers in the outgoing collision channels. The magnitude of the external field determines the splitting between the Stark and Zeeman levels and consequently the amount of the kinetic energy released as a result of the inelastic transitions. If the kinetic energy is larger than the maximum of the long-range centrifugal barriers, the inelastic transitions are unconstrained and efficient. However, at low external fields the centrifugal barriers in the outgoing channels suppress inelastic scattering. Zeeman and Stark relaxation leads to a loss of molecules from magnetic and electrostatic traps. As argued by Tscherbul {\it et al.} in this Special Issue~\cite{Tscherbul09}, the centrifugal barriers in the outgoing collision channels may stabilize a molecular gas of $^3\Sigma$ molecules in shallow magnetic traps and enable evaporative cooling of molecules to temperatures below 1 mK.

The suppression mechanism discovered by Volpi and Bohn can be exploited to explore the effects of external space symmetry on binary inelastic collisions of ultracold molecules in optical lattices, (see Sec.~\ref{sec:current}
and Danzl {\it et al.}~\cite{Danzl09} for discussions of molecules trapped in optical lattices). Consider, for example, a gas of ultracold spin-$1/2$ $^2\Sigma$ molecules confined by optical laser fields to move in two dimensions (2D)~\cite{Bloch05}. Collisions of molecules in 2D are characterized by the rotational angular momentum projection $m_l$ on the quantization $z$-axis directed perpendicular to the plane of the motion.
$S$-wave collisions correspond to $m_l=0$.  Suppose the electron spin of the molecules is aligned by a magnetic field directed along the $z$-axis, so that $m_s = +1/2$ for each molecule.  Then the projection of the total electron spin $S$ of the two-molecule collision system along $\hat{z}$ is $m_S = 1$ before the collision.   Now consider inelastic spin-relaxation collisions, such that after the collision $m_S = -1$. Due to the cylindrical symmetry of the collision problem, the sum of $m_S$ and $m_l$ cannot change, so $m_l$ must change from $0$ to $2$. This must give rise to centrifugal barriers in the outgoing collision channels and suppress the collision processes leading to spin relaxation, as described by Volpi and Bohn.

The symmetry of the problem will dramatically change if the magnetic field axis is rotated with respect to the confinement plane normal \cite{Li08}. In this case, the electron spin is no longer projected on the quantization axis. The Zeeman states can be written in terms of the spin states projected on the $z$-axis as follows:
\begin{eqnarray}
|1/2 \rangle_B = \cos{(\gamma/2)} |1/2 \rangle_z - \sin{(\gamma/2)} | -1/2 \rangle_z
\nonumber
\\
| -1/2 \rangle_B = \sin{(\gamma/2)} | 1/2 \rangle_z + \cos{(\gamma/2)} | -1/2 \rangle_z
\end{eqnarray}
\noindent
where the subscripts indicate the axis of projection and $\gamma$ is the angle between the magnetic field axis and the confinement plane normal. The Zeeman levels are thus superpositions of different projection states in the coordinate system defined by the confinement, and transitions from the Zeeman state $|1/2 \rangle_B$ no longer have to change the orbital angular momentum $m_l$.  The
Zeeman transitions in ultracold collisions of molecules in states with maximum spin projections on the magnetic field axis must therefore be enhanced if the magnetic field axis is directed at a non-zero angle with respect to the confinement plane normal. An experimental measurement of the Zeeman relaxation efficiency in such a system would probe the interactions coupling states of different total angular momentum projections, which can not be achieved in a thermal gas.

Studies of ultracold molecules in confined geometries may result in several other fundamental applications. The energy dependence of cross sections for molecular collisions in confined and quasi-confined geometries is different from Wigner's threshold laws in three dimensions~\cite{Sadeghpour00,Petrov01,Li08}. Chemical reactions and inelastic collisions of molecules under external confinement must therefore be modified. External confinement also changes the symmetry of long-range intermolecular interactions. As explained above, the dipole-dipole interaction averaged over the scattering wave function of polar molecules vanishes in the limit of ultracold $s$-wave scattering in three dimensions, but remains significant in two dimensions. The collision properties of ultracold molecules confined in 2D must therefore be very different for those in an unconfined 3D gas. Measurements of molecular collisions in confined geometries may thus provide a sensitive probe of long-range intermolecular interactions and quantum phenomena in collision physics. Confining molecules in low dimensions may be a practical tool for increasing the stability of ultracold molecular gases~\cite{Li08,Li09}. Finally, measurements of chemical reactions in confined geometries may be a novel approach to study stereodynamics and differential scattering at ultralow temperatures.  These effects should be observable~\cite{Li09} with molecular ensembles confined to oscillate in one or two dimensions in the ground state of a harmonic potential induced by optical lattices, i.e. when the translational energy of the molecules is much smaller that the oscillation frequency in the confinement potential. Experiments with ultracold molecules on optical lattices may therefore lead to the development of a new research direction unraveling ultracold chemistry in confined geometries.

The duration of an ultracold collision is very long.  The scattering dynamics of ultracold atoms and molecules is therefore sensitive to weak interactions between transient multipole moments of the collision system and an external field.  The results of Refs.~\cite{Krems06} and~\cite{Li07}, for example, show that collisions of ultracold atoms can be controlled by DC electric fields with magnitudes that do not perturb the isolated atoms.  When two different atoms collide, they form a heteronuclear collision complex with an instantaneous dipole moment.  The dipole moment function of the collision complex is typically peaked around the equilibrium distance of the diatomic molecule in the vibrational ground state and quickly decreases as the atoms separate. Only a small part of the scattering wave function samples the interatomic distances, where the dipole moment function is significant and the oscillatory structure of the scattering wave function diminishes the interaction of the collision complex with external electric fields. Collisions of atoms are therefore usually insensitive to DC electric fields of moderate strength ($<$ 200 kV/cm).   At the same time, the interaction with an electric field couples states of different orbital angular momenta. The zero angular momentum $s$-wave motion of ultracold atoms is coupled to an excited $p$-wave scattering state, in which the colliding atoms rotate about each other with the angular momentum $l=1 \hbar$. The probability density of the $p$-wave scattering wave function at small interatomic separations is very small and the coupling between the $s$- and $p$- collision states is suppressed. However, the interaction between the $s$-wave and $p$-wave scattering states is dramatically enhanced near scattering resonances. This may be
used to manipulate ultracold collisions and chemical reactions of molecules.  Experimental studies of ultracold atomic and molecular systems near scattering resonances in the presence of superimposed electric and magnetic fields may thus provide a unique probe of subtle intermolecular interactions that cannot be measured with thermal ensembles of molecules or accurately computed with modern \emph{ab initio} techniques.

Research on ultracold atomic gases has been revolutionized by the discovery of magnetically tunable Fano-Feshbach resonances~\cite{Kohler06,Chin09b,Feshbach92,Tiesinga93,Inouye98,Courteille98,Donley02,Herbig03,Regal03b,Zwierlein03,Jochim03,Greiner03,Werner05,Regal04}. Tscherbul {\it et al.}~\cite{Tscherbul09} showed that molecule-molecule scattering at ultralow temperatures is likewise dominated by magnetically tunable zero-energy scattering resonances. The density of Fano-Feshbach resonances in  molecule-molecule collisions is generally much larger and the properties of the resonances dramatically change with internal excitation of molecules. Measurements of the positions and widths of Fano-Feshbach resonances in atomic gases  are routinely used to construct accurate interaction potentials governing the collision dynamics of atoms at ultralow temperatures~\cite{Schunck05,Wille08,Li08b}. Measurements of Fano-Feshbach resonances in molecule-molecule collisions may similarly provide rich information for adjusting intermolecular interaction potentials to reproduce collision observables at ultralow temperatures. The paper of Tscherbul {\it et al.}~\cite{Tscherbul09} calls for experimental studies of magnetically tunable Fano-Feshbach resonances in molecule-molecule scattering.  As the interaction strength between molecules is typically much stronger than the energy of rotational excitation, collision dynamics of molecules must also be influenced by rotational Feshbach resonances~\cite{Bohn02}. Rotational Fano-Feshbach resonances may be tuned by electric fields~\cite{Tscherbul07}, which means that similar types of control should be accessible even to closed-shell species.

Ultracold molecular gases may be ideal for studies of quantum state resolved chemical encounters and the development of novel mechanisms for optical-field control of chemical reactions. In particular,  experiments with cold molecular gases may probe a new class of laser-induced bi-molecular chemical reactions with small endoergicity (absorption of energy).  For example, the chemical reaction of RbCs molecules in the ground rovibrational state $2 {\rm RbCs}  \rightarrow  {\rm Rb}_2 + {\rm Cs}_2$ is endothermic by about $40$ Kelvin.
The RbCs molecules at ultralow temperatures must therefore be chemically inactive. The vibrational frequency of RbCs is about 71 K. Laser excitation of cold RbCs molecules to the first vibrationally excited state may therefore promote the chemical reaction. Tscherbul {\it et al.}~\cite{Tscherbul08} have recently shown that the chemical reaction of two RbCs molecules is barrierless so it must be efficient at low temperatures. RbCs molecules in the ground rovibrational states were created at milliKelvin temperatures by Sage {\it et al.}~\cite{Sage05}. Chemical reactions in cold gases may thus be induced by infrared or even microwave excitation, which may provide new opportunities to study the role of individual rotational and vibrational energy levels in bi-molecular chemical reactions.

Although most of the current research of collision physics at ultralow temperatures focuses on atoms and diatomic molecules, recent work shows that it may be possible to cool large polyatomic molecules to ultralow temperatures as well.
Barletta, Tennyson and Barker~\cite{Barletta09} present in this special issue an interesting proposal
to cool benzene molecules using noble gases as refrigerants.
The techniques described in this special issue by  Lu and Weinstein~\cite{Lu09},
Patterson, Rassmussen and Doyle~\cite{Patterson09}, Meek, Conrad and Meijer~\cite{Meek09}, Salzburger and Ritsch~\cite{Salzburger09}, Motsch {\it et al.}~\cite{Motsch09}, Parazzoli {\it et al.}~\cite{Parazzoli09}, Takase {\it et al.}~\cite{Takase09}, Tokunaga {\it et al.}~\cite{Tokunaga09}, and Narevicius, Bannerman and Raizen~\cite{Narevicius09} can all potentially be applied to produce ultracold polyatomic molecules.
The research field of ultracold chemistry, while still in its infancy, is  therefore projected to expand very rapidly in the next few years.

\subsubsection{Cold Temperature Regime.}

The lowest temperature in interstellar space in the present era is about 3 degrees Kelvin~\cite{Sahai97}. Studies of molecules in the cold temperature regime ($1$ mK - 2 K) thus bridge astrophysics and astrochemistry with the zero temperature limit.  Moreover, the cold temperature regime looks forward in time to an era when the cosmic microwave background has cooled due to the accelerating expansion of the universe.  Therefore this temperature regime of molecular dynamics is very interesting. As the temperature of a molecular ensemble is reduced, the Maxwell-Boltzmann distribution of molecular speeds narrows down. Energy-resolved scattering resonances may therefore have a dramatic effect on elastic and inelastic collision rates in a cold gas. For example, a single shape resonance in collisions of CaH molecules with He atoms at the collision energy $~0.02$ cm$^{-1}$ enhances the rate for collisionally-induced spin relaxation at $0.4$ K by three orders of magnitude~\cite{Krems03}.
Electromagnetic fields of moderate strength shift molecular energy levels by up to a few Kelvin, so molecules at low temperatures can be thermally isolated in magnetic~\cite{deCarvalho99,Campbell07,Sawyer08a,Weinstein98,Sawyer07,Campbell09}, electrostatic~\cite{Bethlem00,Veldhoven05,Veldhoven05b,Meerakker06b,Rieger05,Kleinert07,Kleinert07b} or laser-field traps~\cite{Grimm00}. The development of experimental techniques for trapping cold molecules has opened up exciting possibilities for new research in molecular physics.  For example, confining molecules in external field traps has been used for measuring radiative lifetimes of molecular energy levels with unprecedented precision~\cite{Meerakker05,Gilijamse07,Campbell06}. Magnetically trapped cold molecules have been used as a target for beam collisions, determining absolute collision cross sections and revealing evidence of quantum threshold scattering and resonant energy transfer between colliding particles~\cite{Sawyer08a}. Cooling molecules to sub-Kelvin temperatures allows for spectroscopic measurements with enhanced interrogation time~\cite{Veldhoven2004,Hudson06b}. At the same time, collisions of cold molecules are sensitive to fine and hyperfine interactions, unimportant at elevated temperatures. Scattering measurements at low temperatures may therefore reveal subtle details of molecular dynamics in a new regime of collision physics where intermolecular interactions can potentially be manipulated externally through a variety of mechanisms, in many instances different from the external field control mechanisms in the ultracold temperature regime described above.

Experimental studies of chemical reactions in external field traps at temperatures $\sim 0.2 - 1$ K are particularly interesting as a novel approach to explore cold controlled chemistry~\cite{Krems08, Tscherbul06b, Tscherbul08c}. When molecules are trapped, their magnetic or electric dipole moments are aligned by the confining field. Collisions and chemical reactions of trapped molecules are different from reaction processes in a thermal unconfined gas. The alignment of molecular dipole moments by the trapping field restricts the symmetry of the interaction potential in the entrance reaction channel and limits the number of adiabatically accessible states. This can be exploited to develop mechanisms for controlling chemical reactions with external fields~\cite{Krems08}.
Consider, for example, a chemical reaction between a $^2\Sigma$ diatomic molecule in the rotational ground state ({\it e.g.} CaH) and an atom with one unpaired electron ({\it e.g.} Na) in a $^2$S electronic state in a magnetic trap.  The interaction between a $^2\Sigma$ molecule and a $^2$S atom gives rise to two electronic states corresponding to the total spin values of the reactive complex $S=1$ and $S=0$. If the atom and the molecule are both confined in a magnetic trap, they are initially in the state with the total spin $S=1$. The interactions between $^2$S atoms and $^2\Sigma$ molecules in the triplet spin state are typically characterized by strongly repulsive exchange forces, leading to significant reaction barriers. The interactions in the singlet spin state $S=0$ are usually
strongly attractive, leading to short-range minima and insertion chemical reactions~\cite{Verbockhaven06}. In the absence of non-adiabatic interactions between the $S=0$ and $S=1$ states, chemical reactions of atoms and molecules with aligned magnetic moments should therefore be much slower than reactions in the singlet spin state.
The non-adiabatic coupling between the different electronic states of the A($^2$S) - BC($^2\Sigma$) collision complex may be induced by the spin-rotation interaction and the magnetic dipole-dipole interaction~\cite{Tscherbul06b,Abrahamsson07}. The latter is negligibly small and the $S=1 \leftrightarrow S=0$ transitions are determined by the spin-rotation interaction in the open-shell molecule. The spin-rotation interaction can be effectively manipulated with an external electric field~\cite{Tscherbul06b,Abrahamsson07,Tscherbul06}.

The development of experimental techniques for producing slow molecular beams with narrow velocity distributions~\cite{Bethlem99,Bochinski03,Tarbutt04,Heiner06,Maxwell05,Herschbach01,Deachapunya08,Patterson07}
has opened another exciting research direction to study molecular collisions in the cold temperature regime.  For example, colliding slow molecular beams with trapped molecules can be used for precision measurements of molecular scattering cross sections with high energy resolution. Using an electrostatic or magnetic guide to direct slow molecular beams in collision experiments with trapped molecules may be employed to study stereodynamics of cold collisions and the effects of molecular polarization on chemical reactions. Experiments with cold molecular beams may also be used to explore the effects of external electromagnetic fields on chemical dynamics at low temperatures.  Studies of low-temperature chemical reactions in the presence of external fields may elucidate several fundamental questions of modern chemical physics. As demonstrated in Refs.~\cite{Tscherbul06b, Tscherbul06}, electric fields may induce avoided crossings between states otherwise uncoupled. Measurements of cross sections for molecular collisions and chemical reactions as functions of the electric field magnitude may provide information on the role of avoided crossings and the associated geometric phase in chemical dynamics of molecules.  Measurements of chemical reactions at low temperatures, where molecular collisions are determined by a few partial waves, may elucidate the effects of interference between different scattering states and orbiting shape resonances on chemical dynamics.  Molecules cooled to low temperatures can be confined in unusual geometries~\cite{Crompvoets04,Heiner07}, which may be used to study the effects of complex gauge potentials on molecular structure and collisions.

Unlike at ultralow temperatures, collisions of cold molecules may exhibit differential scattering. Measurements of differential scattering cross sections are often used for the analysis of details, particularly the angular dependence, of intermolecular interaction potentials~\cite{Aquilanti99}. Unlike at thermal temperatures, collisions of cold molecules are determined by a very limited number of partial waves, usually less than 10. Tscherbul has shown in a recent study~\cite{Tscherbul08b} that the scattering wave functions of polar molecules corresponding to different partial wave states can be coupled by external electric fields. These couplings may significantly modify the dynamics of differential scattering at low temperatures, especially near shape resonances. Measurements of cross sections for differential scattering of cold molecules in the presence of electric fields may therefore provide a very sensitive probe of shape resonances and the shape of intermolecular interaction potentials.

\subsection{Precision Measurements}
\label{ssec:precisionmeasurements}

Ultracold atoms and molecules are ideal systems for high resolution spectroscopy, quantum measurements, and precision tests of fundamental laws of nature. The inherent advantages of operation in the ultracold regime arise from the ease of control of molecules and their readiness for quantum state synthesis. For example, the advent of ultracold atomic physics has brought us a new generation of high-accuracy atomic clocks~\cite{Ludlow08,Rosenband08}, quantum sensors~\cite{Fixler2007}, some of the most precise measurements for fundamental constants~\cite{Peters1999,Cadoret2008} and stringent constraints on their possible time-dependent variations~\cite{Fischer2004,Peik2004,Bize2005,Rosenband08,Blatt2008}, probes of the theory of general relativity~\cite{Salomon2001} and physics beyond the standard model of particles and fields~\cite{Amoretti2002}, and tests of fundamental quantum physics, including the structure of the vacuum, quantum statistics, and quantum measurement processes~\cite{Raimond2001,Mabuchi1999,McKeever2004}. With their complex and information-rich level structures, cold and ultracold molecules are poised to take on and further expand the role of cold atomic physics in precision measurement~\cite{Hudson2002,Kawall04,Hudson06b,Koelemeij2007}. This is fueled by the rapid and substantial progress made recently~\cite{Weinstein98,Bethlem99,Bochinski03,Doyle04,Sage05,Merakker08,Ni08,Danzl08} in the control of molecular degrees of freedom and the possibility to prepare molecules in a single quantum state.
The generation and control of ultracold molecules is finally getting ready to play a big role in precision measurement, and the rapid progress in this field truly represents continued advances of the modern measurement science. Precision measurement thus takes on a much broader context with the involvement of molecules.

The use of molecules at low temperatures facilitates precision measurement in several ways. Single state control of molecules allows the most accurate measurements, with potentially complete understanding of systematic errors. In most cases, the relevant measurement can be cast as the determination of a frequency shift $\Delta \omega$ between two sublevels in the system.  The shot-noise limit of frequency sensitivity for such a measurement is given by $\delta (\Delta \omega) = 1/(2 \pi \tau \sqrt{N})$, where $\tau$ is the coherence time for the measurement, and $N$ is the number of detected particles.  The use of \textit{cold} molecules has two obvious attractive features.  First, at finite temperature a large number of molecular internal states are populated, according to the Boltzmann distribution.  For example, at room temperature for a typical molecule with rotational constant $B \sim 10$ GHz, only $\sim \! 10^{-3}$ of the population resides in a single quantum level.  Hence cooling the internal degrees of freedom of the molecule can give a large increase in $N$.  Similarly, cooling the external (motional) degrees of freedom yields molecules at lower velocities, and hence provides the opportunity for longer coherence times $\tau$.  If the molecules are cold enough to be confined in a trap, further increases in $\tau$ could result. As the sample temperatures are lowered, weaker traps can be used, resulting in reduced perturbations to the measurement process.

Molecular systems possess rich structures of electronic, vibrational, and rotational energy levels, as discussed in Sec.~\ref{ssec:structure}. Molecules can thus provide a series of precise frequency or wavelength benchmarks from the microwave and infrared, to the visible spectrum. Historically, molecules played a decisive role in the development of laser physics and high-resolution nonlinear laser spectroscopy, allowing the first spectroscopic observation of photon recoil splitting~\cite{Hall1976}. Molecular systems also facilitated the early development of natural-resonance-based frequency standards and the first molecular clock was built in 1949 using microwave transitions in ammonia~\cite{Townes1951}. Optical frequency chains were constructed to measure molecular transition frequencies which, along with accurate wavelength metrology, led to the accurate measurement of the speed of light~\cite{Evenson1972}, an important fundamental constant. Many secondary molecular frequency standards were developed, providing important wavelength references over a large spectral range until recently when optical frequency combs were developed to allow distributions of high accuracy reference signals anywhere in the visible and infrared spectral domains~\cite{Ye01}. Interestingly, it is the need for higher spectral resolution and measurement accuracy that laid the seed for the first ideas on the use of cold molecules~\cite{Bagayev1989,Chardonnet1994,Ye1998}.

The rich structure and unique properties of molecules can enhance the effect of violation of certain types of discrete symmetries, compared to the analogous cases in atoms, thus making molecules invaluable for tests of fundamental physics. In the search for a permanent electric dipole moment of the electron~\cite{Kozlov2002,Hudson2002}, the measurement sensitivity can be greatly enhanced due to very large internal electric fields of polar molecules.  Diatomic molecules have also been proposed as good candidates for parity violation studies
\cite{Flambaum09,Kozlov1995,Flambaum2006,DeMille2008} due to the enhanced sensitivity of rovibrational spectra to nuclear effects. Furthermore, when electronic and vibrational transitions of a molecule are probed precisely at the same time, one is essentially comparing clocks built from two fundamentally different interactions, one of the origin of quantum electrodynamics, the other of strong interaction. Such cross-system comparisons can be very useful for precision tests of possible time variations of fundamental constants~\cite{Flambaum2007}. Some astronomical measurements indicate that fundamental constants in the early universe may have been different  by as much as 10$^{-5}$ fractionally from the current values, which bears profound implications for cosmology and fundamental physics~\cite{Webb2001,Quast2004,Reinhold2006}. The time-dependence of the fine structure constant and some other fundamental constants can be explored with high accuracy clocks based on molecular transitions that are observable from distant galaxies. It turns out that in the vast inter-galactic space the universe holds a lot of cold molecules, albeit at temperatures of a few Kelvin.

\subsection{Many Body Physics}
\label{ssec:manyBody}

Many body physics with ultracold molecules takes on a different character at different phase space densities.  We divide many-body phenomena into three regimes: classical, semi-classical, and quantum.  These regimes are influenced by degeneracy and reduced dimensionality, as we will explain.  We define $T_{\mathrm{BEC}}$ as the critical temperature for Bose condensation and $T_F$ as the Fermi temperature.  Since these critical temperatures are functions of density, the \emph{quantum degeneracy} $T/T_{\mathrm{BEC}}$ ($T/T_F$) for bosons (fermions) is a measure of phase space density.

\subsubsection{Classical and Semi-Classical Regimes.}
\label{sssec:semiclassical}

The classical regime occurs strictly for $T/T_{\mathrm{BEC}}>1$ ($T/T_F>1$) for bosonic (fermionic) molecules.  In this regime collective excitations of the mean field can be derived explicitly, as has been done for cold atoms~\cite{Guery-Odelin00,Guery-Odelin02}.  Certain of the collective excitation modes and frequencies change as the system becomes quantum degenerate, whether the molecules are bosonic, fermionic, or a mixture thereof~\cite{Cohen-Tannoudji01}.  The full description of the collective modes of molecules above the critical temperature remains an open problem, as does the interface between these modes and the internal molecular degrees of freedom.

Also in the classical regime, one can study the dynamics of ultracold molecular plasmas. The fundamental physics of strongly coupled Coulomb systems underlies a great number of natural and anthropogenic phenomena, ranging from the formation of stars to fusion dynamics and to the plasma processing of nanomaterials. High-temperature plasmas have strongly coupled Coulomb fluid-dynamical properties due to their high densities. The experiments with ultracold atoms and molecules are making it possible to approach the limit of strongly coupled plasmas under highly rarefied conditions. Ultracold neutral plasmas were first created with atoms~\cite{Killian99,Robinson00,Killian07} and recently with molecules~\cite{Morrison08}. Such ultracold plasmas integrate molecular and mesoscopic domains, thereby opening a number of important many-body multiscale problems to fundamental experimental study. For example, ultracold Rydberg gases resemble rarefied amorphous solids. With the excitation of constituent molecules to higher electronic states, such a system can evolve to form an electron-correlated ultracold plasma. The rotational and vibrational degrees of freedom may provide new interaction mechanisms in such a system as well as new degrees of freedom to probe and control the dynamics of ultracold plasmas.

In the semi-classical regime, which occurs for $T/T_{\mathrm{BEC}}<1$ or $T/T_F<1$, one finds either a BEC (BEC) for bosonic molecules, or a Fermi liquid for fermionic molecules.  The bosonic case has been considered in great depth already in the context of the dipolar Bose gas~\cite{Baranov08}; the fermionic case less so.  However, these studies have focused on the long range and anisotropic properties of the dipole-dipole interaction; for the most part they have not considered the rotational, vibrational, and other internal modes innate to heteronuclear polar molecules~\cite{Baranov08}, as described in Sec.~\ref{ssec:structure}.

The simplest mean field picture for a molecular BEC is that of a dipolar Bose gas.  A mean field description is useful when the gas is dilute: $\sqrt{\bar{n} a^3}\ll 1$, where $a$ is the s-wave scattering length and $\bar{n}$ is the average molecular density, as has been derived rigorously for contact interactions alone; and $\sqrt{\bar{n} a_d^3}\ll 1$, where
\begin{equation}
a_d \equiv \frac{\hbar^2}{m d^2}\,.
\end{equation}
is the \emph{dipole-dipole interaction length}, and $m$ is the molecular mass, a diluteness criterion we state by analogy (we are not aware of a precise derivation in the literature along the lines of the Lee-Yang formalism~\cite{Lee57a,Lee57b}).
Such a dilute dipolar gas is described by a nonlocal Gross-Pitaevskii, or \emph{nonlocal nonlinear Schrodinger equation} (NNLS equation), first derived for dipolar gases by Yi and You~\cite{Yi00}, and then developed in a series of papers over the last eight years~\cite{Santos00,Baranov08}.  The NNLS equation takes the form
\begin{equation}
\left[-\frac{\hbar^2}{2m}\nabla^2 +V^{\mathrm{trap}}(\vec{r},t) + g|\psi(\vec{r},t)|^2
+ \int_{\mathbb{V}} d\vec{r}\,'\,V_{\mathrm{dd}}(\vec{r},\vec{r}\,')|\psi(\vec{r}\,',t)|^2\right]\psi(\vec{r},t)=i \hbar \frac{\partial}{\partial t}\psi(\vec{r},t)
\label{eqn:nnls}\,,
\end{equation}
where the dipole-dipole interaction $V_\mathrm{dd}(\vec{r},\vec{r}\,')\propto D^2/|\vec{r}-\vec{r}\,'|^3$ was defined in Eq.~(\ref{eqn:dipole}) in Section~\ref{ssec:chemistry}, except that here we take the average value of the operator $V_{\mathrm{dd}}$.
The NNLS appears in a number of other fields, including optically nonlinear media~\cite{Rotschild2005}.  The order parameter for Bose condensation is given by $\psi(\vec{r},t)$, which contains density and velocity information for the molecular gas: $n(\vec{r},t)=|\psi(\vec{r},t)|^2$ and $\vec{v}(\vec{r},t)= (\hbar/m)\vec{\nabla}\mathrm{Arg}[\psi(\vec{r},t)]$; we have normalized the order parameter to the number of molecules $N$, $\int_{\mathbb{V}}d\vec{r}\, n(\vec{r},t)=N$.  This leads to a quantum hydrodynamic description~\cite{Fetter01} as an inviscid classical fluid with quantum pressure and non-local interactions.  The trapping potential energy $V^{\mathrm{trap}}$ is typically harmonic or periodic, and may depend on time, as indicated; trapping technology is discussed in detail in Sec.~\ref{sec:current}.  In addition to the dipole-dipole interactions, contact interactions give rise to a local nonlinearity with coefficient $g\equiv 4\pi\hbar^2 a / m$, where $a=a(d)$ (see below).

The relative effect of contact and dipole-dipole interaction terms can be quantified in terms of lengths by comparing $a$ to $a_d$.  When $a_d$ is large compared to $a>0$, an untrapped, initially uniform dipolar Bose gas collapses in three dimensions~\cite{Santos00,Lushnikov02,Yi02}, as can be determined by Boguliubov theory~\cite{Boguliubov47}, i.e., linear stability analysis of Eq.~(\ref{eqn:nnls}); for small $a_d$, the system is stable.  Although a precise stability criterion $a_d < 3 a$ has been derived, this does not take into account possible many body tunneling effects or \emph{nonlinear} instabilities.  Therefore it can be taken as an approximate criterion.  For attractive contact interactions $a<0$ in a uniform system, collapse occurs in three dimensions under all circumstances (see Ref.~\cite{Sulem99} and references therein).  However, the introduction of a harmonic trap in 3D leads to metastability, for both contact and dipole-dipole interactions.  Metastability for attractive contact interactions has been studied extensively~\cite{Ruprecht95,Saito02}: the main result is that when the attractive interactions are sufficiently weak the system is stable against collapse on experimental time scales, while for stronger attractive contact interactions the BEC implodes in a ``Bose-nova''~\cite{Sackett99,Roberts01,Donley01}.  In the process of collapse the diluteness criterion breaks down and a higher order quantum theory is required, of which variations on Hartree-Fock-Boguliubov theory offer a first alternative~\cite{Zaremba99,Morgan99,Griffin09}.

In contrast, repulsive contact interactions cause the BEC to spread out.  Sufficiently large repulsive interactions $a>0$ lead to a Thomas-Fermi description, in which the quantum pressure, i.e., the term proportional to $\nabla^2$ in Eq.~(\ref{eqn:nnls}), is neglected~\cite{Dalfovo99}.  What is the effect of dipole-dipole interactions under such circumstances?  First, $a=a(d)$, so the shape of the molecular cloud changes.  This is because of coupling between different scattering channels determined by angular momentum summation rules; specifically, $V_{\mathrm{dd}}$ modifies the short-range part of the intermolecular potential in the $\ell=0$ channel, causing the effective $s$-wave scattering length to decrease.  If the dipoles are prevented from lining up head to tail, through use of a trap and suitable polarizing fields, and there is a sufficiently small number of molecules, then the dipole-dipole interactions remain repulsive, and there is no collapse.  However, if one or more of these constraints are lifted, the dipole-dipole interaction becomes attractive, and the system is unstable to collapse.  Since the dipole strength for polar molecules is a function of the external DC electric field vector, all of these effects are tunable. Dynamic variation of the dipole strength via combined DC and AC electric fields provides further tunability via averaging~\cite{Giovanazzi02}.  A common set-up often discussed in the literature is an axisymmetric harmonic trap of form
\begin{equation}
V^{\mathrm{trap}}(\vec{r})=\frac{1}{2}m(\omega_\rho^2 (x^2 + y^2) + \omega_z^2 z^2)\,,
\label{eqn:axisymmetric}
\end{equation}
and a DC electric field in the $z$ direction.  Then it can be shown through analysis of Eq.~(\ref{eqn:nnls}) that the dipoles cause mildly prolate clouds to be more elongated in the $z$ direction than the trap length ratio $\ell_z/\ell_\rho$ would warrant, where $\ell_i\equiv \sqrt{\hbar/m\omega_i}$, $i\in\{\rho,z\}$.  Moreover, in certain regimes the cloud is not ellipsoidal at all, but bi-concave, a shape reminiscent of a red blood cell~\cite{Ronen07}; even more exotic shapes can occur in non-axisymmetric traps~\cite{Dutta07}.  Thus the intuitive idea that the trap shape should be compressed along the electric field direction, where dipoles tend to line up and are therefore head to tail, is not borne out in the semiclassical many body theory.

Beyond the mean field ground state, one can consider dynamics.  Dipolar BECs exhibit a novel feature in their collapse dynamics, which occurs in a $d$-wave form, as recently observed for chromium atoms~\cite{Lahaye08} utilizing a magnetic dipole moment and a Fano-Feshbach resonance to tune the contact interaction strength $g$ to be small.  In the stable regime, excitations of dipolar BECs can be studied by analyzing the Boguliubov-de-Gennes equations.  The lowest energy collective excitation is the breathing mode, the next lowest the quadrupole mode, etc.  If the BEC is confined in quasi-2D geometry there is a roton-maxon shape to the dispersion relation~\cite{Fischer06} like the well-known one for superfluid helium, except with the advantage that the roton-minimum depth can be adjusted freely.  By \emph{quasi}-2D we refer to squeezing of the mean field to prevent collective excitations in one spatial direction.  An actual 2D system is squeezed at the single molecule level.  In this latter case, as for an actual 1D system, the 3D mean field theory of Eq.~(\ref{eqn:nnls}) does not apply and one must be careful to consider confinement-induced resonances~\cite{Olshanii98,Sinha07}, among other possibilities.

The NNLS equation has another set of collective excitations which do not appear in linear perturbation analysis.  These nonlinear excitations are sometimes called \emph{emergent phenomena}. They have particle-like properties.  In quasi-1D one finds solitons, localized density peaks or dips that persist in time and collide elastically.  In higher dimensions one finds both solitons and vortices.  The nonlocal term in the NNLS modifies the well-known properties of such excitations from atomic BECs~\cite{Kevrekidis08}.  For example, bright solitons in quasi-2D are stabilized by the dipole-dipole interactions~\cite{Pedri05}, and can even modify collapse dynamics~\cite{Nath08}. In 3D dark solitons are stabilized~\cite{Nath08b}.  Similarly, vortices have modified properties: dipole-dipole interactions cause oblate traps to have a lower critical frequency for formation of a single vortex, while prolate traps have a higher one~\cite{O'dell07}. Under stronger rotation, but still in the semi-classical regime, multiple vortices enter the BEC; increasing dipolar interactions pushes the system through a series of different lattice structures, from triangular and square to stripe and bubble phases~\cite{Cooper05}.  Even the shape of the vortex cores becomes a function of the orientation of the dipoles~\cite{Yi06}.  Finally, exploration of the NNLS is fairly new, and there may be emergent particle-like objects which have not yet been identified, for example, in transitional geometries between 3D and quasi-2D/quasi-1D~\cite{Jones82,Brand02,Berloff04}.

Beyond harmonic traps, one can consider periodic traps in the form of optical lattices, as discussed for molecules in Sec.~\ref{sec:current}.  Optical lattices consist of standing waves between two opposing laser beams.  In three dimensions, arbitrary lattice structures can be created, including all crystal structures known from solid state theory~\cite{Marder00}.  An optical lattice applied in one dimension across a dipolar gas can be used to create an array of quasi-2D molecular gases, and applied in two dimensions an optical lattice makes an array of quasi-1D molecular gases; this is equally true for bosonic and fermionic molecules.  The emergent and other collective properties of dipolar bosons in optical lattices in the semi-classical limit of Eq.~(\ref{eqn:nnls}), though a topic hardly touched in the literature, is likely to have a rich phenomenology~\cite{Tikhonenkov08}.  The fully quantum theory of molecular gases in optical lattices is discussed in Sec.~\ref{sssec:fullyQuantum}.

We again emphasize that almost none of the semiclassical treatment of dipolar gases so far has moved beyond point particles with associated dipoles.  We can take a hint of what might occur for molecules from spinor atomic gases, which utilize the atomic hyperfine degree of freedom to create a \emph{pseudo-spin} manifold.  Pseudo-spin has the same operator algebra, and therefore the same observable physics, as spin.  In the case of molecules there is an even richer internal state structure, depending on the molecular species, as discussed in Sec.~\ref{ssec:structure}.  Spinor condensates are described by a vector generalization of the nonlinear Schrodinger equation; we expect a vector NNLS equation to describe molecular condensates in the same way.  Spinor condensates are a very rich area of exploration in atomic BECs, from spin waves in the classical regime~\cite{Mcguirk03} to spin textures~\cite{Mueller04} (vortices utilizing spin-space), domain formation~\cite{Higbie05}, new phases~\cite{Saito06}, and knots~\cite{Kawaguchi08} in the semi-classical regime, among many other phenomena.  We emphasize that in a mean field description the different internal states can interact coherently or incoherently.  We say the interactions are coherent when the relative phase of the order parameter describing the internal states enter the NNLS in a non-trivial way: for example, for an $F=1$ hyperfine manifold a term of the form $\psi_{-1}^*\psi_{+1}^*\psi_0$ appears in a vector nonlinear Schrodinger equation~\cite{Ho98}, where the subscript refers to $m_F$ in $|F,m_F\rangle$.  An incoherent term would be of form $|\psi_1|^2\psi_{0}$, for example.  The coherence of the interactions gives rise to substantially different phenomena, since in the incoherent case the number of molecules in each internal state is fixed, while in the coherent case atoms can redistribute themselves across internal states.

The theory of ultracold molecular Fermi gases has been much less thoroughly explored than the theory of bosons.  One principal difference between Fermi liquid theory in solid state materials and in dilute quantum gases is that in the latter the Fermi surface is essentially a free parameter, instead of material-specific.  So, in a harmonic trap, the Fermi surface can be manipulated~\cite{Jin99,Giorgini08} by changing the trap frequency, by adjusting the number of atoms, and by changing the strength and character of the interactions.  In an optical lattice, the system can be driven from conductor to insulator by adjusting the filling factor~\cite{Kohl05}.  Also, fermions with attractive interactions are more stable against collapse due to Fermi pressure.  All this was already known and demonstrated with atoms, and also applies to molecules.  What new features appear in dipolar and molecular systems?  First, because of the Pauli exclusion principle, only odd partial wave contact interactions are allowed for spin polarized fermions.  Therefore, in the spin-polarized dipolar Fermi gas only dipole-dipole interactions occur.  Second, below a critical temperature $T_{\mathrm{BCS}}<T_F$, $p$-wave Cooper pairing can occur in the polarized gas~\cite{Baranov2002}.  This pairing is anisotropic and has a critical temperature which is a function of the trap anisotropy~\cite{Baranov2004}, making BCS in trapped dipolar fermions significantly different from electrons in metals~\cite{Leggett75}.  One can see this already from the non-symmetric Fermi surface even above $T_F$~\cite{Miyakawa2008}.  One of the more exciting predictions for fermionic molecules is a ferroelectric Fermi liquid~\cite{Iskin07}.

Several advances in the semiclassical many body physics of ultracold molecules are presented in this special issue.  Klawunn and Santos~\cite{Klawunn09} demonstrate a second order phase transition in a vortex in a dipolar Bose gas in a 1D optical lattice, that is, in a series of oblate clouds speared by a vortex.  The rotonization of the Kelvin wave spectrum~\cite{Saffman92} of vortex line excitations leads to a helically distorted vortex.  Sogo \emph{et al.}~\cite{Sogo09} present a derivation of collective excitation frequencies of a polarized (one-component) dipolar Fermi gas in a harmonic trap.  They also show how the dipoles affect the expansion of the Fermi cloud when the trap is dropped, a key observation technique in experiments.  Metz \emph{et al.}~\cite{Metz09} explore collapsing dipolar BECs in Chromium, matching experiment to simulations of Eq.~(\ref{eqn:nnls}) in oblate and prolate geometries.  They demonstrate phase coherence by interfering multiple copies of collapsing BECs.  Because mean field semiclassical theory is only applicable to a phase-coherent BEC, this provides excellent justification for the use of Eq.~(\ref{eqn:nnls}).  Finally, Xu \emph{et al.}~\cite{Xu09} propose a completely new way to pump vorticity into a dipolar BEC.  The usual method is to rotate the system with an off-resonant laser beam, either by distorting an axisymmetric trap or by focusing two beams into the condensate and stirring.  Xu \emph{et al.} describe a modified Ioffe-Pritchard trap based on the interaction between the electric dipole and a spatially varying electric field, in analogy to the magnetic case already in use.  Adiabatic flipping of the axial bias of this trap pumps vorticity into the molecular BEC.

\subsubsection{Fully Quantum Regime.}
\label{sssec:fullyQuantum}

In this regime, semi-classical methods such as the NNLS equation are insufficient, because all molecules do not reside in a single mode or even in a few modes of the single particle density matrix.  There are two ways to achieve the fully quantum regime.  The first is by introducing massive degeneracy, either with an optical lattice potential or by rotating a harmonically trapped molecular cloud near the frequency of the confining potential.  In the case of a lattice potential, the degeneracy is spatial.  In the case of a rotating system, the degeneracy is rotational.  Both such systems are well known in condensed matter physics.  The second way to achieve the fully quantum regime is through strong interactions, violating the dilute gas criteria of Sec.~\ref{sssec:semiclassical}.  This can be accomplished via a Fano-Feshbach resonance.  In principle it is also possible to compress the molecular gas until it reaches a high density.  However, experimentally accessible ultracold molecular gases are metastable, as they prefer to be solid at such low temperatures.  Even if new traps could be designed, a high density would lead to three body processes which could nucleate a phase transition to a liquid or a solid.

Rather than rederiving the description of ultracold molecules in lattices from scratch, we begin with a minimal model of quantum lattice physics well known in condensed matter, called the \emph{Hubbard model}~\cite{Hubbard63,Fisher89}.  The Hubbard Hamiltonian in its simplest form in the canonical ensemble is given by
\begin{equation}
\hat{H}=-t\sum_{\langle i,j\rangle,\sigma} (\hat{a}^{\dagger}_{i\sigma} \hat{a}_{j\sigma}+\mathrm{h.c.})
+\frac{1}{2}\sum_{\sigma,\sigma'}U_{\sigma\sigma'}\sum_{i}\hat{n}_{i\sigma}\hat{n}_{i\sigma'}\,,
\label{eqn:hubbard}
\end{equation}
where all constants preceeding sums are in units of energy: $t$ represents hopping, or tunneling\footnote{Here $t$ does not represent time; the hopping energy is denoted as $J$ in some papers on ultracold atoms in optical lattices, but continues to be denoted as $t$ in the larger body of literature, including condensed matter journals.}; $U_{\sigma\sigma'}$ represents interactions; and we have neglected disorder.  The site indices are given by $i,j$, and $\langle i,j \rangle$ means the sum is only over nearest neighbors.  The indices $\sigma,\sigma'$ refer to an internal degree of freedom; originally, in the case of electrons, this was spin.  The ``hat'' symbol refers to the fact these are second quantized operators: $\hat{a}_{i\sigma}$ destroys a particle on site $i$ in internal state $\sigma$; $\hat{n}_{i\sigma}\equiv \hat{a}^{\dagger}_{i\sigma}\hat{a}_{i\sigma}$; and $[\hat{a}^{\dagger}_{i\sigma} ,\hat{a}^{\dagger}_{j\sigma'}]_{\pm}=\delta_{ij}\delta_{\sigma\sigma'}$ and $[\hat{a}_{i\sigma} ,\hat{a}^{\dagger}_{j\sigma'}]_{\pm}=[\hat{a}_{i\sigma} ,\hat{a}_{j\sigma'}]_{\pm}=0$ for fermions (+) or bosons (-), respectively.  The coefficients $t$ and $U_{\sigma\sigma'}$ are material parameters in solid state systems, but can be derived more directly from first principles for ultracold gases.  The coefficients $t$ and $U$ are calculated from overlaps of wavefunctions convoluted with kinetic and lattice potential energy for $t$ and interaction energy for $U$.  The Hubbard model is a useful description for low filling factor (average number of particles per site): if the filling becomes much larger than unity then a semiclassical theory can provide a better description of the system in the superfluid or supersolid phase.

\begin{figure}[t]
 \begin{center}
\includegraphics[scale=0.8]{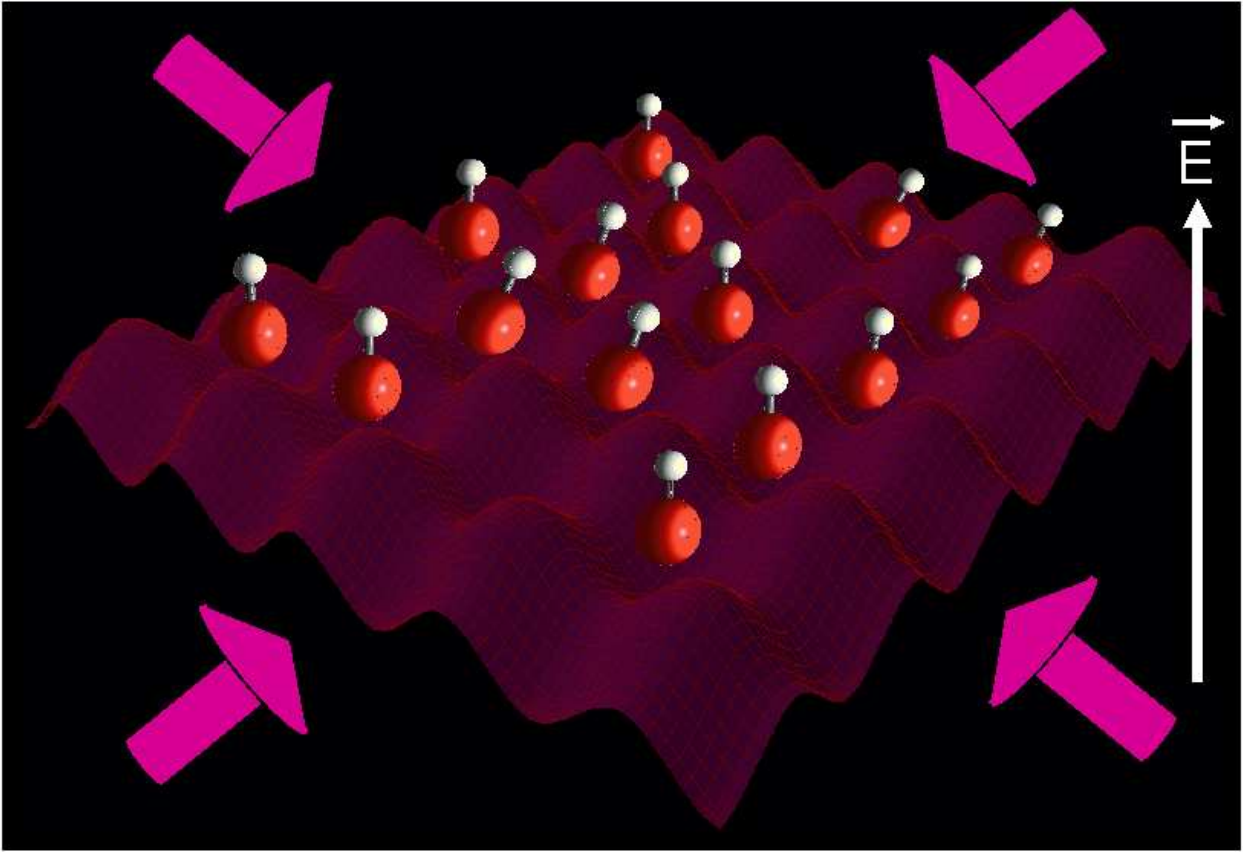}
\caption{Ultracold polar molecules in an optical lattice potential. The arrows represent opposing pairs of laser beams which interfere to make an optical standing wave. A low filling factor of one molecule per site is shown, which is the fully quantum regime discussed in Sec.~\ref{sssec:fullyQuantum}. All molecules, or a specific subset of molecules, can be addressed and their dipole moments are controlled. The strong and long-range interactions between molecules of different lattice sites will introduce a rich set of quantum phases, and allow the control of interesting dynamics needed for quantum simulators.}
\label{fig:lattice}
 \end{center}
\end{figure}

Equation~(\ref{eqn:hubbard}) represents a minimal model for two reasons. First, a lattice requires coupling between sites in order to not factorize trivially; nearest neighbor hopping is the minimal form of motion between sites.  In fact, the mean field limit of Eq.~(\ref{eqn:hubbard}) maps onto the kinetic energy, or second spatial derivative in the minimal discretized version of Eq.~(\ref{eqn:nnls}), and the hopping integral $t$ contains kinetic and trap potential energy for the case of ultracold molecules in optical lattices.  Second, a many body problem requires interaction to not reduce trivially to a product of single particle wavefunctions, and on-site interaction is the minimal form of such interaction.  Equation~(\ref{eqn:hubbard}) can be extended to include disorder, nearest-neighbor interactions, next-nearest-neighbor hopping, multiple bands, etc.  But even in its simplest form, Eq.~(\ref{eqn:hubbard}) remains unsolved for repulsive interactions~\cite{Micnas90}, $U_{\sigma\sigma'}>0$.  In fact, it is even proposed as a model for high-$T_c$ superconductivity.
The description by the Hubbard Hamiltonian has proven effective for cold quantum gases, where, for instance, the dynamics of quantum revivals has been both observed experimentally~\cite{Greiner02,Greiner02b} and described theoretically~\cite{Jaksch98,Lewenstein07} by a modified version of Eq.~(\ref{eqn:hubbard}).  Such complex quantum dynamics offer exciting possibilities of new research with ultracold molecules.  Moreover, significant experimental steps towards transferring trapped ultracold molecules into optical lattices are described in this special issue~\cite{Danzl09}. A sketch of polar heteronuclear molecules in an optical lattice is shown in Fig.~\ref{fig:lattice}.

A substantial difference between polar molecules and atomic systems is the possibility for strong correlations due to the long range nature of the dipole-dipole interaction.  A point not often remarked on in the literature is that this is only true in 3D.  Therefore, when speaking of a 1D or 2D system, we clarify that, if we wish for the interactions to remain long-range, the transverse trapping lengths should be much larger than $a_d$.  In a truly 2D or 1D system, the dipole-dipole interaction is not long range and can be folded into the contact interaction.  We reserve the terms quasi-2D and quasi-1D for lattice systems in which the dipole-dipole interaction has the possibility to remain long-range.

The first paper to appear on dipolar Bose gases in the fully quantum regime~\cite{Goral02} found the same phases in quasi-2D as the extended Bose Hubbard Hamiltonian: Mott insulator, checkerboard insulator, superfluid, and supersolid, as well as a collapse regions in the phase diagram.  However, these phases depend on the ratio of the effective on-site trapping frequencies, as well as the direction of polarization of the dipole.  Therefore a novel aspect of dipolar Bose gases is that squeezing a quasi-2D $x$-$y$ square lattice in the $z$ direction drives them through quantum phase transitions~\cite{Sachdev99}.  Later study of the same system indicates a multitude of nearly degenerate excited states, making the dipolar boson system in the lattice effectively disordered~\cite{Menotti07}. Use of molecules in an optical lattice for construction of a quantum simulator, i.e., a system designed to reproduce known but unsolvable condensed matter Hamiltonians~\cite{Feynman82,Pupillo08}, will be discussed in Section \ref{ssec:quantumInfo}.

Beyond optical lattices, polar molecules can self-assemble into a crystalline structure if their long-range interactions are made to be sufficiently repulsive.  This is the strong-interaction route to the fully quantum regime (as opposed to degeneracy).  Then the lattice concept is closer to that of solid state systems, because the molecular crystal displays phonon modes.  This can occur in quasi-1D, quasi-2D, and 3D~\cite{Gorshkov08}.  An atom-molecule mixture in which the crystallized molecules play the role of ions and leftover atoms and molecules play the role of electrons has been proposed~\cite{Pupillo08b}.  This is a natural situation to consider because transfer efficiency under a Fano-Feshbach resonance from atoms to molecules is not 100\%, as discussed in Sec.~\ref{ssec:photo}.

Finally, we briefly discuss an alternative route to a fully quantum system, namely, strong rotation of a dipolar gas in a harmonic trap at a rate near the trap frequency (see the work of Xu {\it et al.}~\cite{Xu09} in this special issue for a new route to achieve this goal).  The basic idea is that rotation pushes the single particle states of the harmonic potential until they are degenerate or nearly degenerate. Then ultracold quantum gases in such a trap, whether atomic or molecular, map onto the fractional quantum Hall effect~\cite{Bhongale04,Cooper04}.  In these rotating systems the filling factor is given by the number of atoms per vortex.  At exactly half filling, and in the absence of dipole-dipole interactions, bosons are in a bosonic analog of the Laughlin state, a highly correlated state which can be written down exactly as a product wavefunction.  As the dipole-dipole interactions are increased, the system reverts to stripe and bubble phases which can be described by a mean field theory~\cite{Cooper05}.  In the strongly interacting regime, where the lowest Landau level approximation is no longer valid, the basic features of bubble and stripe phases persists; triangular and square lattices also occur, as well as a collapse region in the phase diagram~\cite{Komineas07}.  These results depend on the trap ratio, as is generically the case for dipolar gases.  For fast-rotating fermions, an additional non-harmonic confining potential is required to stabilize the system.  Then at one third filling and in the lowest Landau level approximation, one again obtains the (originally antisymmetric) Laughlin wavefunction~\cite{Baranov2005,Osterloh2007}.  For sufficiently small filling factor, the dipole-dipole interactions give rise to a Wigner crystal phase.  Increasing the filling pushes the dipolar gas through a quantum phase transition to a Laughlin liquid~\cite{Baranov2008b}.

There are three contributions in this special issue on the fully quantum regime.  First, a complementary question to ``How can ultracold molecules in optical lattices be used as a quantum simulator?'' is, ``What is the minimal lattice Hamiltonian expected for near-term experiments with ultracold molecules in optical lattices?''  Such a minimal set up should involve a uniform polarizing DC electric field in order to create the dipoles, and possibly a uniform AC electric field to drive the system, as well as of course an optical lattice in at least one direction.  As an answer to this question, the Molecular Hubbard Hamiltonian (MHH) has been presented in this special issue by Wall and Carr~\cite{Wall09}:
\begin{eqnarray}
\hat{H}&=&-\sum_{JJ'M}t_{JJ'M}\sum_{\langle i,i'\rangle}\left(\hat{a}_{i',J'M}^{\dagger}\hat{a}_{iJM}+\mbox{h.c.}\right)\nonumber\\
&&+\sum_{JM}E_{JM}\sum_i\hat{n}_{iJM}-\pi\sin\left(\omega t\right)\sum_{JM}\Omega_{JM}\sum_{i}\left(\hat{a}_{iJ,M}^{\dagger}\hat{a}_{iJ+1,M}+\mbox{h.c.}\right)\nonumber\\
&&+\frac{1}{2}\sum_{\tiny{\begin{array}{c} J_1,J_1',J_2,J_2'\\ M,M'\end{array}}}U_{dd}^{\tiny{\begin{array}{c} J_1,J_1',J_2,J_2'\\ M,M'\end{array}}}\sum_{\langle i,i'\rangle}\hat{a}_{iJ_1M}^{\dagger}\hat{a}_{iJ_1'M}\hat{a}_{i'J_2M'}^{\dagger}\hat{a}_{i'J_2'M'}.
\label{eqn:molecularHubbard}
\end{eqnarray}
In Eq.~(\ref{eqn:molecularHubbard}) the hopping $t$ depends on rotational modes $|J M\rangle$ due to use of a dressed (rotational plus DC electric field) basis, and the dependence of the optical lattice potential energy on the molecular polarizability has been taken into account.  The other three terms in Eq.~(\ref{eqn:molecularHubbard}), in order, represent the DC electric field, the AC electric driving field, and the dipole-dipole interactions, which have been taken as hard core.  When used to describe bosons, Eq.~(\ref{eqn:molecularHubbard}) can be viewed as a modified, multiband, driven, extended Bose Hubbard Hamiltonian; with no driving and only a single rotational state occupied, it is exactly the extended Bose Hubbard Hamiltonian~\cite{Kuhner98}.  We present Eq.~(\ref{eqn:molecularHubbard}) to give one example of how much more complicated MHHs are compared to the original Hubbard model of Eq.~(\ref{eqn:hubbard}).  Thus advanced simulation techniques, as described in Sec.~\ref{ssec:numericalMethods}, can be required~\cite{Vidal04} in order to understand experiments on these systems and yield new insight into complex quantum dynamics and emergent properties.

The second contribution is by Ortner {\it et al.}~\cite{Ortner09}.  In previous work, atom-molecule mixtures were described as dipolar crystals in which the extra particles moving in the crystal are dressed by crystal phonons, leading to a polaron picture.  In this special issue, Ortner {\it et al.} review this idea and show how it can be derived in the master equation context~\cite{Gardiner97,Jaksch97}, where the acoustic phonon modes of the crystal are treated as a thermal heat bath.  They provide a second example of a Hamiltonian for ultracold molecules, describing atoms moving in a dipolar crystal and coupled to crystal phonons.  In this Hamiltonian, phonons appear as operators explicitly coupled to atomic operators in the lattice.  Finally, the third contribution is from Roschilde {\it et al.}~\cite{Roschilde09}.  Quantum polarization spectroscopy is proposed as a sensitive probe of quantum phases, in particular the Fulde-Ferrell-Larkin-Ovchinnikov phase~\cite{Larkin64,Fulde64}, which is a BCS-like pairing between two different Fermi surfaces.  Quantum polarization spectroscopy is a non-destructive measurement that imprints quantum fluctuations on to light polarization.  For a balanced Fermi gas, Roschilde {\it et al.}~\cite{Roschilde09} address the question of how spin-spin correlations transform through the BCS to BEC crossover; the crossover occurs in the preliminary stage of magneto-association of atoms into molecules into a highly excited vibrational state.

\section{State-of-the-art Technology in Experiments}
\label{sec:current}

To fulfill the goals of scientific explorations with cold and ultracold molecules, several major efforts have been undertaken to achieve large phase-space densities for ground-state molecules, especially polar ones. As explicitly discussed in the previous section and also shown in Fig.~1(b), a large phase space density, made possible with a large spatial density of molecules at ultracold temperatures, is a prerequisite for some research directions such as many-body physics and is a great benefactor to varying degrees for many aspects of our explorations including novel collisions and ultracold chemical reactions, quantum information science, and precision measurement.

The complex structure of molecular energy levels has so far precluded a straightforward extension of laser cooling to molecules, although several proposals have emerged for direct~\cite{Bahns1996,Rosa2004,Stuhl08} or cavity-enhanced laser cooling~\cite{Domokos02,Andre06,Morigi07,Lev08}. Instead, over the last dozen years, efforts for the production of cold and ultracold molecules have been focused on two general categories with successful results. The first is the direct manipulation of stable ground-state molecules via either sympathetic cooling (such as buffer gas cooling~\cite{Weinstein98}) or phase-space-conservative deceleration of molecular beams using external electric~\cite{Bethlem99,Bochinski03}, magnetic~\cite{Merkt2007,Raizen08}, or optical fields~\cite{Barker06}. The general concept of molecular deceleration is depicted in Fig.~\ref{fig:decelerator} while buffer gas cooling is schematically described in Fig.~\ref{fig:buffer}. Both techniques will be discussed in more detail in Section 3.1 below. The second widely used approach is to ``assemble'' ultracold molecules from pairs of ultracold atoms. When a dual-species atomic gas is used, the result is heteronuclear polar molecules~\cite{Wang04,Bigelow04,Sage05}. This technique builds on a strong tradition of photoassociation in ultracold atomic physics~\cite{Thorsheim87,Wang00,HeinzenScience2000,Dion01,Weiner99,Jones06}.

The results of these various approaches are summarized in Fig.~\ref{fig:schematic}. The direct methods have so far produced molecules only at low temperatures ($T \sim 10$ mK - 1 K), with no experimentally demonstrated routes to the ultracold regime. The phase space densities produced with these methods are generally limited to below 10$^{-12}$. The direct methods, however, are versatile and have produced a large diverse set of cold molecular species~\cite{Campbell07,Maussang05,Maxwell05,Merakker08,Sawyer07,Sawyer08a,Sawyer08b}. These methods could ultimately lead to superior samples of ultracold molecules for many applications, when combined with followup techniques such as laser cooling, or sympathetic or evaporative cooling, as shown in Fig.~\ref{fig:schematic}. Some other techniques that have produced lower-density cold molecules include counter-rotating supersonic nozzles~\cite{Herschbach01} and crossed molecular beam collisions~\cite{Chandler03,Takase09}, as described in Fig. \ref{fig:buffer}. In the single collision approach, two beams are crossed typically at around 90$^o$ and a fraction of collisions produce molecules nearly at rest.

The indirect cooling methods have focused mainly on bialkali molecules, both homonuclear and heteronuclear. These molecules are in the ultracold regime, usually about tens or hundreds of microKelvin. However, they are typically formed in states of high vibrational excitations. These states are susceptible to destructive inelastic collisions~\cite{Zirbel08a,Zirbel08b,Hudson08} and in addition have negligibly small electric dipole moments when they are heteronuclear; both aspects preclude the use of such molecules for many interesting applications. It has been shown that it is possible to transfer bialkali molecules into their absolute rovibrational ground state~\cite{Sage05}. More recent work further demonstrates that a highly efficient and phase-coherent transfer to the absolute ground state is possible, largely preserving the phase space density of the original atomic gas. The resulting dipolar gas of ground state KRb molecules is near quantum degeneracy~\cite{Ni08}. For homonuclear molecules, a similar coherent transfer approach has also been demonstrated for both Rb$_2$~\cite{Lang08} and Cs$_2$ molecules~\cite{Danzl08,Danzl09}.

\subsection{Direct Cooling Methods}
\label{ssec:direct}
\begin{figure}[t]
 \begin{center}
\vspace*{12mm}
\includegraphics[scale=0.7]{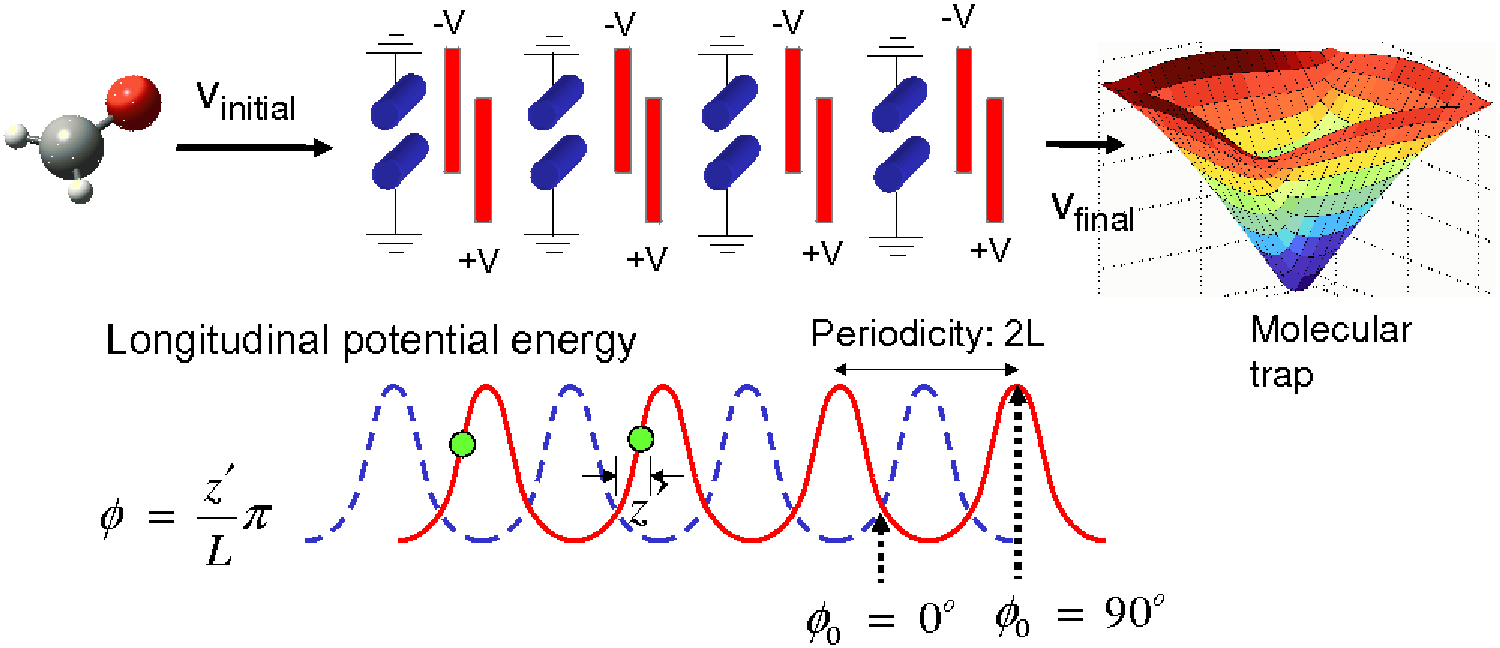}
\caption{A diagram of Stark-effect-based molecular deceleration via time-varying inhomogeneous electric fields arranged in a linear array. A Zeeman-effect-based decelerator employs the same operational principle, except it uses inhomogeneous magnetic fields.  Deceleration produces slow molecules in the laboratory frame, but it does not enhance the phase space density, hence the importance of injecting a dense and monochromatic molecular beam into the decelerator. Typically the incident beam is prepared via supersonic expansion to cool both the internal and external degrees of freedom. The spatially inhomogeneous fields (acting via either the Stark or Zeeman energy shifts of the traveling molecules) are switched synchronously with the position of the molecular beam along the deceleration path. The phase angle of 0$^o$ corresponds to phase-space bunching at a fixed packet velocity without deceleration or acceleration. Isolated packets of cold molecules are produced at the end of the decelerator at various deceleration phase angles. An increasing phase angle leads to greater deceleration, along with a reduced phase-stable area, which results in a decreasing number, and also a decreasing temperature, of cold molecules contained in the isolated packet. The complex dynamics governing the evolution of the molecules within the decelerator are now well understood. At the output of the decelerator, the slowed molecules can be loaded into an electric or magnetic trap. In current experiments researchers typically accelerate/decelerate a supersonic beam of molecules to a mean speed adjustable between 600 m/s to rest, with a translational temperature tunable from 1 mK to 1 K, corresponding to a longitudinal velocity spread from a few meters to a hundred meters per second. These velocity-manipulated stable packets contain 10$^4$ to 10$^6$ molecules (depending on the translational temperature) at a density of $\sim$10$^5$ to 10$^7$ cm$^{-3}$ in the beam and $\sim$10$^6$ cm$^{-3}$ in the trap.}
\label{fig:decelerator}
 \end{center}
\end{figure}
Decelerated molecular beams can be used for a variety of studies of molecular scattering and reactions at low energies, as well as for loading molecular traps. For the most efficient operation of any molecular deceleration experiments, whether using Stark~\cite{Bethlem99,Bochinski03}, Zeeman~\cite{Merkt2007,Raizen08}, or optical potentials~\cite{Barker06}, a well-characterized and relatively high phase-space density source of molecules in the ground state represent a critical initial condition. Supersonic expansion provides a direct and convenient approach towards this goal as it represents the only true cooling step during the entire deceleration process. The initial pulsed molecular beam injected into a decelerator is thus typically rotationally cold, with a narrow, boosted velocity distribution centered around a mean longitudinal velocity of several hundred meters per second~\cite{Lewandowski04}.

In both Stark and magnetic field decelerations, a narrow pulse of molecules is slowed by rapidly switching and dynamically controlled inhomogeneous electric or magnetic fields. In optical deceleration, laser beams forming the optical potential are switched~\cite{Barker06}.
In the paper by Kuma and Momose in this special issue, infrared lasers tuned to vibrational resonances are to be used to form deceleration optical potentials~\cite{Kuma09}.
As molecules prepared in their weak-field seeking states propagate into a region with an increasing field strength, their longitudinal kinetic energy is converted to potential energy; thus they are slowed down. Before the molecules exit the high field region, the field is rapidly extinguished.  The net effect is to remove energy from the molecules while returning them to the same initial condition of no applied field. This process is repeated with successive stages of electrodes until the molecules have been slowed to the desired velocity. Figure \ref{fig:decelerator} shows an example of Stark deceleration with the longitudinal potential energy distribution produced by spatially-alternating oriented electrode stages. The solid red line corresponds to the potential produced by the active set of electrodes and the dashed line indicates the other (off) set.  Switching the electric field between successive stages leads to a change in the molecular kinetic energy per slowing stage that depends critically on the position of the molecules at the switching time. This dependence is conveniently characterized by a spatial coordinate $\phi$ as the deceleration phase angle, which is defined at the instant of electric field switching (Fig. \ref{fig:decelerator}).

A synchronous molecule is always positioned at exactly the same phase angle $\phi_o$ at the moment of field switching, thus losing an identical amount of energy per stage. Operating in the range 0$^o < \phi_o < 90^o$ results in phase-stable deceleration of a molecular packet. The restoring force experienced by non-synchronous molecules when the slower is operated in the proper regime is the underlying mechanism for the phase stability of the molecular packet trapped in moving potential wells. For example, given that the spacing between the successive stages is uniform in the decelerator, if the electric field switching time is also uniform, then the position of the synchronous molecule will be $\phi_o$ = 0$^o$, leading to the so-called "bunched" molecular packet traveling at a constant speed down the slower. As $\phi_o$ increases, the equation that describes the phase-stable area for deceleration is the same as the one describing an oscillating pendulum with an offset equilibrium position of $\phi_o$. The longitudinal phase-space distribution of the non-synchronous molecules will rotate inside an asymmetric oscillator potential. The most relevant information for the operation of the decelerator is the rapidly decreasing area of stable evolution~\cite{Bethlem99,Hudson04}. Larger values of $\phi_o$ correspond to more energy loss per stage; however, at the same time the stable phase-space area decreases, resulting in a decrease in the spread of capturable velocities and hence in the number of slowed molecules.  This sets a practical, observable limit to the ultimate temperature of the molecular packet.

There is also a net confining potential in the transverse plane created by the local minimum of the transverse electric field in between a pair of electrodes, and the successive stages alternate in their orientations. Some more recent work~\cite{Meerakker06,Sawyer08b} has studied in detail the effect of transverse guiding within a Stark decelerator and found that the transverse motion plays an important role in determining the total efficiency of deceleration. Two specific loss mechanisms during the slowing process have been identified~\cite{Sawyer08b}. The first mechanism involves distributed loss due to coupling of transverse and longitudinal motion, while the second is a result of the rapid decrease of the molecular velocity at the final few stages of the decelerator. Modified electric-field-switching sequences have been implemented to address the transverse over-focusing effect, but the success is limited to the intermediate velocity regime, and loss at very low final velocities remains problematic~\cite{Sawyer08b}.  New designs have been proposed that could improve the efficiency of the decelerator, albeit with modest gains~\cite{Sawyer08b}. The paper by Parazzoli \textit{et al}. in this special issue also discusses improved energy resolutions from a molecular Stark decelerator~\cite{Parazzoli09}. In current experiments, polar molecules can be decelerated to a mean speed adjustable between about 500 m/s and 20 m/s, with a longitudinal velocity spread from 100 m/s to about 10 m/s around the mean laboratory velocity, corresponding to a translational temperature tunable from a few hundred mK to 20 mK. For more massive molecules that do not possess weak-field-seeking ground states, deceleration via an alternate-gradient electric field has been implemented~\cite{Tarbutt04}, although the significantly larger loss has so far prevented deceleration to speed below 200 m/s. We also note that a chip-based miniature Stark decelerator has also been reported by the group of G. Meijer, for example in this special issue~\cite{Meek09} where a microstructured array of electrodes is used to decelerate CO molecules directly from a molecular beam with an initial velocity of 360 m/s to a final velocity of 240 m/s.

A variety of polar molecules (CO~\cite{Bethlem99}, ND$_3$~\cite{Merakker08}, OH~\cite{Bochinski03}, YbF~\cite{Tarbutt04}, H$_2$CO~\cite{Hudson06a}, NH~\cite{Merakker08}, and SO$_2$~\cite{Jung06}) have been decelerated and have already been employed in precision spectroscopy and crossed-beam collision experiments. A new species, LiH, has been Stark decelerated, as reported in this special issue~\cite{Tokunaga09}. The sufficiently decelerated molecules can also be readily loaded into various molecular traps and have been held there for a duration of about 1 s. Hydroxyl free radicals (OH) present a good example, with their importance for studies ranging from physical chemistry to astrophysics to atmospheric and combustion physics.  After the initial deceleration~\cite{Bochinski03,Bochinski04}, OH molecules have been loaded into and trapped by either static~\cite{Meerakker05} or time-varying electric fields~\cite{Merakker08}, as well as static magnetic fields~\cite{Sawyer07,Sawyer08a}. The depths of these traps vary with type and molecular dipole moments. Typically traps will hold molecular samples with maximum temperatures ranging from 10--1000 mK. For example, electrostatic trapping can provide a trap depth near 1 K, ideally suited for these decelerated molecules. Traps using time-varying electric fields provide a much shallower trap depth (a few mK or less), but can trap both weak- and strong-field-seeking polar molecules. Free radicals are the best candidates for magnetic trapping because of the presence of unpaired electron(s)~\cite{Sawyer07,Sawyer08a}. Confining these molecules in a magnetic trap leaves open the freedom of applying electric fields to electrically polarize the sample at different orientations.  The lifetime in the trap is usually limited by background gas collisions or optical pumping by blackbody radiation~\cite{Meerakker05}. Trapped samples are ideal for reaction and collision studies because of the possibility of long interaction times and the ease of monitoring the trap loss.

\begin{figure}[t]
 \begin{center}
\includegraphics[scale=0.7]{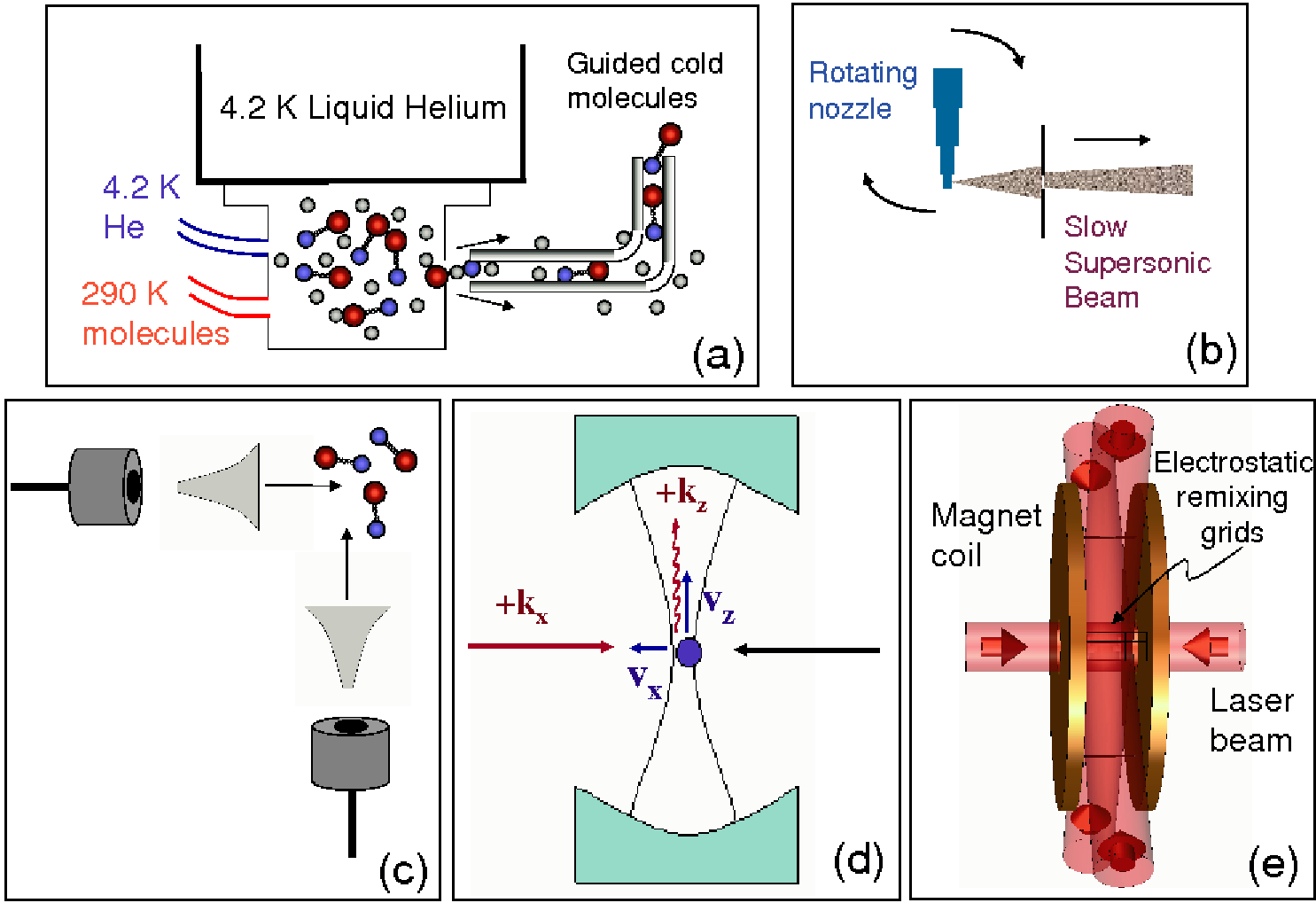}
\caption{Various technical approaches for producing cold molecules. (a) Buffer-gas cooling is perhaps the most versatile technique for the production of a great variety of cold molecules. Buffer-gas-cooled molecules have been loaded into magnetic traps and cold beams can be routinely produced with low temperature (around 4 K), low velocity (about 100 m s$^{-1}$), and high intensity (up to $10^{14}$ s$^{-1}$ sr$^{-1}$); these beams are useful for loading into both electric and magnetic guides to be delivered for collision and trapping studies.  (b) A nozzle rotating in the direction opposite to an emerging supersonic molecular beam allows a slowed molecular beam to be produced in the laboratory frame~\cite{Herschbach01}. The drawback of this technique is the limit of achievable low velocity at about 100 m/s. Regardless of the final beam velocity, its spread is fixed by the original supersonic beam. (c) A technique of producing cold molecules by colliding two perpendicular molecular beams~\cite{Chandler03}. The technique is generally applicable but the production efficiency and the resulting temperatures are limited. (d) A proposed technique of laser cooling molecules via cavity-enhanced Rayleigh scattering~\cite{Domokos02,Andre06,Morigi07,Lev08}. (e) A proposed technique of laser cooling molecules in free space, in a configuration named as electrostatically remixed magneto-optical trap (MOT)~\cite{Stuhl08}. A pair of anti-Helmholtz electromagnet coils produces a quadrupole magnetic field similar to that in a standard atomic MOT. Six beams of the cooling laser are converged on the center with their polarizations oriented as usual for a MOT, but a set of four open-mesh grids are added. The grids are pulsed in pairs to produce the dipole electric fields needed to remix the magnetic sublevels in the ground state polar molecules to remove the dark states.
}
\label{fig:buffer}
 \end{center}
\end{figure}

Another powerful and versatile method for preparing cold molecules is that of buffer-gas cooling of molecules via collisions with cryogenically cooled He atoms (Fig. \ref{fig:buffer}(a)), a technique pioneered by J. Doyle's group~\cite{Weinstein98}. This technique has been used to cool a large variety of atomic and molecular species~\cite{Doyle04,harris2004}, such as CaH, CaF, NH, Cr, Mn, and N. The temperature range of the cold molecules is typically around 1 K, with molecular densities reaching values larger than 10$^9$/cm$^3$. Molecules with magnetic moments of a few Bohr magneton can be easily confined in a large-depth magnetic trap created by a pair of anti-Helmholtz superconducting coils. This has allowed studies of various collisional regimes and dynamics between molecules and He atoms~\cite{Maussang05}. The background He atoms can present a problem in certain experiments of molecular collisions. For molecules with relatively large magnetic moments, the magnetic trapping force can be sufficiently strong to withstand the removal of He atoms from the trapping region via vacuum pumps. It is also worth noting that the preparation of large samples of cold molecules may allow evaporative cooling of these molecules to ultralow temperatures~\cite{nguyen2005two,nguyenMn07}, as we discuss further in Sec.~\ref{sec:conclusions}.

An important new direction of buffer gas cooling is the creation of cold molecular beams~\cite{Maxwell05}, and the paper by Patterson \textit{et al}. in this special issue describes the production of continuous and intense free-space atomic beams and guided molecular beams~\cite{Patterson09}. Molecules of the desired species are introduced into a cold chamber containing cryogenic noble buffer gas, usually helium or neon~\cite{Patterson07}. After a sufficient number of collisions with the buffer gas, the molecules are cooled to very near the temperature of the buffer gas, both rotationally and translationally. The buffer gas and the molecules are both allowed to exit the cell via an orifice. To produce a pure molecular beam, curved electric or magnetic field guides can be placed at the output of the cell such that only molecules (with either electric or magnetic dipole moments) that are moving sufficiently slowly will be guided to a desired final location. The paper in this special issue by Motsch \textit{et al} presents a detailed study on velocity-dependent collisional effects on the formation of guided cold polar molecular beams, providing useful guidelines for experimental designs~\cite{Motsch09}. The properties of the beam and the extraction efficiency of molecules from the cell are determined by the ratio of the molecular diffusion lifetime in the cell and the dumping time of the cell. In the case where the diffusion lifetime is shorter, the hydrodynamic flow inside the cell leads to very high extraction efficiency of molecules out of the cell. Extraction efficiencies of 10\% are easily achieved and efficiencies as high as 50\% ~\cite{Patterson07} have been observed, leading to high intensities. Buffer-gas beams are now routinely produced in a number of laboratories with low temperature (around 4 K), low velocity (about 100 m/s), and high intensity (up to $10^{14}$ s$^{-1}$ steradian$^{-1}$)~\cite{Patterson07}. Similar to Stark deceleration, this technique can be used for almost all molecules.

Buffer-gas-cooled molecular beams can be used to load traps, for example via a dynamically operated trapping door, or through optical pumping for dissipative loading~\cite{vandeMeerakker01,DeMille04,Narevicius09}, or using a single collision beam approach~\cite{Chandler03} with a low initial temperature. The paper by Takase \textit{et al}. in this special issue proposes a new way to produce molecules at mK temperatures via kinematic collisions with cold atoms, with a detailed discussion on the effectiveness of such an approach~\cite{Takase09}. The paper by Narevicius \textit{et al}. in this special issue discusses a proposal of producing cold trapped molecules via a single photon cooling, which is implemented with a single event of absorption followed by a spontaneous emission~\cite{Narevicius09}. Usually an electric- or magnetic-field guide is used between the cryogenic cell and the trap. As the temperature of the beam is near the depth of the guide in these proposals, nearly 100\% of the guided molecular beam would be loaded into the trap, with a potential density of $\approx 10^{9}~{\rm cm}^{-3}$.  This density is near the region where collisions between molecules should be observable.

The use of buffer-gas cooled molecular beams, in combination with Stark deceleration and trapping, can also lead to a very promising direction for molecular collision studies. In a recent experiment, an open structure of a permanent magnetic trap allowed for low center-of-mass energy collision studies between the trapped and polarized molecules and an incoming beam of another molecular species~\cite{Sawyer08a}. By bombarding a magnetically trapped molecular sample with a packet of cold polar molecules at varying collision energies, one can monitor the trap loss, and thus the scattering rate, as a function of the collision energy and the magnitude of an applied external electric field.  This collision apparatus is superior to conventional crossed beam setups. First, a cold, state-selected polar molecule sample confined within a permanent magnetic trap reduces the center-of-mass collision energies well below those observed in conventional crossed-beam experiments. Second, having the trapped molecules as a collision partner and observing the trapped molecular lifetime yield a new method for determination of absolute collision cross sections, which is notoriously difficult for crossed-beam experiments. In the case of an incident beam of buffer gas cooled molecules, the incident energy for collision can be tuned in the range of a few Kelvin and the beam intensity is larger than that produced by a Stark decelerator. The trap loss can then be monitored to determine the collision cross sections under various conditions of collision energy, molecular density, molecular polarization and isotopes, and different internal quantum states. Molecular collisions that exhibit explicit dipolar characteristics under external electric fields can be studied.

Production of high-density samples of ultracold polar molecules will be very important to permit explorations of novel collisions in different energy regimes. To this end, neither Stark deceleration nor buffer-gas cooling are satisfactory in their current state of development.  However, the large density of cold molecules produced by a buffer-gas-cooled molecular beam may enable further development of direct laser cooling of molecules. For example, cavity-assisted Doppler laser cooling of molecules can be utilized to enhance photon-molecule scattering via a high-finesse cavity~\cite{Domokos02,Andre06,Morigi07,Lev08}. A challenging problem identified in initial work is the insufficient number of molecules inside the cavity mode volume~\cite{Lev08}. This molecular density problem might be addressed by sending a buffer-gas-cooled high density molecular beam into the intracavity region. Once the molecular density is sufficiently high for the collective effect to cross the cavity-cooling threshold, the sample temperature can be expected to be quickly cooled from tens of milliKelvins to $<$100 $\mu$K, improving the phase-space density by at least a factor of 100. The paper by Salzburger and Ritsch in this special issue discusses light-induced transverse collimation and phase space compression of a fast molecular beam traversing a high finesse optical cavity~\cite{Salzburger09}. It is noted that collective enhancement can lead to significant transverse cooling and collimation above the self-organization threshold.

The buffer-gas-cooled beam will also be a staging platform for the recently proposed free-space laser cooling of polar molecules~\cite{DiRosa04,Stuhl08}.  Direct, free-space laser cooling and trapping would be the ideal method for producing a large variety of ultracold molecules, just as it is for atoms. For optical cooling of molecules, a most pressing issue is to identify a suitably closed transition structure that supports a sufficient momentum exchange between molecules and photons. A class of molecules has been identified to be exceptionally good candidates for laser cooling: they have good Franck-Condon overlaps (closing the vibration ladder), and their ground or lowest metastable state has a higher angular momentum than the first accessible electronically excited state (thus the rotational ladder is closed without any leakage). A number of molecules satisfy these requirements. TiO and TiS are both satisfactory in their absolute ground states, while electronically metastable FeC, ZrO, HfO, ThO, and SeO; and $\mathcal{R}=1$ states of CaF, SrF, BaF are also promising.  The oxides and carbides listed here have the additional advantage of no net nuclear spin (thus no hyperfine complexity).  The paper by Lu and Weinstein in this special issue reports the production of large numbers of TiO molecules at a translational temperature of 5 K produced by laser ablation and cryogenic helium buffer-gas cooling~\cite{Lu09}. Using buffer-gas-beam TiO or SrF sources, the scattering rate of photons and repumping mechanisms have been studied and found to permit sufficient cycling for laser cooling. A key to the success of this scheme is the recognition that, for polar molecules, the presence of an electric dipole moment provides a way to continually remix the ground-state sublevels so that all the molecules spend some fraction of their time in bright states. Such remixing of the ground-state magnetic sublevels allows the building of a new kind of trap, the electrostatically remixed magneto-optical trap (ER-MOT)~\cite{Stuhl08}. Another idea is to use microwave electric fields near resonance with a rotational transition to provide the required destabilization of dark Zeeman sublevels.

Finally, collisional (i.e., evaporative and sympathetic) cooling, an essential tool in the production of quantum degenerate atomic samples~\cite{Anderson95,Davis95,Bradley95,Truscott01,Schreck01}, may be extended to polar molecules, with the goal of producing quantum-degenerate gases with strong dipolar interactions~\cite{Ospelkaus09}. This cooling step can take place once a sufficient density of molecules is accumulated inside a trap and the conditions are favorable for elastic collisions to dominate over inelastic collisions. The internal structure of polar molecules introduces qualitatively new features that need to be explored in detail in order to implement and optimize collisional cooling in these systems.  Here, the synergy between precision spectroscopy, indirectly-cooled molecules (such as ground-state bialkali molecules described in the next subsection), and directly-cooled molecules will be important in order for insights gained from one system to guide parallel efforts in the other. Molecule-atom collisions will be explored with the aim to use ultracold atoms as a heat bath for cooling molecules. For example, a promising research path undertaken by the Doyle group is to buffer-gas cool both NH molecules and N atoms in a common magnetic trap, then study collisions between NH and N and explore evaporative cooling of N along with sympathetic cooling of NH~\cite{Hummon2009}. The paper by Barletta et al. in this special issue discusses in theory the possibility of sympathetic collisional cooling of large molecules such as benzene with cold rare gas atoms such as He or Ne~\cite{Barletta09}. Of course, a particular focus will be given to collisional properties unique to ultracold polar molecules, where electric fields are expected to produce dramatic changes in collision rates.  These effects arise from two distinct causes: direct dipole-dipole interactions~\cite{Ticknor05,Ticknor08} and Fano-Feshbach resonances due to rotational structure~\cite{Volpi02}. Both effects can lead to unprecedentedly large elastic collision rates, even in fermionic molecules.

\subsection{Indirect Cooling and Manipulation with External Electromagnetic Fields}
\label{ssec:photo}

One of the most successful methods for creating ultracold molecules has been to assemble them from constituent atoms that are already ultracold.  These techniques take advantage of the enormous power of existing methods in laser cooling and evaporative cooling\cite{Anderson95,Davis95,Bradley95} for producing samples of ultracold atoms.  Remarkably, it has proven possible to induce the binding of atoms into molecules without significant motional heating, despite the fact that the binding energy of molecules can exceed the atomic temperature by roughly 10 orders of magnitude.  The challenges associated with these techniques arise from the difficulty in controlling the \textit{internal} degrees of freedom of the molecules during and after their formation. \footnote{Note that with atoms, ultracold generally refers only to the external/motional degree of freedom; this is because the internal structure of laser-cooled atoms typically consists only of hyperfine structure that is easily controlled via optical pumping, or electronic structure at such high energy that it is always frozen out.}  However, recent developments in the field have almost completely surmounted these difficulties~\cite{Ni08,Lang08,Danzl09,Ospelkaus09}.  It now appears likely that it will be possible to create, in the very near future, quantum-degenerate gases of molecules in any selected internal state.

It should be emphasized that these ``indirect cooling'' techniques have a fundamental limitation: namely, they can only be used to create ultracold molecules whose constituent atoms can be laser cooled and trapped.  Thus, for the conceivable future it will still exclude many chemically relevant species such as hydrides, nitrides, oxides, fluorides, etc.  However, the properties of molecular species that can be created using these methods provide a wide range of applications and opportunities.  Although a growing number of experiments of this type use ``second-generation'' laser cooling species (e.g. alkaline earths), to date most indirect cooling experiments have been done with bialkali molecules.  Hence we focus here on the properties of these species.

It is useful to classify the electronic states of bialkali molecules according to the atomic states to which they asymptotically connect at long range.  Two ground-state ($s_{1/2}$) alkali atoms can combine to form either $^3\Sigma$ or $^1\Sigma$ states.  Due to exchange interactions, the former has a strongly repulsive potential curve $V(R)$ at intermediate internuclear distances $R$ while the latter is attractive until short range.  At long range, where exchange effects are small, both potentials are governed by the same van der Waals forces and hence merge together and scale as $V(R) \propto R^{-6}$.  Therefore, $^3\Sigma$ states have shallow potential curves, with a typical binding energy of $\lesssim 0.05$ eV, while $^1\Sigma$ states are deeper, with typically $\sim 0.5$ eV of binding energy.  In the $^1\Sigma$ state, the heteronuclear bialkalis all exhibit permanent electric dipole moments $D$, whose magnitude is correlated with the difference in atomic number between the two atoms~\cite{Igel-Mann86,Aymar05} (note that a DC electric field is required to make use of these dipole moments, as described in Sec.~\ref{ssec:chemistry}).  For example, LiCs (with the largest difference) has the largest dipole moment $D \sim 5.5$ Debye, while LiNa (smallest difference) has $D \sim 0.5$ Debye. A note of caution is that the magnitude of the external electric field needed to polarize these polar molecules depends not only on the dipole moment, but also on the energy gap between the two relevant opposite parity states (for example, between two adjacent rotational levels). Homonuclear species have $D=0$ due to the symmetry between constituent atoms, which imposes an additional quantum number ($u/g$, gerade for symmetric and ungerade for antisymmetric) corresponding to the sign of the electronic state under exchange of the two identical nuclei.

The first excited states of bialkalis correlate to atomic states with one atom still in its $s_{1/2}$ ground state and the other in its first excited $p$ orbital.  This leads to a large number of states, which at short range can be described in terms of their total spin $S$ and their value of $\Lambda$; hence there are $^{1,3}\Sigma$ and $^{1,3}\Pi$ states in this overlapping manifold of potentials.  These states all have substantial binding energy $\sim 0.1-1$ eV.  At long range (such that atomic spin-orbit splitting in the $p$ state is large compared to $V(R)$) the internal angular momenta recouple as described in section \ref{ssec:structure}, leading to multiple avoided crossings and associated couplings between short-range singlet and triplet states.  For heteronuclear species the long-range behavior of the potentials again satisfies $V(R) \propto R^{-6}$, although the proportionality coefficient is many times larger than for the ground state.  For homonuclear species, these states have long range potentials $V(R) \propto R^{-3}$, since here the virtual dipole-dipole interaction leading to van der Waals forces has a resonant term.

The earliest experiments in forming ultracold molecules from laser cooled atoms used the technique of photoassociation (PA) to produce and probe bound but short-lived excited states of the molecules~\cite{Stwalley99,Vanhaecke04,Jones06}.  In PA, a laser is tuned to resonance with a transition from the free (scattering) state of two ultracold ground ($s$) state atoms, to a bound level of the excited ($s + p$) potential.  These transitions are hence at wavelengths to the red of the atomic transitions.

It is useful to characterize photoassociation by the rate of excitation per atom in the sample, $R_{\mathrm{PA}}$.  For sufficiently low PA laser intensity $I_L$ (such that the PA rate is not saturated~\cite{Junker08}), in general $R_{\mathrm{PA}} \propto I_L F_{fv'} \Omega T$, where $F_{fv'}$ is the Franck-Condon factor describing the overlap between the initial free state and the final bound state, $\Omega$ is the phase space density, and $T$ is the temperature~\cite{Bohn99b,Drag00}.  Several features combine to make $F_{fv'}$ generally very small.  First, at typical densities in ultracold atomic samples ($n \sim 10^{11-14}~\mathrm{cm}^{-3}$), the average interatomic spacing is $R \sim 2000-20,000$ \AA, well beyond the classical outer turning point $R_C$ (the ``Condon radius'') of even the weakest bound states.  Hence, even in the best case the scattering-state wavefunction will be spread over a large range of $R$ compared to the bound state.  In addition, the kinetic energy associated with atoms colliding in the ground state potential at $R=R_C$ leads to two more effects that suppress $F_{fv'}$.  This is easily seen in a WKB picture of the scattering-state wavefunction, where the amplitude is diminished and the oscillation period is short when the kinetic energy is large.  This should be compared to the vibrational wavefunction of the excited state, which is peaked in the vicinity of $R_C$; hence the overlap integral largely cancels.

Early PA work focused on the study of homonuclear molecules~\cite{Wagshul93,Miller93}, where the $R^{-3}$ dependence of the excited-state potential leads to bound states with large values of  $R_C$ and hence the largest values of $F_{fv'}$. PA of heteronuclear species, where $F_{fv'}$ is smaller~\cite{Wang98}, was demonstrated a few years later~\cite{Shaffer99,Kerman04a}.  In both cases, with sufficiently large values of $I_L$ applied for long times (possible exactly because the atoms are trapped), $R_{\mathrm{PA}}$ can be often be made large enough to excite a large fraction of the precursor atoms, even from heteronuclear samples with the fairly low phase space density $\Omega$ associated with magneto-optical traps~\cite{Prodan03,Kerman04a}.  In many circumstances, $R_{\mathrm{PA}}$ can reach its largest possible value $R_{\mathrm{PA}}^{\mathrm{max}}$, determined by the unitarity bound for inelastic collisions~\cite{Bohn99b}: $R_{\mathrm{PA}}^{\mathrm{max}} = \Omega \frac{k_B T}{2 \pi \hbar}$. Note that for sufficiently deeply bound levels, $R_C$ (and hence the molecular moment of inertia) is small, and the rotational constant $B$ is larger than the natural linewidth of the transition.  In this case individual rotational levels can be resolved using PA; these methods have been used extensively for precise spectroscopy of weakly-bound excited-state levels, which are usually difficult to probe by other means.

%introduce E_{b'}, E_{b''}?

%
\begin{figure}[t]
 \begin{center}
\vspace*{8mm}
\includegraphics[scale=0.45]{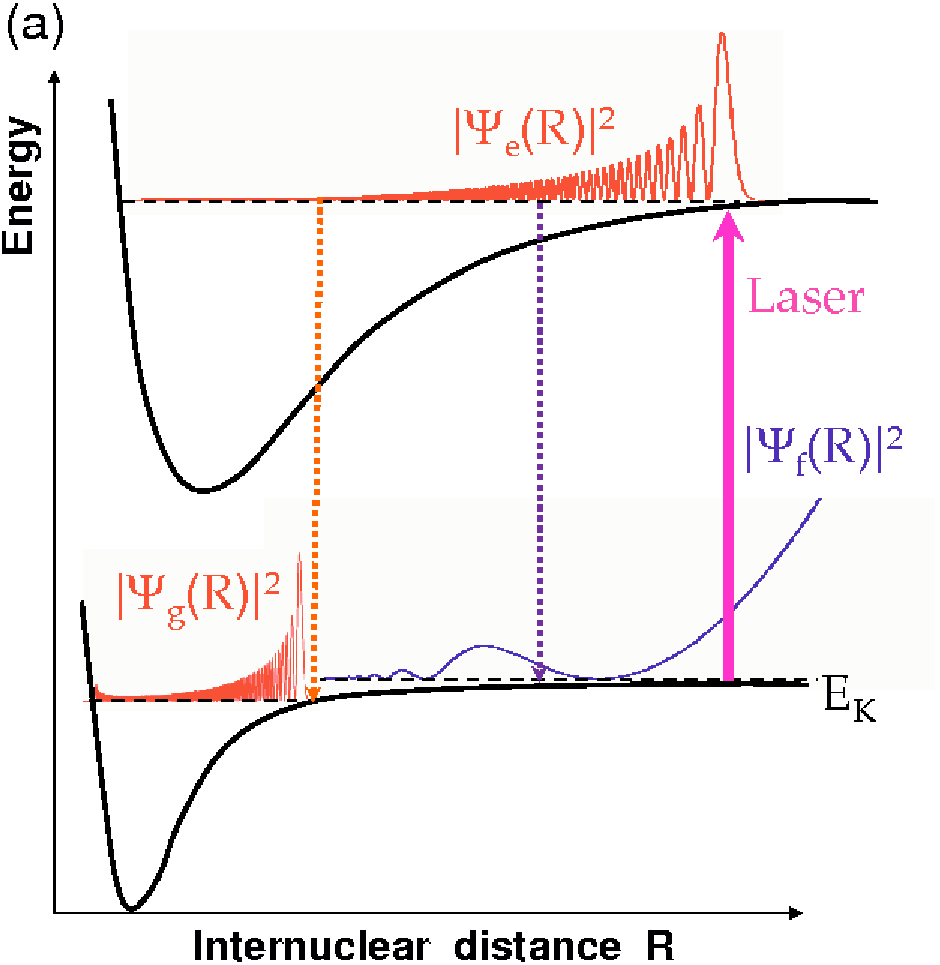}
\hspace*{12mm}
\includegraphics[scale=0.45]{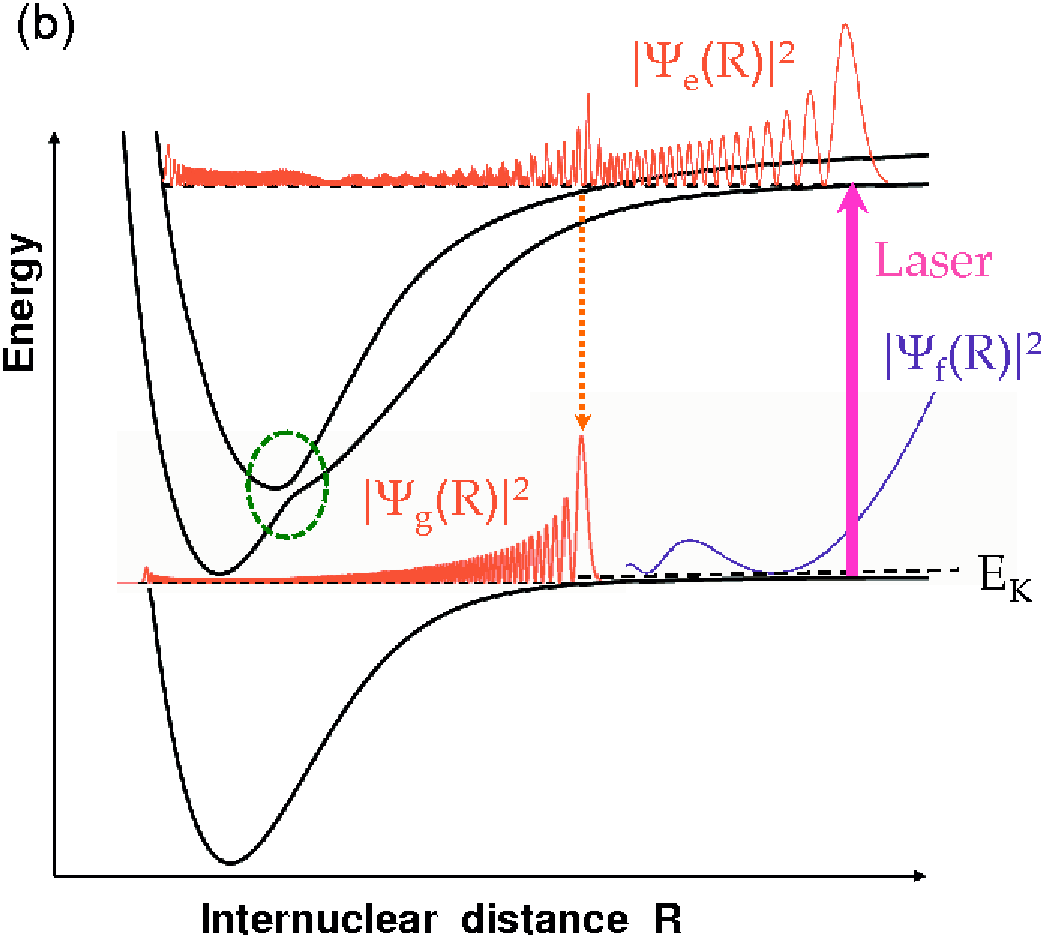}
\caption{Optical methods for indirect production of ultracold molecules.
Here a laser drives a transition from the free (unbound) state of an atom pair, into an electronically excited but vibrationally bound state. This state then decays rapidly back to both bound and free states of the ground potential. The transition strengths for both the laser-driven upward transition and the spontaneous downward transition are determined primarily by Franck-Condon overlaps between initial and final states. (a) Simple photoassociation (PA) scheme. Here the Franck-Condon overlap is poor both for free-bound excitation and bound-bound decay, leading to small rates of bound molecule production. (b) Enhanced production of bound molecules via resonant coupling of excited-state potentials. Here the coupling between two excited state potentials (evident from the avoided crossing in the green oval) causes the excited-state vibrational wavefunction to be a mixture of states from the two potentials; one is weakly bound but the other strongly bound. The free-bound transition has strength similar to that of the simple PA scheme, but bound-bound transitions are greatly enhanced by the intermediate- and short-range peaks from the strongly bound state contribution to the excited-state wavefunction.}
\label{fig:em}
 \end{center}
\end{figure}

Formation of molecules in the ground-state potential can also be accomplished using PA, as illustrated in Fig.~\ref{fig:em}.  Such states can be populated by spontaneous radiative decay from the short-lived upper state of the PA transition (the typical lifetime is $\sim 10-30$ ns)~\cite{Fioretti98,Takekoshi98,Mancini04,Kerman04b}. In this special issue and their earlier work~\cite{Bigelow04}, Haimberger \textit{et al}. report the production of ultracold NaCs molecules in the electronic ground-state potential~\cite{Haimberger09}. A crucial feature of this ``radiative stabilization'' process is that the probability for decay into a given ground-state level is determined by the bound-bound Franck-Condon factor $F_{v'v''}$.  Since $F_{v'v''}$ is maximized when the states share the same classical turning points, and since the excited-state potentials are always deeper than the ground-state potential for the same value of $R$, decays are typically into levels bound by less than the original PA level.  In this simple picture, there is a clear tradeoff between $R_{\mathrm{PA}}$ and the efficiency of forming deeply bound ground-state levels.  High efficiency (of order unity) in conversion from free atoms to bound ground-state molecules has so far only been possible into states with binding energy of $\lesssim 10^{-4}$ eV.

\begin{SCfigure}
  \centering
\vspace*{4mm}
\includegraphics[scale=0.45]{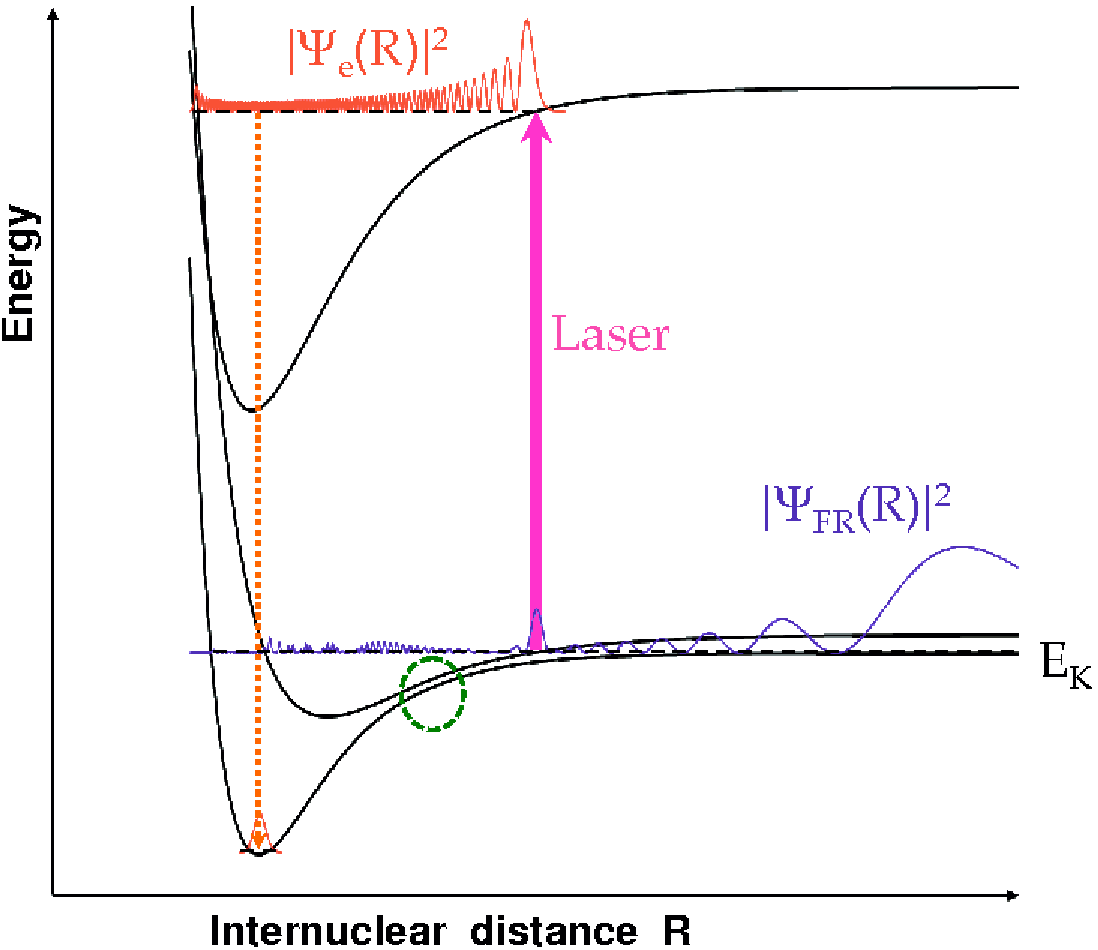}
\caption{Feshbach-optimized photoassociation (FOPA)~\cite{Pellegrini08,Tolra03}. Here a Fano-Feshbach resonance between two coupled groundstate potentials causes the initial-state wavefunction to be a mixture of states from the two potentials, enhancing the intermediate and even short-range part of the wavefunction. This enhances the free-bound transition rate to deeply bound excited-state levels, which can in turn decay efficiently to deeply bound ground state levels.}
\label{fig:em2}
\end{SCfigure}

In order to efficiently produce very deeply bound ground-state molecules, including the absolute rovibrational ground state X$^1\Sigma(v=J=0)$, methods beyond the simplest version of PA seem to be necessary.  A variety of techniques have been proposed and demonstrated to circumvent the contradiction between large $R_{\mathrm{PA}}$ and large $F_{v'v''}$.  For example, in the ``R-transfer'' method~\cite{Band95,Nikolov00}, a second laser is used to transfer population from the excited level of PA to a higher intermediate level, where the potential curve allows good Franck-Condon overlap with both the (primarily) long-range PA state and the (short range) deeply-bound ground vibronic levels.  Another method takes advantage of accidental (but fairly common) resonances between bound states in coupled excited-state potentials (see Fig.~\ref{fig:em}).  In such cases~\cite{Dion01,Kerman04a,Kerman04b} the excited-state vibrational wavefunction can have amplitude peaks (corresponding to effective classical turning points) simultaneously at both long range (as needed for large $R_{\rm PA}$) and at short range (as needed for large $F_{v'v''})$. A promising new method is to use a Fano-Feshbach resonance to enhance the amplitude of the scattering state at short range~\cite{Courteille98, Junker08, Pellegrini08,Tolra03}. In this special issue, a pair of papers from the C\^{o}t\'{e} group further expand the theoretical model of this ``Feshbach-optimized PA'' (FOPA) process, discussing the possibility of reaching the unitarity limit for relatively
small laser intensities and its application for producing ground-state polar molecules~\cite{Pellegrini09,Kuznetsova09}. The FOPA method makes it possible to achieve large $R_{\mathrm{PA}}$, possibly even for PA into deeply-bound excited levels with good Franck-Condon overlap with the X$^1\Sigma(v=0)$ state (see Fig. \ref{fig:em2}).  A similar mechanism for enhancement, due to an accidental near Fano-Feshbach resonance in the scattering state, is reported in this special issue~\cite{Deiglmayr09}. Here, it was demonstrated that even absolute ground state [X$^1\Sigma(v=J=0)$] LiCs molecules could be produced at detectable levels by PA to a very deeply bound excited level~\cite{Deiglmayr08}. Even with this enhancement only $R_{\mathrm{PA}} \sim 3\times 10^{-5}~\mathrm{s}^{-1}$ (which may be compared to $R_{\mathrm{PA}}^{\mathrm{max}} \sim 1~\mathrm{s}^{-1}$) was achieved; however, higher rates may be possible by operating at higher phase space density and/or laser intensity than in this demonstration experiment.  This opens the interesting possibility for continuous formation of rovibronic ground state molecules.

A few other features of radiative stabilization as a means for producing ground-state molecules are notable.  For example, the dissipation associated with the spontaneous emission event means that molecules can be produced continuously.  It is also not crucial to begin with atoms in a quantum degenerate state; rather, even at finite temperature it is possible (in principle) to convert all available atoms into ground-state molecules.  In this sense PA can be thought of as a type of cooling, which removes entropy associated with the relative atomic motion.  The PA process itself induces negligible heating, with momentum transfer of only two photon recoils.  The bound-bound decay typically results in population of a large number of vibrational levels, according to the distribution of $F_{v'v''}$ values.  However, in many cases one particular level $v''^*$ receives a substantial fraction of the population, with $F_{v'v''^*} \gtrsim 10\%$.  In addition, it was demonstrated in this special issue~\cite{Sofikitis09} and elsewhere~\cite{Viteau08} that, by using broadband laser pulses spectrally tailored to excite vibrational levels other than a selected level $v''$, it is possible to optically pump from higher levels into the selected level. Finally, we note that the strict selection rules associated with angular momentum states involved in the decay mean that at worst a small number of rotational sublevels -- in some cases, even a single rotational level -- are populated in each vibrational level from the decay following PA.

An alternative route to producing ground-state molecules in states of high vibrational excitation uses stimulated processes rather than radiative stabilization.  In principle, these methods do not cause any heating of the sample.  This is crucial since here, for high efficiency of molecule production, it is necessary to begin with a quantum-degenerate sample.  This is because entropy is conserved in stimulated (i.e., reversible) processes; hence, only a single state of relative atomic motion can be converted into a single target molecular state, and the maximum possible atom-molecule conversion efficiency with a stimulated process is given roughly by the phase-space density $\Omega$ (in units of $\hbar^3$) ~\cite{Hodby05}.  With that in mind, these methods have proven extremely powerful and general. An increasingly common technique of this type involves sweeping a magnetic field through a Fano-Feshbach resonance, such that atom pairs are adiabatically transferred from an unbound state (in one hyperfine/Zeeman spin configuration) to a weakly bound state (in a different configuration)~\cite{Kohler06}.  Using this ``magneto-association'' process, atom-molecule conversion has been demonstrated with near-unit efficiency for both homonuclear and heteronuclear species~\cite{Kohler06,Chin09b,Regal03b,Xu03,Herbig03,Zwierlein03,Jochim03,Greiner03,Durr04,Regal04,Ospelkaus06,Zirbel08a,Zirbel08b,Klempt08,Weber08,Voigt09}.  An interesting alternative is the use of optical fields rather than magnetic fields for control.  In a dressed-state picture, optical coupling of a free-atom state to a bound state can play a role similar to that of the inter-potential couplings responsible for ``natural'' Fano-Feshbach resonances, while the detuning of the optical field from resonance plays a role similar to that of the magnetic field, i.e., for tuning through the resonance.  Although formation of molecules via adiabatic transfer using such an optical Fano-Feshbach resonance has not been demonstrated, preliminary work is encouraging~\cite{Wynar02,Drummond02}.  It is also possible to use direct, on-resonance transitions (e.g. in the method of ``two-color PA'')~\cite{Abraham95,HeinzenScience2000}, although for efficient molecule formation adiabatic passage is generally more favorable.

Stimulated photo- and magneto-association processes are already being used to probe weakly-bound states of molecules larger than diatomic species, and may be used to produce detectable numbers of such ``large'' molecules in the near future.  In a series of experiments, the Innsbruck group has demonstrated the existence of both Cs$_3$ and Cs$_4$ molecular states by observing resonant features in the collisional loss of ultracold Cs atoms or ultracold Cs$_2$ molecules~\cite{Chin05,Kraemer06,Knoop09,Ferlaino09}.  Similarly, photoassociation of ultracold atoms in mixtures of Bose-Einstein condensates can be enhanced by exploiting the principles of coherent control developed in chemical physics for manipulating molecular photodissociation \cite{Shapiro03, Vardi97}. Jing and coworkers \cite{Jing08a, Jing08b} have recently shown that constructive multi-path interference in the atom-molecule conversion in $^{87}$Rb - $^{40}$K-$^6$Li mixtures allows for the creation of triatomic molecules almost with the perfect yield. The idea of Jing and coworkers is based on exploiting an atom-diatom dark state to enhance the molecule creation rate. The same idea had previously been used in the experiments of Winker and coworkers \cite{Winker05} to produce a quantum degenerate gas of Rb$_2$ molecules in specific ro-vibrational states.

\begin{figure}[t]
 \begin{center}
\vspace*{10mm}
\includegraphics[scale=0.4]{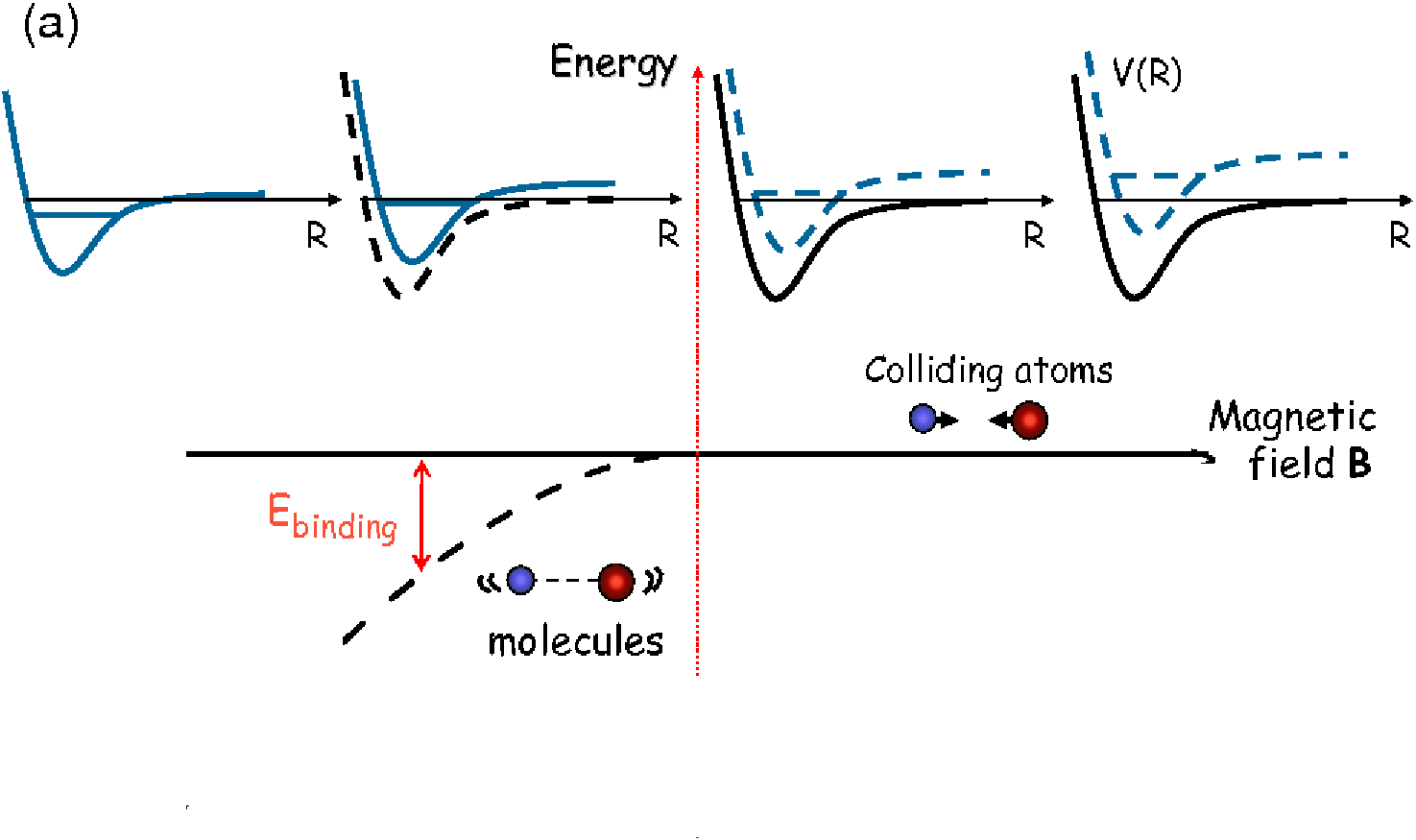}
\hspace*{24mm}
\includegraphics[scale=0.4]{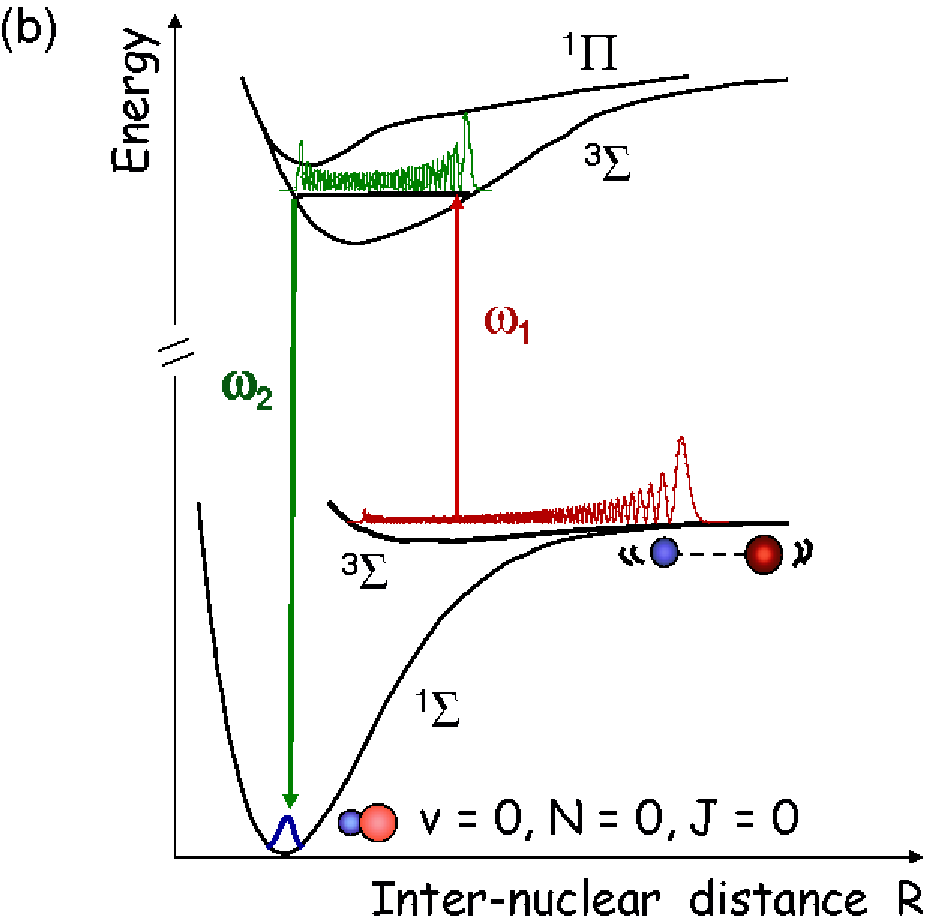}
\caption{Creation of high-density ground-state polar molecules from dual-species, ultracold atoms. The conversion process is fully coherent and preserves the phase space density of the original atomic gases in order to reach the quantum degenerate regime for the polar molecular gas. (a) The colliding atom pairs from a near degenerate, dual-species atomic gas are converted to weakly bound Feshbach molecules by sweeping the magnetic field near an inter-species Fano-Feshbach resonance. This process allows the initial scattering state to be converted into a single bound level, albeit at a limited efficiency of about 10 - 20\% in free space. (b) In the second step, an optical Raman transfer scheme is employed to coherently transfer the weakly bound Feshbach molecules to the rovibration ground level of the electronic ground potential. Rovibrational levels in the ground electronic state are first mapped out using coherent two-photon spectroscopy before implementing the actual state transfer. The actual population transfer is made via Stimulated Raman Adiabatic Passage (STIRAP). The efficiency of the transfer process, shown to be above 90\%, is crucial for maintaining the molecular phase space density. This high efficiency is enabled both by precise phase control of the pair of Raman optical fields (via connection to a phase-stable optical frequency comb) and by systematic spectroscopic studies of the relevant transition pathways for optimized transition strengths.}
\label{fig:comb}
 \end{center}
\end{figure}
Molecules formed in high vibrational states can also be transferred to deeply-bound levels using stimulated processes, as shown in Fig.~\ref{fig:comb}.  This was first demonstrated with an incoherent ``pump-dump'' method, where a single, high-lying vibrational state of RbCs (formed by PA) was transferred to the vibronic ground state~\cite{Sage05}.  Despite the incoherent nature of the process, efficiency of $\sim 6\%$ for atoms addressed by the transfer lasers was demonstrated. For coherent transfer, reasonable strengths for both up and down transitions are necessary since optical nonlinear processes will limit the highest laser intensities that could be used for the transfer process. In such scenarios, a recent theory proposal advocates the use a weak-field control via coherent accumulation~\cite{Marian04} of a train of weak pulses to achieve a strong-field effect -- a complete population transfer~\cite{Peer07}. This conversion process by a coherent pulse train was later identified in the context of KRb molecules to be equivalent to a coherent Stimulated Raman Adiabatic Passage (STIRAP) process in a piece-wise manner~\cite{Shapiro08}. The paper by Ghosal \textit{et al}. in this special issue also discusses the use of wavepackets to aid the efficiency of a two-color conversion process~\cite{Ghosal09}, pointing out some important differences in the wavepacket dynamics for heteronulear and homonuclear molecules.

Recently, a spectacular series of papers have demonstrated highly efficient transfer to deeply-bound levels, using the  STIRAP process based on a pair of CW lasers~\cite{Ni08,Ospelkaus09,Ospelkaus08b,Danzl08,Lang08}, including in an optical lattice in this special issue~\cite{Danzl09}. As shown in Fig. 8, the coherent transfer process takes two steps. First, atomic pairs are converted into loosely bound molecules using a Fano-Feshbach resonance. Then a coherent two-photon Raman process transfers the population from the initial, highly excited, molecular level to a much more deeply bound state. Thanks to good transition strengths, here it has been possible to transfer molecules to a variety of different states, including the $v=0$ level of the metastable a$^3\Sigma$ potential of Rb$_2$ and KRb, and the absolute rovibronic ground state X$^1\Sigma$($v$=$J$=0) of KRb.  The latter case~\cite{Ni08,Ospelkaus09,Ospelkaus08b} is particularly significant: it represents the formation of a strongly dipolar molecule that should be stable even in the presence of collisions (see below).  Moreover, the resulting KRb sample is extremely cold and dense, with phase space density $\Omega \sim 10^{-1} \hbar^3$ near the conditions of quantum degeneracy. The overall STIRAP process proves to be very efficient, reaching nearly 95\% for a single-pass transfer, owing to the long lifetime of the initial and target states and the good transition strengths~\cite{Ni08,Ospelkaus09}. We note that in this coherent transfer regime, it is no longer necessary to discuss the production rate of the molecules -- in principle, 100\% conversion efficiency can be achieved in a single coherent step. As described in Ref.~\cite{Inouye09} of this special issue, Inouye and his colleagues are planning to follow this approach to produce a quantum degenerate gas of bosonic $^{41}$K$^{87}$Rb polar molecules. Towards this end, also reported in this special issue is the work of Thalhammer \textit{et al}.~\cite{Thalhammer09} where recent spectroscopic measurements of two weakly bound molecular levels and newly observed narrow d-wave Fano-Feshbach resonances can be used to improve the collisional model for the bosonic mixture of $^{41}$K$^{87}$Rb.

We point out several relevant features of this two-step process for creating ultracold ground-state molecules~\cite{Zirbel08a,Hudson08,Staanum06,Zahzam06}, in which one first forms a high vibrational level, then transfers it to a lower level.  One is that the high vibrational levels are in general susceptible to loss by collisions with remaining atoms and other molecules.  Although these collisions are suppressed for the most weakly-bound states of homonuclear fermionic pairs, in most other cases inelastic collisions appear to proceed at nearly the unitarity-limited rate.  Although the stimulated transfer to lower levels can occur on very short times scales (determined e.g. by the excited state lifetime, or for STIRAP by the optical Rabi frequencies), the production of vibrationally-excited states is often much slower.  In magneto-association the time to sweep through the Fano-Feshbach resonance depends inversely on the width of the resonance, while in PA the maximum production rate is also limited by unitarity.  Hence, significant collisional loss can occur during this stage, although under typical conditions $10-50\%$ of the molecules can survive until they are transferred to deeply bound states.  Once in the rovibronic ground state, two-body inelastic processes are energetically forbidden.  It has been noted that the problem with collisions can be entirely circumvented by starting with atoms in an optical lattice, if each lattice site contains no more than two atoms~\cite{Jaksch02,Danzl09}.  Tight confinement of free atoms in the lattice can also enhance the strength of free-bound optical transitions.

For the STIRAP transfer process, it is also crucial to identify a suitable intermediate state.  Efficient and rapid STIRAP requires coupling of the initial and final states to an intermediate state with large Rabi frequencies in each step.  Hence, the transition dipole matrix element must be large for the intermediate electronically excited state to couple both to the primarily long-range vibrationally excited initial state and the short-range vibrational ground state.  This requires significant Franck-Condon factors for both steps, which can be found only for certain shapes of the relevant potentials.  This is complicated by the fact that both PA and magneto-association often lead to vibrationally excited molecules that are primarily triplet states, and singlet-triplet transitions are nominally forbidden.  (This does have the advantage that the ground $^3\Sigma$ potential has a classical inner turning point at intermediate range, enhancing the vibrational wavefunction in this region; see Section \ref{ssec:structure}).  Hence, the intermediate state in such cases must have a mixed singlet-triplet character to couple to both the initial and final states.  Miraculously, intermediate states with the required potential shape and mixing, due to spin-orbit interactions, appear to exist not only in the demonstrated cases of RbCs~\cite{Bergeman04,Sage05} and KRb~\cite{Ni08,Kotochigova09}, but in fact for all of the heteronuclear bialkali species~\cite{Stwalley04}.  The situation is further complicated in homonuclear species by the additional $u/g$ symmetry; here it appears that at least two STIRAP transfers will be necessary to reach the X$^1\Sigma(v=J=0)$ state.

Now that ultracold, absolute ground-state bialkalis have been produced, it is necessary to also consider the hyperfine substructure (hfs) of these states~\cite{Aldegune08}.  For $^1\Sigma$ states, where no unpaired electron spin is present, hfs arises due to nuclear spin-spin interactions (with splittings at the kHz level) and, for rotational states with $N \geq 1$, due to electric quadrupole interactions (with splittings in the MHz regime).  In order to produce a quantum-degenerate gas of molecules, it will be necessary to control these nuclear spin degrees of freedom.  This in turn will require spectroscopic resolution of this very fine substructure, in addition to the use of angular momentum selection rules during the coherent population transfer process~\cite{Ospelkaus09}.

A crucial advantage of reaching the ultracold regime in these experiments is that this enables the use of optical trap technology~\cite{Takekoshi98}, including both regular optical dipole traps and optical lattices.  These traps for ultracold molecules are in many ways analogous to those now used commonly for atoms.  In particular, as for atoms, conservative traps for ultracold molecules can be formed by tuning a laser to a frequency well below a strong electronic transition; here the real part of the polarizability function $\alpha_{\gamma v J}(\omega)$ (see Section~\ref{ssec:structure}) is generically positive, so that molecules experience a negative energy shift $\Delta E = -\alpha_{\gamma v J}(\omega) \mathcal{E}^2/2$ due to the electric field $\mathcal{E}$ oscillating at frequency $\omega$.  Hence molecules are attracted to the most intense part of the optical field.  Much like for atoms, with typical laser power ($\sim 1\mathrm{W}$), focused beam diameter ($\sim 100~\mu \mathrm{m}$), and detuning ($\sim 0.1$ eV), the trap depth is on the order of $\sim 1$ mK and elastic (Rayleigh) photon scattering rates are acceptably low for long-term trapping. These optical traps are thus ideal for ultracold molecules, but not for cold molecules produced by direct cooling methods discussed in Section~\ref{ssec:direct}. However, a few key features of optical traps are different for molecules than for atoms.  First, inelastic (Raman) scattering occurs at much higher rates in molecules than in atoms, due to the rotational and vibrational structure: typically for molecules inelastic and elastic rates are roughly equal~\cite{Demille02}, while for atoms the inelastic rates are orders of magnitude smaller.  In addition, the tensor nature of $\alpha_{\gamma, v, J}(\omega)$ leads to splittings between $m_J$ sublevels for $J > 1/2$; while these are negligibly small in most laser-cooled atoms, in molecules these differential splittings can be comparable to the trap depth.  It is worth noting that the rotational and vibrational transitions in molecules contribute little to $\alpha_{\gamma, v, J}$ at optical frequencies; this is because the response of an oscillator at frequencies $\omega$ far above its resonance frequency $\omega_0$ falls off rapidly ($\propto \omega_0^2/\omega^2$), while at frequencies below resonance the response falls more slowly ($\propto 1/(\omega_0 - \omega)$).  All of these concerns must be taken into account when using optical lattices to achieve the fully quantum regime discussed in Sec.~(\ref{sssec:fullyQuantum}).

\subsection{Molecular Ions}
\label{ssec:ions}

An exciting new direction in the field involves the use of molecular ions rather than neutral species~\cite{Roth08}.  Here, trapping is straightforward due to the large depth of ion traps.  Cooling of the motional states of ions in a trap down to temperatures sufficient for Coulomb crystallization ($T < 100$ mK) has been demonstrated by sympathetic cooling with co-trapped, laser-cooled atomic ions~\cite{Molhave00,Roth05,Ostendorf05}.  Simpler but less extreme motional cooling can also be accomplished by equilibration with a cryogenically-cooled resistor coupled to the trap electrodes, or with cryogenic buffer gas~\cite{Pearson95}.  Cooling the internal states of molecular ions is more challenging: due to Coulomb repulsion, co-trapped atomic ions cannot interact with molecules at sufficiently short range to significantly affect their rotation or vibration.  However, several ideas have recently been proposed for addressing this issue. For example, it may be possible to sympathetically cool the internal states to very low temperature using laser-cooled neutral atoms~\cite{Hudson08b}.  Use of cryogenic neutral buffer gas such as He is also possible, but cannot provide sufficiently low temperatures to ensure population of the lowest $N=0$ rotational level for a typical molecule.  For polar species, the long trapping times possible with ions should allow relaxation to the ground vibrational state $v''=0$ simply via spontaneous emission; the vibrational energy for most molecules is large compared to $k_BT$ even at room temperature, so black-body-radiation-induced vibrational excitation should be insignificant.  Other proposals for rotational cooling include use of optical pumping on vibrational transitions, again taking advantage of trapping times long compared to the vibrational lifetime~\cite{Vogelius04}, or optical pumping via a coupling between rotational and trap motion~\cite{Vogelius06}.

There are a variety of intriguing prospects for use of ultracold molecular ions.  Applications to precision measurements can likely benefit from the long trapping time and low motional temperature.  Moreover, because the trapping potential is completely independent of the internal state, one can speculate that long coherence times may be possible for various types of superpositions.  Experiments to study time variation of constants~\cite{Schiller05}, the electron electric dipole moment~\cite{Sinclair05,Meyer06}, and parity-violating interactions~\cite{DeMille08b,Roth08} have been proposed using molecular ions.  In each case, the potential for long coherence times and narrow spectral lines make molecular ions potentially extremely sensitive to these effects.  It has also been recently proposed~\cite{Schuster09} to use molecular ions, rather than neutral species, for coupling to microwave stripline resonators in a hybrid quantum information processor (see Section \ref{ssec:quantumInfo}).  In all these cases, interesting questions arise regarding the effect of the ion trap itself on internal state coherence.  To investigate these questions, methods must be devised for state-selective detection of molecular ions~\cite{Bertelsen06}. In the paper by Hojbjerre \textit{et al}. in this special issue, rotational state-selective dissociation spectra were presented for translationally and vibrationally cold molecular ions with an aim toward rotational cooling~\cite{Hojbjerre09}.

An interesting demonstration of applying molecular ions to high-precision spectroscopy was recently performed with trapped HD$^+$ ions~\cite{Koelemeij07}.    A vibrational overtone line was excited with a diode laser.  State-selective detection of the excited state was performed by resonance-enhanced multiphoton dissociation, resulting in a change of shape in the co-trapped atomic ion cloud (which is easily imaged using cycling fluorescence).  The resulting spectral line had a width of $\sim 40$ MHz, dominated by Doppler shifts due to uncontrolled micromotion in the rf ion trap.  The energy splitting was determined with an accuracy of 0.5 MHz, over 100 times more precise than in previous measurements.

Molecular ions also open the prospect for new and interesting studies of molecular collisions~\cite{Roth08,Willitsch08a,Willitsch08b,Staanum08}.  Chemical and charge-exchange reactions with neutral species, photofragmentation, and similar processes are being studied at low temperatures for the first time.  Collisions of these type are relevant to many astrophysical problems, and also -- similar to the work described earlier with neutral species -- may provide precise benchmarks for theoretical studies.

\section{Applications and New Directions}
\label{sec:applications}

As demonstrated in the previous section, the technological progress of the field of cold and ultracold molecules has been extremely rapid.
We are therefore confident that, while technical challenges remain, a diverse selection of high density, ultralow temperature, stable molecules will soon become widely available in many laboratories worldwide. It is thus the proper time to look ahead for important scientific directions and applications involving these new forms of matter and technologies. In this section we focus on exciting applications that, in our opinion, are either well within our reach and already being pursued or will soon form new fields of research. Our discussion here will follow the structure of Section 2 where three major scientific directions are identified.

\subsection{Controlled Molecular Dynamics}
\label{ssec:detection}

As explained in Sections~\ref{sec:fundamental} and~\ref{sec:current} above, experiments with molecular gases at low and ultralow temperatures allow for studies of controlled microscopic interaction processes involving binary collisions of molecules.  This opens up numerous possibilities for new fundamental research.  We envision that theoretical and experimental studies of cold and ultracold molecules in external fields will in the near future give rise to the following research directions:

(i) {\it controlled cooling} including the development of new experimental techniques for producing dense ensembles of ultracold molecules, i.e., the techniques that will bridge the regimes of low and ultralow temperatures and lead to the production of ultracold polyatomic molecules;

(ii)  {\it ultracold chemistry} including the study of resonance-mediated reactions, and the role of quantum effects in determining chemical reactivity and reaction processes in confined geometries;

(iii) {\it cold controlled chemistry} including the study of external field effects on molecular interactions at temperatures near 1 K, shape resonances and differential scattering at low temperatures;

(iv) {\it coherent control} of ultracold molecular collisions;

(v) {\it quantum degenerate chemistry} including the study of collective many-body dynamics leading to molecular transformations in BECs and degenerate Fermi gases.

The need to develop new experimental techniques for the production of dense ensembles of ultracold molecules is clear and urgent. As described in Section~\ref{sec:current} of this article, existing experimental techniques generate ultracold molecules either by linking ultracold atoms (indirect methods) or by cooling thermal gases (direct methods).  The indirect methods are largely restricted to derivatives of alkali metal atoms, which limits the variety of reaction processes that can be currently studied in the limit of zero temperature. Some of the direct methods may produce large ensembles of cold molecules. However, cooling cold molecular gases to the ultralow temperature regime remains a significant challenge. The number of papers focusing on the development of new experimental methods for the production of ultracold molecules presented in this special issue~\cite{Lu09,Patterson09,Meek09,Salzburger09,Motsch09,Parazzoli09,Takase09,Tokunaga09,Narevicius09} speaks for itself. The existing  experimental techniques will also have to be extended to cool large polyatomic molecules. The production of ultracold polyatomic molecules will enable the study of new dimensions of ultracold chemistry as well as the synthesis of novel quantum materials.

Ten years after the first experiments on cooling and trapping molecules, it is safe to say that the creation of ultracold molecules has opened a new era of chemistry. As the dynamics of ultracold molecules is entirely determined by quantum effects, measurements of chemical reaction rates at zero temperature provide unique information about the role of tunneling, zero-point energy and quantum reflection effects in determining chemical reactivity. Research on chemistry at cold and ultralow temperatures may in the near future result in both practical and fundamental applications. When molecules are cooled to low temperatures, inelastic and reactive collisions become extremely state-selective and propensities for populating specific rovibrational states are enhanced~\cite{Balakrishnan01,Bodo04,Weck04,Weck05,Krems02,Krems05}. This could be used for the efficient production of atoms or molecules with inverted populations of internal energy levels and potentially the development of new atomic or molecular lasers. Photodissociation of molecules in a BEC may be used for controlled preparation of entangled pairs of radicals~\cite{Moore02}. Entanglement of spatially separated molecules is necessary for the study of quantum information transfer and the development of quantum computing schemes based on atomic and molecular systems.
The production of entangled molecules may also be used for the realization of coherent control of bi-molecular chemical reactions~\cite{Shapiro03}. As described in Section 3.1, chemical reactions in a cold buffer gas can provide a rich source for slow molecular beams~\cite{Patterson09}. Slow molecular beams may find a lot of applications in chemistry research~\cite{Hudson06a}, ranging from high-precision spectroscopy, to novel scattering experiments~\cite{Gilijamse06,Sawyer08a}, to studies of collective dynamics of strongly interacting systems at low temperatures.

Fano-Feshbach scattering resonances are ubiquitous in collision dynamics of ultracold atoms and molecules~\cite{Tscherbul09,Hutson07,Weck05,Cvitas05,Cvitas05b,Ticknor05a,Mack06,Yang06,Lara06,Chin05,Simoni09}.
 The mechanisms and the effect of Fano-Feshbach resonances on molecular collision properties at ultralow temperatures are well understood. By contrast, the effect of shape scattering resonances, which occur in a multiple partial wave scattering regime, on dynamics of cold molecules remains to be investigated.  Slow molecular beams and the molecular synchrotron~\cite{Heiner07} may allow for the study of shape resonances. Tunable shape resonances may offer new routes to control molecular dynamics and explore the role of long-range interactions in determining chemical reactions at low temperatures. The effect of shape resonances on thermally averaged chemical reaction rates remains a significant question in chemical dynamics research.

The ultracold temperature regime of intermolecular interactions may allow for the realization of coherent control of bi-molecular collision processes with laser fields. In its bare formulation, coherent control of molecular processes is based on quantum interference between distinct interaction processes leading to the same outcome. Schemes for coherent control of molecular scattering rely on the creation of coherent superpositions of internal states of the molecules moving with different relative momenta corresponding to the same momentum state of the center-of-mass motion~\cite{Shapiro03}. It is practically impossible to create such coherent superpositions in a gas with random molecular motion. The thermal motion of molecules becomes insignificant when the molecular gas is cooled to ultralow temperatures.
Herrera has recently proposed a method of coherent control of atomic and molecular scattering in ultracold gaseous mixtures~\cite{Herrera08}. The approach is based on creating coherent superpositions of different Zeeman or Stark states in the presence of a magnetic or electric field. At certain magnitudes of the fields, the energy splittings between the different angular momentum states of the collision partners become equal, which enables coherent control of collisions between distinct particles. This could be used for precision spectroscopy measurements of molecular structure in the presence of external fields and for detailed studies of matter-light interactions. This scheme of coherent control can be implemented at elevated collision energies using the molecular synchrotron~\cite{Heiner07}. The translational motion of molecules in the molecular synchrotron can be precisely controlled by external electric fields so it should be possible to create molecules in coherent superpositions of different Zeeman states moving with the same speed.

Nanodeposition of atoms and molecules on solid materials has recently become a rapidly growing research field. Lithography is a key technology which has enabled enormous progress both in fundamental research and in commercial applications \cite{O'Dwyer05,Ohmukai09}. Progress in lithography is usually discussed in the context of Moore's law, which expresses the trend in the semiconductor industry that the number of transistors on a chip approximately doubles every two years. In an effort to keep up with this rapidly shrinking transistor size, engineers and scientists explore alternative lithographic techniques such as atom or molecule  nanolithography. In contrast to conventional optical lithography, which uses a monochromatic beam of light, in atom lithography the roles of
light and matter are interchanged. Here, the beam of atoms or molecules is manipulated by optical fields
before being deposited on a solid surface. As the de Broglie matter wavelength of atoms and molecules is measured in picometers, diffraction effects are not a limiting factor for atomic or molecular nanolithography and its sub-optical-wavelength resolution approaches extremely short length scales, pushing the boundaries of nanofabrication to new limits. At the same time, masks made out of light are very easy to switch, move or modify quickly, which offers unprecedented flexibility to nanodeposition technology. As the interaction of light with matter is usually weak, successful implementation of molecular nanolithography relies on the preparation of low-energy, preferably ultracold, molecules. The existing technology of atomic nanolithography is presently limited to manipulating only a small selection of atoms amenable to laser cooling \cite{O'Dwyer05,Ohmukai09}. The experimental techniques discussed in this paper and in this special issue may provide new sources of molecular ensembles for nano-deposition of a wide variety of  molecules.

As described in Section~\ref{sec:current}, counterpropagating laser beams create standing waves that can be used to trap ultracold atoms and generate an optical lattice.  Photo- or magneto-association of ultracold atoms on an optical lattice will produce a lattice of molecules suspended in three dimensions by laser fields, which can be used to study collisions of individual molecules. Studies of molecular interactions in confined geometries may result in many fundamental applications~\cite{Krems08,Li08,Li09}. For example, the energy dependence of cross sections for both elastic and inelastic molecular collisions in confined geometries is different from the usual three dimensional behavior~\cite{Sadeghpour00,Li08}.
As a result, chemical reactions and inelastic collisions of molecules under external confinement are expected to be greatly modified. External confinement also changes the symmetry of long-range intermolecular interactions.  Studies of molecular collisions in low dimensions may therefore provide a sensitive probe of long-range intermolecular interactions and quantum phenomena in collision physics. Because the symmetry of a collision event can be completely destroyed by the combined effect of the confining laser fields and an external static electric field~\cite{Li08}, studies of chemical reactions in confined geometries may prove to be a novel way of probing stereochemistry. Measurements of the collision threshold laws for molecules in quasi-1D and quasi-2D geometries will provide ground-breaking results. Atomic and molecular ensembles in optical lattices can be used  as model systems for fundamental studies of quantum condensed matter physics~\cite{Ortner09} as well as quantum optics phenomena such as exciton polaritons in microcavity semiconductors~\cite{Lidzey99,Deng02}. The results of collision experiments with molecular systems in confined geometries may therefore be used to elucidate the dynamical behavior of excitons and exciton polaritons.

The efforts of many research groups are currently directed to producing molecular BECs and degenerate Fermi gases. Quantum degenerate gases of molecules will have unique
properties,  different from the properties of both atomic quantum degenerate gases and thermal
ensembles of molecules. Dynamics of molecules in quantum degenerate gases are determined by
collective effects and quantum many-body statistics. The many-body coherence makes the microscopic single-molecule
dynamics inseparable from the dynamical behavior of the entire quantum degenerate gas.
Of particular interest are non-linear collective phenomena such as
Bose-enhancement or Pauli blocking that may enable novel mechanisms for controlling molecular processes.
Bose-enhancement, first discussed in the context of chemistry by Heinzen and coworkers \cite{Heinzen00} and Moore and Vardi~\cite{Moore02}, amplifies
the probability of microscopic chemical reactions to produce macroscopic quantities of reaction products with perfect yield.
If confirmed experimentally, this idea may lead to the development of the research field of Bose-enhanced chemistry.

The idea of Bose-enhanced chemistry builds on the photoassociation spectroscopy experiments performed with
atomic BECs. As described in Section~\ref{sec:current}, photoassociation of ultracold atoms produces ultracold
diatomic molecules.  A BEC of atoms may thus be transformed into a BEC
of molecules. In the presence of laser light that couples the isolated atoms with molecules, the molecules may form and then coherently
dissociate back into their constituent atoms. This gives rise to oscillations between the atomic and molecular BECs.
The oscillation frequency is proportional to $\sqrt{N}$, where $N$ is the number of particles in the condensates.

Consider now a polyatomic molecule that dissociates under laser radiation into several distinct products. For example, a molecule ABC
may yield upon photodissociation the atom A and the molecule BC {\it or} the atom C and the molecule AB. For typical molecules, the branching ratios
of the photodissociation into the different products are normally on the order of unity. A major effort in chemical dynamics research has been devoted in the past three decades to
finding mechanisms for manipulating the photodissociation branching ratios  with external laser or static fields. Moore and Vardi showed that the branching ratios, if only slightly different from unity at thermal temperatures, are dramatically enhanced when molecules form a BEC before photodissociation.
The equations of motion describing the photodissociation dynamics of ABC molecules in a BEC are equivalent to
those of a parametric multimode laser. The process of photodissociation is therefore equivalent to parametric superfluorescence in quantum optics and
Bose-amplification leads to an exponential increase of the branching ratios. Unlike laser systems, where photons are always in the field of photons, Bose-enhanced dissociation of molecules represents a process of transforming complex bosons into bosons of different kind. The study of Bose-enhanced chemistry may therefore probe new interaction mechanisms in quantum degenerate systems.

While it remains to be seen if Bose-enhancement of photodissociation ratios can be realized experimentally, it is certainly worthwhile to explore the effects of
Bose statistics on bi-molecular chemical reactions in mixtures of molecular BECs. The effects of Pauli blocking may similarly enhance the photodissociation branching ratios by prohibiting certain reaction channels.  Molecular processes in quantum degenerate gases steered by Pauli blocking effects do not have analogs in quantum optics so the study of Pauli-blocked chemistry may uncover new fundamental processes and phenomena.

\subsection{High Resolution Spectroscopy and Quantum Control}

Ultracold molecules provide an ideal platform for ultrahigh resolution molecular spectroscopy. We are entering a qualitatively new regime where molecular transitions can be studied at the highest spectral resolution and coherence time, limited only by the natural lifetimes of relevant molecular energy levels~\cite{Meerakker05}. This will lead to knowledge of molecular structure and dynamics at an unprecedented level of precision. The high-resolution capability may also prove vital for some stringent tests of fundamental physical laws and symmetries. Conversely, detailed knowledge of molecular structure will be important for a thorough understanding and control of dynamics in the emerging field of cold molecule chemistry. For example, as we prepare molecules in the rotational and vibrational ground state of the ground electronic potential, the only possible degree of freedom left unaddressed is the nuclear spins, which can play an important role in molecular collisions. Understanding the nuclear spin states will allow us to establish effective state control and give us unprecedented knowledge of molecular hyperfine energy structures. Another example is the the paper by Kim \textit{et al}.~\cite{Kim09} in this special issue where state-selective detection methods can be used to explore high-lying Rydberg states of ultracold molecules opening a possible route to efficient formation of ground state molecular ions in low-lying rovibrational levels.

Precision laser tools are being developed to make further advances in spectroscopy based on samples of cold molecules, and detailed spectroscopic knowledge will enable developments in novel sensing and control of quantum properties of both molecular internal and external degrees of freedom. These advanced capabilities will include coherent control of their production processes, excitation pathways, spatial confinement, coupling to the environment, and sensitive trace detections. Cold molecular samples have already started to be used in collision and chemical reaction studies~\cite{Gilijamse07,Sawyer08a}. The precision control in both internal and external degrees of freedom will allow us to enter a qualitatively new regime of studying chemical reactions, providing unprecedentedly detailed and precise investigations of some of the most fundamental molecular interactions and chemical reaction processes. The knowledge gained will be of critical importance to the understanding of more complex molecular processes and will facilitate the development of novel sensing of trace molecules in both steady and transient states. However, an outstanding challenge does exist in this exciting scientific path. As molecular interactions and reactions take place at low temperatures, we need to monitor a multitude of quantum states with exquisite detection sensitivities. Optical frequency comb systems have been used to perform detailed molecular structure studies~\cite{Thorpe06,Thorpe08}. The unique properties of optical frequency combs -- precise spectral selectivity and high resolution, combined with a very large spectral bandwidth--led recently to the realization of direct frequency comb spectroscopy and high-resolution quantum control. Integrating direct frequency comb spectroscopy with cold molecules can indeed provide an efficient solution, allowing enhanced molecular manipulation and detection capabilities. In the first demonstration of this endeavor, a detailed tomographic mapping of the thermodynamic properties of a pulsed supersonic molecular jet was obtained by frequency comb spectroscopy, resulting in high-speed acquisition of information on the rovibrational and translational temperature distributions, as well as the spatial density distribution of the molecular beam in all three dimensions~\cite{Thorpe09}. An important scientific goal is to sensitively and quantitatively monitor multi-species molecular spectra over a large spectral bandwidth in real time, providing a new spectroscopic paradigm for studying molecular vibrational dynamics and chemical reactions at ultralow energies where spectral resolution and spectral coverage are both desired.

Further, cold and ultracold molecules provide excellent new opportunities to implement high-resolution quantum control in molecular systems, permitting efficient population transfer among different quantum states or molecular wave packets~\cite{Peer07,Shapiro08}. This will provide unique opportunities to study precisely controlled quantum chemistry. This new era in precise understanding and control of molecular interaction processes are based on two reasons. First, for ultracold molecules the external motional degree of freedom is effectively frozen out. Second, with the development of the concept and technology for high-resolution quantum control using highly stable optical fields in the form of phase-coherent pulse trains, we can selectively, precisely, and efficiently control coherent evolutions among different internal energy states~\cite{Marian04,Peer07}. Molecular quantum wave packets, superpositions of quantum states in molecules, can be manipulated with phase-coherent femtosecond pulse trains. The spectroscopic selectivity of a molecular excitation can be enhanced by exploiting quantum interferences between multiple molecular states, i.e. by targeted excitation of complex wave packets rather than incoherent mixture of single molecular states. The efficiency of such selective excitation can be dramatically boosted through the effect of coherent accumulation of quantum amplitudes if the excitation is executed by a sequence of ultra-short, mutually coherent laser pulses, rather than a single laser pulse~~\cite{Peer07,Shapiro08}. These ideas represent a new direction in quantum control of molecules with strong spectrally broad laser pulses.  Maximum transfer of the molecular population or maximum coherence can be achieved between well defined wave packet states with high frequency resolution and robustness.

 The vibronic X$^1\Sigma^+(v\! =\! J\! =\! 0)$ ground state of bialkali polar molecules possesses small but complex hyperfine structure (HFS) arising from nuclear spins. As described in the previous section, rovibronic ground-state molecules are produced by transferring population from states near the dissociation threshold. The intermediate states used for two-photon transfer can have complex HF-Zeeman structure in electronically-excited potentials. The precise state composition of these intermediate levels determines which HFS sublevels of the ground state molecules can be efficiently formed~\cite{Ni08}. Detailed structural understanding of excited state molecules is thus key for efficient state preparation and coherent quantum control of ultracold polar molecules. Understanding the excited state structure requires a precise study of the complex dynamics between the electron and nuclear spins, as well as the electron orbital and molecular rotational angular momenta. Under the assumption of a definite Hund's case (a) or case (c) coupling scheme, an effective spin-coupling treatment that describes the HFS and Zeeman structure of each rovibrational level of a single potential can be developed. The strength of HFS couplings can only be determined with precision spectroscopy that can be used to check the theory model. The understanding of the molecular structure will allow optimization of transition pathways.

To understand the collision dynamics and the behavior of the bialkali ground-state molecules under arbitrary external electric and magnetic fields, we need to develop a complete understanding of their HFS. The typical HFS energy scales in bialkali ground-state molecules are $\sim 10$ kHz ($\approx 500$ nK) for $J=0$ levels and $\sim 1$ MHz ($\approx 50$ $\mu$K) for $J=1$ levels~\cite{Aldegunde08}. The energy released in inelastic HFS-changing collisions between the ground-state molecules is likely insufficient to cause loss from a trap where the molecules are confined, but it is significant on the temperature scale of a quantum degenerate gas. Understanding of HFS will also allow us to polarize the ground-state molecules to a single HFS sublevel, a step that is necessary to reach quantum degeneracy and it is important for many applications including collision studies and quantum information science.

In the case of ground state molecules prepared by Stark deceleration, high-resolution spectroscopy of the ground-state structure has already been performed for NH$_3$~\cite{Veldhoven2004} and hydroxyl radicals (OH)~\cite{Hudson06b,Lev06}. For example, Stark deceleration of OH can prepare a molecular beam at a mean speed adjustable between 550 m/s to rest, with a translational temperature tunable from 1~mK to 1~K~\cite{Bochinski03,Bochinski04}. These velocity-manipulated stable "bunches" contain 10$^4$ to 10$^6$ molecules at a density of 10$^5$ to 10$^7$ cm$^{-3}$ and they are ideal for high resolution microwave spectroscopy using Rabi or Ramsey techniques. This has enabled an order of magnitude improvement in the precision microwave measurement of the ground state structure of OH, including the $\Lambda$-doublet and hyperfine splittings. As will be elaborated in the next subsection, comparing the laboratory results to those from OH megamasers in interstellar space will allow a sensitivity of 10$^{-6}$ for measuring the potential time variation of the fundamental fine structure constant $\Delta\alpha$/$\alpha$ over 10$^{10}$ years. Understanding of  the molecular structure has been enhanced through such high resolution spectroscopy work. For example, the study of the low magnetic field behavior of OH in its $^2\Pi_{3/2}$ ro-vibronic ground state precisely determines a differential Land\'{e} $g$-factor between opposite parity components of the $\Lambda$-doublet. The energy level perturbations require a new theoretical model as a subtle consequence of higher-order angular momentum perturbations beyond current molecular hyperfine Zeeman theory.

These results highlight the ability of cold molecules not only to enhance our understanding of unexplored regimes of molecular coupling, but also to contribute to research on dipolar gas physics, quantum information processing, and the precision measurement of variations in fundamental constants. In particular, the influence of small magnetic field on hyperfine structure and cold collisional properties of polar molecules must be understood with high precision for the control of ultracold dipolar matter---for instance, via magnetic- and electric-field enhanced resonance effects---and to perform quantum computation with the well-suited polar molecules. Through high precision spectroscopy one can find that these subtle angular momentum perturbations surprisingly provide us with transitions uniquely suited to function as practical qubits in polar molecules such as OH.  Specifically, the refined hyperfine Zeeman theory allows identification of magnetic fields at which specific transitions are extraordinarily insensitive to field fluctuations.  In addition, these transitions are between magnetically trappable states of equal magnetic moment---which suppresses trap heating---and have large, equal-but-opposite electric dipole moments well suited for inducing robust quantum logic gates.

\subsection{Tests of Fundamental Physical Laws}
\label{ssec:DiscreteSymmetries}

High resolution spectroscopy work paves the way for high precision measurement. Some recent work has focused on tests of the variation of the fine structure constant $\alpha$~\cite{Hudson06b} and the proton-to-electron mass ratio $\mu\equiv m_p/m_e$~\cite{Flambaum2007,Schiller2007,Zelevinsky08,Demille08a,Shelkovnikov2008}. The values of these constants may drift monotonically over time, or vary periodically with the Sun-Earth distance in case of the existence of gravitational coupling. State-of-the-art optical atomic clocks have set the most stringent limits on both types of $\alpha$ variations~\cite{Rosenband08,Blatt2008}, but atoms generally lack transitions that can reveal relative $\mu$ variations ($\Delta\mu/\mu$) in a model-independent way. Molecules can play a powerful role here. If $\mu$ changes, the vibrational (and rotational) energy levels of molecules move relative to their electronic potentials.  The effects on the weakly bound vibrational levels (near dissociation) and on the most deeply bound levels (near the potential minimum) are expected to be much smaller than that on the moderately bound levels at intermediate internuclear distances~\cite{Demille08a,Zelevinsky08}. This can allow accurate spectroscopic determinations of $\Delta\mu/\mu$ by using the least sensitive levels as frequency anchors.  Two-color optical Raman spectroscopy of vibrational energy spacings within a single electronic potential takes advantage of the entire molecular potential depth in order to minimize the relative measurement error. Similar gains in sensitivity can be achieved when two molecular potentials cross such that a moderately bound vibrational level is nearly degenerate in energy with a weakly bound vibrational level from a different potential. This allows a differential measurement in the microwave frequency domain for the relative energy shifts between these two vibrational levels~\cite{Demille08a}. Similarly, fine-structure splitting can be matched with vibrational splitting and hence the microwave probe would be sensitive to both $\alpha$ and $\mu$ variation simultaneously~\cite{Flambaum2007}. Ideas on the use of Fano-Feshbach resonances to enhance the sensitivity to $\Delta\mu/\mu$ in energy levels near the scattering threshold have also been proposed~\cite{Chin09} in this special issue. In this same issue, Kajita proposes to study pure vibrational transition frequencies of magnetically trapped cold XH molecules to measure the time-dependent variations in $\mu$~\cite{Kajita09}.

The $J=3/2$ $\Lambda$-doublet of OH~\cite{Hudson06b,Lev06}, including both the main lines ($\Delta$F = 0) and the magnetically sensitive
satellite-lines ($\Delta$F = $\pm$1), are at the focus of intense astrophysical research because they can provide a measurement of the variation of the fine-structure constant and the electron-proton mass ratio by comparing the energy levels of Earth-bound and cosmological OH molecules~\cite{Darling03,Kanekar03,Kanekar05}.  Urgently needed are more precise measurements of these transition frequencies to compare to high-resolution radio telescope data expected in the near future. Precise Earth-bound transition data are now possible to obtain in this new era of high-resolution microwave spectroscopy using cold molecules.  The improved understanding of the subtle effect of small magnetic fields on the transition frequencies, as well as the slow, monoenergetic packet of OH from the Stark decelerator, allows for a tenfold precision improvement on all of these Lambda-doublet lines. The comparison of different atomic clock systems has provided tight constraints on the time variation of various fundamental constants during the modern epoch. However, an observation of absorption lines in distant quasars~\cite{Webb2001,Quast2004} provides conflicting results about
possible $\alpha$ variation over cosmological time. Recently, there has been much interest in using OH megamasers in interstellar space
to constrain the evolution of fundamental constants~\cite{Darling03,Kanekar03,Kanekar05} with several key advantages. Most importantly, the multiple lines (that have different dependence on the fundamental constants) arising from one of these localized sources differentiate the relative Doppler shift from the actual variation in the transition frequency. Furthermore, it can been shown that the sum and difference of the $\Delta$F = 0 (F is the total angular momentum) transition frequencies in the ground $\Lambda$-doublet of OH depend on $\alpha$ as $\alpha^{0.4}$ and $\alpha^{4}$, respectively~\cite{Darling03}. Thus, by comparing these quantities as measured from OH megamasers to laboratory values, it is possible to remove the Doppler shift systematics and constrain $\alpha$ over cosmological time.  All four lines must be detected from the same source, and the closure criterion, i.e., the zero difference of the averages of the main- and satellite-line frequencies, provides a critical systematics check. Because of the unique properties of the $\Lambda$-doublet, the $\Delta$F = 0 transitions are extremely insensitive to magnetic fields, while the $\Delta$F = $\pm$1 satellite transitions can be used to calibrate the B-field. However, for the current limits on $\Delta\alpha/\alpha$ the change in the relevant measurable quantities is on the order of 100 Hz, which is the accuracy of the best laboratory based measurement to date~\cite{terMeulen72}. Moreover, an astrophysical measurement of OH megamasers, scheduled for in the near future, expects a resolution better than 100 Hz, and thus better laboratory measurements of the OH $\Lambda$-doublet are imminently needed to allow for tighter constraints on $\Delta\alpha/\alpha$.

Another interesting and important example where molecules hold significant promise is in the search for parity (P) and time reversal (T)-violating permanent electric dipole moments (EDMs)~\cite{Landau57}.  Here, the effect of interest is an EDM $\mathbf{D} =  D \mathbf{S}/S$ along the spin axis $\langle\mathbf{S}\rangle$ of any particle.\footnote{This conventional definition ensures that for the state of the particle with maximum spin projection $m_S = S$, the expectation value of $\mathbf{D}$ has magnitude $|\left\langle\mathbf{D}\right\rangle | = |\left\langle S,m_S | \mathbf{D} |S,m_S\right\rangle | = D$.}  This gives rise to a Hamiltonian $H_{\mathrm{EDM}} = -\mathbf{D} \cdot \hbox{\boldmath{$\mathcal{E}$}} \propto \mathbf{S} \cdot \hbox{\boldmath{$\mathcal{E}$}}$.  (Note that this differs essentially from the ``permanent'' EDM of a polar molecule, which has no net component along the molecular angular momentum $\mathbf{J}$ and hence does \textit{not} violate P or T symmetry.)  Such P- and T-violating EDMs can be induced in fundamental particles such as the electron, by the virtual emission and reabsorption of particles of very high mass~\cite{Khriplovich97}. Hence the search for EDMs effectively probes physics at very high energy scales.  Searches for EDMs are widely considered to be key measurements for understanding the observed asymmetry between matter and antimatter observed in the universe~\cite{Huber07}, since the only known explanations for this asymmetry require new types of T-violating effects associated with new particles at high energies.

Consider the case of the electron EDM, $d_e$~\cite{Commins98}.  It has long been known that, when an external electric field $\mathcal{E}_{\mathrm{ext}}$ is applied to an atom or molecule, a much larger internal \textit{effective} field $\mathcal{E}_{\mathrm{eff}}$ can act on the EDM of the electrons bound in the system~\cite{Sandars65,Sandars66}.  This enhancement occurs due to the relativistic motion of the bound electrons, and hence increases rapidly with the nuclear charge $Z$ (for molecules, of the heaviest constituent atom). It is convenient to write the effective field in the form $\mathcal{E}_{\mathrm{eff}} = Q\mathcal{P}$, where $Q$ is a factor that includes the relativistic effects as well as details of atomic (or molecular) structure, while $\mathcal{P}$ is the degree of polarization of the atom or molecule by the external field~\cite{DeMille00}. For typical paramagnetic atoms or molecules, $Q \approx 4 \cdot 10^{10}\:\mathrm{V}/\mathrm{cm} \times \left[\frac{Z}{80}\right]^{3}$. The main difference between atoms and molecules occurs in the factor $\mathcal{P}$. For typical atoms $\mathcal{P} \approx 10^{-3}$ even when subjected to the maximum attainable laboratory fields $\mathcal{E}_{\mathrm{ext}} \approx 100~\hbox{kV/cm}$. By contrast, in a typical polar diatomic molecule, nearly complete polarization $(\mathcal{P} \approx 1)$ can be achieved with relatively modest external fields $\mathcal{E}_{\mathrm{ext}} \approx 10 - 10^{4}$ V/cm.  Thus, $\mathcal{E}_{\mathrm{eff}}$ is approximately 3 orders of magnitude larger in molecules than the maximum attainable value with atoms~\cite{Sandars67}.  A similar amplification effect for molecules versus atoms occurs for the nuclear Schiff moment, a P- and T-odd effect closely related to the EDMs of nuclei~\cite{Sandars67,Khriplovich97,Cho91}. (See Ref. \cite{Flambaum02} for a particularly clear discussion of the relationship between EDMs and the Schiff moment.)

In addition to their intrinsic sensitivity, certain features of molecules can be used to very deeply suppress systematic errors in this type of experiment, relative to previous work in atoms.  For example, in electron EDM experiments, the electron's magnetic moment $\hbox{\boldmath{$\mu$}}  \propto \mathbf{S}$ can couple to magnetic fields $\hbox{\boldmath{$\mathcal{B}$}}_{\mathcal{E}}$ correlated with $\hbox{\boldmath{$\mathcal{E}$}}$, leading to spurious signals.  One source of correlated magnetic fields is relativistic: an atom or molecule moving through $\hbox{\boldmath{$\mathcal{E}$}}_{\mathrm{ext}}$ at velocity $\mathbf{v}$ experiences a motional magnetic field $\hbox{\boldmath{$\mathcal{B}$}}_{\mathrm{mot}} = \frac{\mathbf{v}}{c} \times \hbox{\boldmath{$\mathcal{E}$}}_{\mathrm{ext}}$.  Effects due to $\hbox{\boldmath{$\mathcal{B}$}}_{\mathrm{mot}}$ appear only in $2^{\mathrm{nd}}$ order in molecules, due to the cylindrical symmetry of the molecular state together with the fact that $\hbox{\boldmath{$\mathcal{B}$}}_{\mathrm{mot}}$ is perpendicular to $\hbox{\boldmath{$\mathcal{E}$}}_{\mathrm{ext}}$~\cite{Player70,Hudson02}.  Another source of $\hbox{\boldmath{$\mathcal{B}$}}_{\mathcal{E}}$ is leakage currents.  These can be suppressed by using molecules with $\Omega$-doublet structures~\cite{Kawall04}.  In the presence of $\hbox{\boldmath{$\mathcal{E}$}}_{\mathrm{ext}}$, these closely-spaced levels mix completely, resulting in a pair of levels with $\mathcal{P} = \pm 1$.  Hence, in these states $\hbox{\boldmath{$\mathcal{E}$}}_{\mathrm{eff}}$ has opposite sign, while their magnetic moments are very nearly equal.  In the difference between EDM energy shifts in both states, systematic shifts due to leakage currents (and many other spurious effects) cancel to high precision, typically $\sim 10^{-3}$ or more.  Even more improvement can be obtained by using molecules in states where the magnetic moment due to orbital motion nearly cancels that due to the electron spin, e.g. $^3\Delta_1$ states~\cite{Sinclair05,Meyer08,Shafer-ray06}.  Finally, the small values of $\hbox{\boldmath{$\mathcal{E}$}}_{\mathrm{ext}}$ needed to achieve $\mathcal{P} \approx 1$ give smaller leakage currents and also yield EDM energy shifts that are independent of $|\hbox{\boldmath{$\mathcal{E}$}}_{\mathrm{ext}}|$ over a wide range, unlike most systematic effects.  Taken together, the best molecular systems promise control over systematic errors at a level many orders of magnitude beyond that achieved in atoms. Because of these many advantages, several experiments are now underway to search for the electron EDM $d_e$ using molecules.  At present the most mature experiments use hot molecules (PbO~\cite{Kawall04}) or cold but fast-moving molecules (YbF~\cite{Sauer06}); a next generation of experiments plan to use cold, slow buffer-gas beams (ThO~\cite{Vutha08}) or trapped molecular ions (HfF$^+$~\cite{Sinclair05}).

The structure of molecules also leads to the enhancement of effects due to electroweak forces~\cite{Labzovskii78,Sushkov78}.  In particular, molecules have been identified as useful for the study of nuclear spin-dependent (NSD) parity violating (PV) effects~\cite{Flambaum85,Kozlov95}.  Such effects arise primarily from two underlying mechanisms~\cite{Flambaum80,Ginges04}.  One is the exchange of virtual $Z^0$ bosons between electrons and a constituent nucleus.  The other is a two-step process: exchange of $Z^0$ and $W^{\pm}$ bosons betweens nucleons deforms the structure of the nucleus to produce a so-called ``anapole moment'', and then the molecular electrons probe this deformation.  The direct $Z^0$ exchange interaction is described precisely by standard electroweak theory, but the NSD part of this process is numerically small and essentially unmeasured so far.  The anapole moment effect probes the way that electroweak interactions are modified by strong interactions within the nuclear medium.  The measurement and understanding of these modifications have been an elusive goal of the nuclear physics community for several decades~\cite{Haxton01}. Only one observation of a nuclear anapole moment has been reported~\cite{Wood97}, and the measured value is in poor agreement with predictions based on the few other measurements of electroweak effects in nuclei~\cite{Haxton01,Ginges04}.

Within molecules, the NSD-PV effects can be described by a simple Hamiltonian: $H_{\mathrm{NSD-PV}} = \kappa C (\mathbf{S} \times \mathbf{I})\cdot \hat{e}_n$. Here $\kappa$ is a dimensionless parameter describing the strength of the underlying electroweak physics; $C$ is a constant calculated from the electronic wavefunction; $\mathbf{S}$ is the electron spin; $\mathbf{I}$ is the nuclear spin; and $\hat{e}_n$ is the direction of the internuclear axis.  $H_{\mathrm{NSD-PV}}$ mixes adjacent hyperfine/rotational states of opposite parity in molecules.  By contrast, in atoms the same NSD-PV effects can only mix \textit{electronic} states of opposite parity, which are typically about four to five orders of magnitude more distant in energy.  Hence, by simple perturbation theory the mixing due to NSD-PV is enhanced by a large factor in molecules versus in atoms.  However, even greater enhancements are possible, by using a magnetic field $\hbox{\boldmath{$\mathcal{B}$}}$ to Zeeman-shift the opposite parity states to near degeneracy~\cite{Kozlov91,Flambaum85}. With typical splittings between rotational levels, this can be accomplished with applied fields of $\mathcal{B} \sim 1$ T.  The effective energy separation is then determined either by magnetic field inhomogeneities or by the level width due to finite interaction time.  In practice, this can enhance the NSD-PV mixing in molecules by a factor of roughly $10^{11}$ compared to in atoms.  This mixing can be measured e.g. by observing an AC Stark shift between the near-degenerate levels, on application of an oscillating electric field $\hbox{\boldmath{$\mathcal{E}$}}_{\mathrm{osc}}$ parallel to $\hbox{\boldmath{$\mathcal{B}$}}$~\cite{DeMille08b}.  Due to interference between the Stark- and NSD-PV induced mixings of the level, the AC Stark shift depends on the phase of $\hbox{\boldmath{$\mathcal{E}$}}_{\mathrm{osc}}$ and on the tunable separation of the levels. An experiment based on this principle is underway, with the initial goal to measure the nuclear anapole moment of $^{137}$Ba in the molecule BaF~\cite{DeMille08b}.

A few other tests of discrete symmetries have focused in the past on molecules, and may benefit from the emerging technologies to reach ultralow temperatures.  For example, parity-violating electroweak interactions should cause small energy differences between different enantiomeric states of chiral molecules~\cite{Quack08}.  Although it seems unlikely, there has been speculation that this difference could have provided the bias needed to seed the observed handedness of biological molecules.  To date no experiment has reached the energy resolution needed to observe this difference, but sources of cold chiral molecules could improve these measurements.  Another example is the search for violations of the spin-statistics relation.  In molecules with constituent nuclei that are identical $I=0$ bosons (e.g. $^{16}\mathrm{O}_2$), half of all possible rotational states are forbidden to exist (e.g. in the $^3\Sigma^-_g$ ground state of $^{16}\mathrm{O}_2$, states with even values of the rotational quantum number $\mathcal{R}$ are absent).  Sensitive searches for population in these forbidden states can be interpreted as limits on the possible degree to which bosons must exist in exchange-symmetric states~\cite{deAngelis96,Hilborn96}. Use of cold molecules could enhance population in low-lying levels, and also improve detection sensitivity through the use of narrower lines with no Doppler broadening.

\subsection{Quantum Information}
\label{ssec:quantumInfo}

Ultracold polar molecules have a number of properties that make them attractive for quantum information processing.  Like atomic systems, molecules have a rich internal state structure, including long-lived internal states which can be used to encode quantum information.  However, the large ($\sim$ 1 $ea_0$) electric dipole moment associated with rotational structure in polar molecules provides new means for quantum control beyond those available in atoms.  For example, it is easy to induce and manipulate ``permanent'' molecular electric dipole moments using DC or microwave-frequency electric fields.  The technology for controlling such fields to high precision is routine, and provides a natural means for integration with microelectronic circuits.  Moreover, the electric dipole-dipole interaction makes it possible to couple the internal states of molecules.  The long-range nature of dipole-dipole interaction, together with the strength of \textit{electric} dipoles, makes the coupling effective even at moderate distances.  The speed of conditional logic operations is proportional to the strength of this interaction, and hence with molecules it appears possible to engineer fast logic gates between remote qubits.  This provides a promising path towards scaling to large networks of coupled qubits.  We note as well that the rich internal structure of molecules--including many long-lived spin, rotational, and vibrational levels--makes them natural candidates for use as ``qudits'', with more than one bit of quantum information stored in each molecule~\cite{Glenn06,Zadoyan01,Bihary02}.  However, here we focus on the simpler (and more easily scalable) use of molecules as qubits.

Many of the interesting features of polar molecules as qubits--as well as the issues that arise when considering their use--can be illustrated with a specific example: the system of polar molecules trapped in an optical lattice~\cite{Demille02}.  Here molecules with a simple rigid rotor structure are used, in the presence of a static, polarizing electric field $\hbox{\boldmath{$\mathcal{E}$}}_{\mathrm{ext}} = \mathcal{E}_{\mathrm{ext}}\hat{z}$.  Qubits are encoded into the two lowest levels of the rotational manifold of states, as follows.  Logical qubit state $\left| 0 \right\rangle$, defined as the state correlated with the lowest rotational level $\left| J=0 \right\rangle$ when $\mathcal{E}_{\mathrm{ext}} = 0$, has net electric dipole moment $\mathbf{D}_{\left| 0 \right\rangle} = d_{\left| 0 \right\rangle}\hat{z}$, where $d_{\left| 0 \right\rangle} > 0$.  (Recall that the rotational eigenstates $\left| J, M_J\right\rangle$ are mixed by the electric field.) Logical qubit state $\left| 1 \right\rangle$, correlated with the first excited rotational level $\left| J=1,M_J=0 \right\rangle$, has net electric dipole moment $\mathbf{D}_{\left| 1 \right\rangle} = d_{\left| 1 \right\rangle}\hat{z}$, where $d_{\left| 1 \right\rangle} < 0$.  This two-level subsystem is well-isolated from other internal states of the molecule, because of the large anharmonicity of the rotational structure and its small scale compared to vibrational and electronic splittings.  For a single qubit, arbitrary superpositions of $\left| 0 \right\rangle$ and $\left| 1 \right\rangle$ can be prepared by applying pulses of microwave electric field, with frequency $\omega$ tuned to resonance with the energy needed to flip the dipole, i.e. such that $\hbar \omega = \hbar \omega_0 + d_{eff}\mathcal{E}$, where $\hbar\omega_0 = 2B$ is the field-free rotational splitting ($B$ is the rotational constant) and $d_{eff} \equiv d_{\left| 0 \right\rangle} - d_{\left| 1 \right\rangle}$ is the effective dipole moment, and $\mathcal{E}$ is the total electric field experienced by the molecule.  In general $\hbox{\boldmath{$\mathcal{E}$}} = \hbox{\boldmath{$\mathcal{E}$}}_{\mathrm{ext}} + \hbox{\boldmath{$\mathcal{E}$}}_{\mathrm{int}}$, where $\hbox{\boldmath{$\mathcal{E}$}}_{\mathrm{int}}$ is the electric field produced by other molecules in the system. Under typical conditions  $\mathcal{E}_{\mathrm{int}} \ll \mathcal{E}_{\mathrm{ext}}$.

A memory register can be created by trapping individual molecules in the sites of an optical lattice.  For a sufficiently cold sample (with internal temperature $T \ll B/k_B$), all qubits initially will be in the state $\left| 0 \right\rangle$.  Loading the lattice requires high phase space density $\Omega \sim \hbar^3$ of molecules, corresponding to number density $n \approx (\lambda/2)^3$ (where $\lambda$ is the optical lattice period) and motional temperature $T \ll V_{\mathrm{trap}}/k_b$ (where $V_{\mathrm{trap}} \lesssim 1$ mK is the optical lattice depth).  Individual sites in the lattice can be addressed spectroscopically, if $\mathcal{E}_{\mathrm{ext}}$ has a gradient along the real lattice vector(s) so that the resonant frequency $\omega_0$ depends on the spatial location of the molecule; this suffices for single-bit gates.  Conditional logic gates (such as controlled-NOT) can also be performed spectroscopically.  This is because the resonant frequency for a given molecule depends on the total electric field $\mathcal{E}$, and the internal field $\hbox{\boldmath{$\mathcal{E}$}}_{\mathrm{int}}$ depends on the orientation (i.e. logical state) of neighboring dipolar qubits.  For molecules traped in a plane normal to $\hat{z}$, $\hbox{\boldmath{$\mathcal{E}$}}_{\mathrm{int}}$ has a simple form: the internal field at molecule $a$ due to molecule $b$ located a distance $R_{ab}$ away is $\hbox{\boldmath{$\mathcal{E}$}}_{\mathrm{int}ab} = -d_b/R^3_{ab}\hat{z}$.  Thus $\mathcal{E} = \mathcal{E}_{\mathrm{ext}} + \mathcal{E}_{\mathrm{int}}$, the resonant frequency of qubit $a$ depends in a simple way on the state of its neighbors, and rotations of $a$ can be performed conditionally on these states.

It may appear that conditional logic operations will be impossibly complex in this system, due to the ``always on'' couplings between all qubits in the system.  However, in principle this is not a difficulty.  Unwanted couplings can be effectively removed from the system by the technique of ``refocusing'' as used in NMR-based quantum computing implementations~\cite{ike-book}.  Refocusing is essentially a procedure to average away couplings to unwanted bits, by deliberately flipping their states during the interaction time.  This makes it possible to engineer effective couplings between a single pair of qubits.  This method has been shown to be computationally efficient (in the sense that only polynomial resources are needed to perform the refocusing), at least in principle~\cite{Leung00}.  Hence, we consider a simplified scheme in which only interactions between nearest neighbor qubits are taken into account. Then, performing CNOT operations requires the ability to spectroscopically resolve qubit transitions differing in energy by $\Delta E = d_{\mathrm{eff}}^2/(\lambda/2)^3$, where $\lambda$ is the optical lattice period and hence $\lambda/2$ is the minimum value of $R_{ab}$.  This in turn means that conditional logic operations require a time of at least $\tau \sim \hbar/\Delta E$.  For typical conditions ($d_{\mathrm{eff}} \approx 1~ea_0$, $\lambda \approx 1~\mu$m), $\tau \sim 10~ \mu$s.

Decoherence in this system can arise from a number of sources.  Spontaneous (and even blackbody-induced) emission lifetimes are quite long, typically $\gtrsim\! 10^2$ sec.  A variety of technical noise sources must be controlled very carefully.  For example, noise in the value of $\mathcal{E}$ or the intensity of the trap laser leads to decoherence of the qubit states.  The latter effect is due to the fact that there is in general a difference between the optical trapping potentials for the $\left| 0 \right\rangle$ and $\left| 1 \right\rangle$ states, which can be traced to the anisotropic polarizability tensor $\alpha^{qq'}_{\gamma v J}(\omega)$ for a diatomic molecule.  More fundamentally, the optical trap itself is an irreducible source of decoherence in this system: the molecule can be excited to higher rotational or vibrational states by inelastic Raman scattering of trap laser photons.  (This effect also arises from the polarizability anisotropy.) For sufficiently large detuning of the trap laser from an electronic resonance, the Raman scattering rate $R \propto \lambda^{-3}$.  Hence, in this limit the maximum number of possible conditional gate operations, $N_g = R \tau$, is independent of the trap wavelength $\lambda$.  Under realistic conditions (described in detail in~\cite{Demille02}), decoherence lifetimes limited by Raman scattering rates in the regime $R^{-1} \sim 1-10$ s may be achievable, corresponding to $N_g \sim 10^5-10^6$.

A variety of technical issues can degrade the fidelity of gate operations.  One obvious potential problem comes from empty or multiply-occupied sites in the lattice filling.  However, the strong dipolar interactions between molecules should deeply suppress such defects: the naturally strong and repulsive (dipolar) interactions between molecules in their initial state $\left|0\right\rangle$ can lead to a robust Mott insulator phase~\cite{Pupillo04} and make multiple occupancy energetically impossible.  Even with a perfectly filled lattice, the modest difference in trap potentials for the qubit states (due both to the polarizability anisotropy, and to the force from the electric field gradient) means that single-bit gates can induce unwanted motional heating of the molecules.  This, and any other mechanism leading to finite temperature, then gives rise to uncertainties in qubit transition frequencies.  Most unwanted effects can be deeply suppressed by performing gate operations slowly enough that they are adiabatic with respect to the motion of molecules in their lattice site.  Typical motional frequencies in a lattice site are $\lesssim 100$ kHz; hence this adiabatic condition may limit the speed of single-bit operations, but has little effect on anticipated two-qubit gate times.

At the end of any computation with this system, the final state of each qubit must be read out.  Efficient detection and resolved imaging of molecules spaced by $\lambda/2$ will be challenging, but plausibly achievable.  For example, one likely approach uses state-selective photoionization, electrostatic magnification of the resulting array of ions, and subsequent imaging detection of the output ions.  (This is of course a fully destructive measurement: it not only collapses the qubit states, but actually rips the qubits apart.) However, the apparent condition on spatial resolution of the readout can be significantly relaxed with a shift-register readout strategy.  Here, by using the same spectroscopic addressing ideas as for gates, qubit state $\left| 1 \right\rangle$ can be transferred to an auxiliary storage state (such as the next rotational level) at lattice site number $i, i+n, i+2n$, etc. by applying appropriate microwave pulses.  This auxiliary state can be selectively detected as before.  Readout of states at these regular, selected lattice sites requires spatial resolution of only $n\lambda/2$.  The process can be repeated $n$ times, with $i$ incremented by 1 each time, to read out the entire array.

This system provides concrete examples of some of the key advantages of molecular dipolar qubits. It is expected that neutral polar molecules can be assembled in large numbers and at high density into regular structures~\cite{Greiner02}, similar to neutral atoms.  This may provide advantages relative to other more mature systems such as trapped atomic ions, where scalability may require the ability to move ions between a large number of remote trapping sites~\cite{Kielpinski02}.  In neutral atoms many proposals have been made to engineer interactions with collisions induced by moving two atoms into the same trap site~\cite{Jaksch99}.  In this special issue, a related scheme is proposed, in which the interaction is engineered using, instead of collisions, radiofrequency magnetoassociation of Li and Cs atoms into weakly-bound states of LiCs~\cite{Brickman09}.  By contrast, the long-range nature of the dipolar interaction in deeply-bound polar molecules eliminates the need for any motion.  A similar advantage arises in proposals to excite neutral atoms to Rydberg states in order to amplify long-range interactions~\cite{Jaksch00}.  However, here, unlike in molecules, the excited state has a short lifetime and generally feels dramatically different trapping potentials than the ground state sublevels used for storage.  Moreover, the amplitude of DC or microwave electric fields used to manipulate molecules are easy to apply and control dynamically, as compared to the intensity of lasers used for control of ions and neutral atoms. In many neutral-atom systems, addressing remains a difficult problem as compared to the simple solution outlined here for molecules.  Finally, as compared to condensed-matter systems, polar molecules provide many of the same advantages as atomic systems, e.g. exactly reproducible qubits, (potentially) long decoherence times, etc.

Considering the details of this system also highlights several disadvantages that arise from the use of deeply-bound, dipolar molecules as qubits.  For example, the non-negligible coupling between the external (trapping) potential and the internal (qubit) states leads to conflicting requirements for the optical lattice: on the one hand, to enable fast gates one would prefer shorter lattice wavelengths and deeper traps (higher optical intensity), while on the other hand one prefers longer wavelengths and lower intensity to reduce decoherence.  This puts a premium on using molecules with the lowest possible temperature, so that the trap depth can be reduced; in practice, the molecular sample must have phase space density near the regime of quantum degeneracy.  Similarly, the nice feature of qubit addressability comes with a $0^{\mathrm{th}}$-order sensitivity to electric field noise.  In addition, the need to employ refocusing methods to effectively control the ``always-on'' dipole-dipole interaction significantly increases the complexity of the system.  Finally, any realistic readout procedure will be destructive, i.e. at the least it will remove the molecule from the trap. A variety of methods have been proposed to alleviate these types of problems.

One promising idea is to encode the logical qubits into hyperfine sublevels of molecules with internal spin substructure, instead of directly into rotational states~\cite{Andre06}. This is of course similar to the encoding used in both neutral atoms and ions. The dipole moments and optical polarizability of such sublevels are nearly identical: residual differences are suppressed, compared to the case of rotational encoding, by roughly the ratio of hyperfine splitting to rotational splitting, typically $\gtrsim 10^{-3}$.  Moreover, the dipole-dipole interactions, while still present, are essentially independent of the qubit states, and hence provide only an overall energy offset.  In this scheme, single-qubit gates are performed by microwave Raman transitions between the qubit states, with an excited rotational level as a (virtual) intermediate state.  This yields gate speeds comparable to those in the original case.  To perform two-qubit gates--which are needed only a small fraction of the time for any individual qubit--one logical state can be transferred to the rotational excited state with a different dipole moment, where the interactions needed for conditional logic are present as before.  In a similar vein, it has been proposed to encode qubits in molecular states with negligibly small dipole moments, and to transfer these to ``active'' dipolar states only when needed for gate operations~\cite{Kuznetsova08a,Lee05,Yelin06,Charron07}.  Non-polar states are commonly available, e.g. in high vibrational levels of the ground state potential (near the dissociation limit), where the electronic wavefunctions of the separated atoms have not substantially hybridized.  Transfer to and from the strongly polar $v=0$ state can be accomplished via a laser-driven stimulated Raman process.

\begin{figure}[t]
 \begin{center}
\includegraphics[scale=1.0]{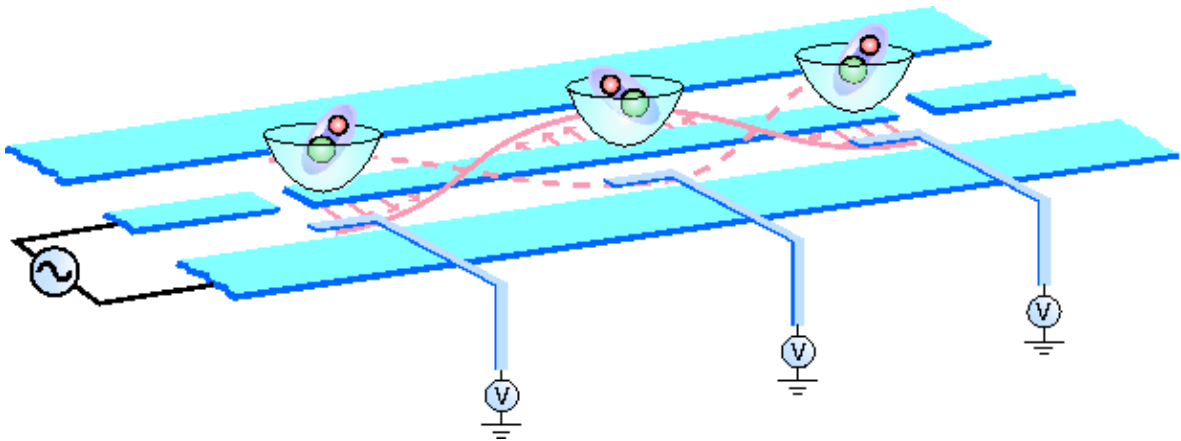}
\caption{Schematic representation of a hybrid quantum device. Molecules are trapped just above the surface of mesoscopic features patterned on the surface of a substrate (the ``chip''). The proximity of molecules to the surface can lead to strong couplings between the internal and/or motional states of the molecules, and the quantum states of objects on the chip. In the example shown here, molecules are trapped by static electric fields formed by on-chip electrodes. The rotational states of the molecules couple strongly to microwave photons confined in the stripline geometry.}
\label{fig:hybrid}
 \end{center}
\end{figure}

A radically different architecture using polar molecule qubits has also been proposed.  Here, the idea is to trap molecules, using inhomogeneous DC or microwave electric fields, near mesoscopic electrodes patterned on the surface of a chip (see Fig. \ref{fig:hybrid})~\cite{Andre06,Rabl06,Sorensen04}.  As before, qubits can be encoded into rotational states with their associated electric dipole moments, and single-bit gates can be accomplished with microwave pulses.  However, boundary conditions associated with the nearby conductors can significantly alter the dipole-dipole interaction.  In particular, if two molecules a distance $L$ apart are trapped a distance $h$ from conductors with transverse dimension $w \approx h$, then the dipole-dipole interaction scales as $L^{-1}w^{-2}$ instead of as $L^{-3}$ as in free space.  For sufficiently small trap and conductor dimensions (typically $w \sim 0.1-10~\mu$m are envisioned), this can make it possible to either enhance the interaction between molecules, or to keep the same interaction strength even for molecules held at large separation.  The interaction can be achieved either through DC capacitive coupling to a thin wire joining the qubits, or via exchange of (virtual) microwave photons in a stripline geometry.  In both cases making the electrodes from superconducting material limits resistive losses; in the stripline case it is possible to construct a high-$Q$ resonator to minimize photon loss. The latter architecture seems particularly favorable, for a few reasons.  For example, here one can envision coupling a large network of molecular qubits via ``flying'' photonic qubits (which can propagate over large distances while retaining quantum coherence).  In addition, it becomes possible to perform a near-ideal projective measurement of the qubit state, by observing the phase shift of off-resonant photons propagating through the stripline cavity; such a measurement leaves the molecule intact, with the only effect being to collapse any superposition of the internal qubit-encoding states.  This type of hybrid quantum processor combines the transparent scalability of microfabricated circuits with the reproducibility and inherently long lifetime of single-particle qubits.

The use of low-frequency electric fields instead of an optical field for trapping the qubits solves certain problems but creates others.  In this system, couplings are enhanced by trapping as near the surface of the chip as possible.  A fundamental limit $w \gtrsim 0.1~ \mu$m is placed by the attraction of the molecule to its image dipole in the substrate.  At such small distances, the requirement for stability in laser intensity (to control decoherence) is now replaced by a requirement for small local electric field fluctuations, since the trapping field couples strongly to the molecular internal states.  Such fluctuations can be caused by various noise sources (e.g. random motion of charges trapped in the substrate, thermal blackbody photons), which make it necessary to cool the chip to deep cryogenic temperatures ($T \lesssim 100$ mK).  Even here, however, electric field noise can be a dominant source of decoherence.  In addition, effective field fluctuations can also be caused by random thermal motion of the molecule itself in the trap potential.  Hence, it becomes necessary to cool the molecule's motion very near the ground state of the trap.  Electric field noise-induced decoherence can again be suppressed by encoding into spin or hyperfine states, or by biasing the trap to a ``sweet spot'' where the differential polarizability of the qubit states is identical~\cite{Andre06}. Note also that the phase space density occupied by a single molecule in such a state is again essentially the same as for quantum degeneracy.  However, initial loading of such traps may be easier than for the optical lattice: although the volumes are roughly comparable, the depth of a static trap can be significantly greater.  After loading the trap at modest temperature, the motion can be cooled by employing the coupling between internal states and motion from the trapping fields, together with the dissipation associated with decay into the microwave cavity.  Together, this suggests the intriguing possibility that the entire system--trapping, cooling, gate operations, and state readout--can be accomplished purely with electrostatic and microwave signals, i.e. without any need for lasers.

It may be possible to further enhance the strength of qubit couplings--and hence the speed of gate operations--by using ensembles of molecules rather than individual particles.  A suitable nonlinearity in response can cause a group of $N$ molecules to respond as a single, collective qubit with dipole moment $d_{\mathrm{coll}} = \sqrt{N} d_{\mathrm{eff}}$.  One possibility source of nonlinearity is the ``dipole blockade''~\cite{Lukin01}. Here, because of dipolar interactions within the ensemble, the energy required to excite two molecules in the ensemble is more than twice the energy needed to excite a single molecule.  Hence if the molecules are all localized in a region small compared to the wavelength of the (microwave) excitation, they respond collectively (much like a phased array of classical antennas) as a two-level system.  In an alternate vision, the microwave photons connecting two ensembles interact with a nearly-resonant superconducting qubit~\cite{Rabl07}.  Such Cooper-pair box (CPB) qubits~\cite{Schoelkopf08} act effectively as a two-level system with a very large dipole moment, causing dispersive shifts in the stripline cavity resonance that allow it to support only a single photon at a time.  This again limits the molecular ensemble to a single excitation and enhances the coupling in a similar manner.

The use of collective qubits raises the new question of decoherence due to collisions between molecules in the ensemble.  Even at modest phase-space density $\Omega \sim 10^{-7} \hbar^3$, estimated collisional decoherence rates ($R \sim 1$ kHz) are small enough to make molecular ensembles useful e.g. as a quantum memory for storage of information processed rapidly by by CPB qubits~\cite{Rabl06}.  However, at sufficiently high phase space density $\Omega \gg 1~ \hbar^3$, molecules confined in a planar trap can form a dipolar crystal~\cite{Buchler07}.  In this new phase of dipolar matter, collisions are eliminated by the repulsive dipolar interaction and molecules become tightly localized~\cite{Rabl07}.  A dipolar crystal ensemble may represent the ultimate version of a molecule-based qubit; arrays of such ensembles could provide a robust, scalable system for quantum information processing.

\subsection{Quantum Simulations and Quantum Simulators}
\label{ssec:numericalMethods}

In order to understand many body aspects of ultracold molecular systems it is often necessary to perform simulations on a classical computer.  We distinguish ``quantum simula\emph{tions}'' from the ``quantum simula\emph{tors}'' discussed later in this section.  New algorithmic development is needed for quantum simulations for molecules as compared to ones already developed for atoms.  In the following, we discuss some of the available simulation techniques and how they might be modified or advanced to treat spatial, temporal, internal molecular, and quantum degrees of freedom.

Let us first consider the semi-classical regime for bosons.  The NNLS equation~(\ref{eqn:nnls}) is not analytically tractable with a harmonic trap even in the scalar case~\cite{Baranov08}.  Studies of the dipolar gas have required (3+1)D simulation methods~\cite{Lahaye08,Metz09}.  These methods include the split-step algorithm, finite-difference methods, and pseudo-spectral adaptive time-step Runge-Kutta, among other variations.  Although formal and rigorous convergence studies have been made of these methods for the NLS~\cite{Taha84,Perezgarcia03}, the introduction of nonlocal nonlinearity in the NNLS equation necessitates new and careful studies of implicit and explicit methods for nonlinear partial differential equations.  Going beyond the scalar NNLS equation to a vector NNLS, as will be necessary to treat many rotational, vibrational, and other internal molecular states, can lead to an effective fourth dimension.  To treat such systems with present computing platforms parallelization is desirable.  What is the appropriate parallel algorithm for a vector NNLS with many vector components (internal states)?  Given the availability of desktop multiprocessor workstations, parallelization for common shared memory is an obvious first step.  Beyond this, parallelization for a high-performance computing cluster has been demonstrated in the finite element discrete variable representation (FEDVR) method with the real space product algorithm (RSP), proposed by Schneider \emph{et al.}~\cite{Schneider06}.  Such methods bear further exploration: one possibility is to use multi-core processors to deal with internal states of molecules, while retaining inter-processor parallelization over configuration-space according to the FEDVR/RSP prescription.  However, even in this case, one must determine how to implement the non-local nonlinearity in a computationally efficient way.  For fermions, one kind of semi-classical approach involves a separate partial differential equation for each fermion in the system.  Then these equations are coupled appropriately, including mean field terms.  Solving such a system without additional internal states, beyond spin-1/2, is already challenging.  Again, to extend this to computations for fermionic molecules with interesting internal states will require significant parallelization.

For the fully quantum regime, one can use quantum Monte Carlo simulations to obtain phase diagrams and other static properties.  Here there have been many recent advances in areas such as cluster methods for dissipative systems~\cite{Werner05} and path integral methods extended to the grand canonical ensemble and finite temperature~\cite{Boninsegni06}.  Such methods can be useful to determine the effect of loss processes in optical lattices, as are likely to be a significant factor in early experiments, more so than for atoms~\cite{Danzl09}.  However, we are not aware of a general formulation of quantum Monte Carlo for a system with a large number of internal states.  Therefore code development in this direction could be helpful.

To study dynamics, there are very few methods besides mean field theories.  Exact diagonalization~\cite{Albuquerque07} is not very useful for molecules because the internal degrees of freedom take up most of the computationally accessible Hilbert space, and one is left with only a few sites and/or a few molecules, at best.  A lattice Hamiltonian such as the Molecular Hubbard Hamiltonian, Eq.~(\ref{eqn:molecularHubbard}), can be simplified through partial use of mean fields, as in the Gutzwiller ansatz, which replaces terms like $\hat{a}_{i+1,\sigma}^{\dagger}\hat{a}_{i\sigma}$ in Eq.~(\ref{eqn:hubbard}) with $\langle\hat{a}_{i+1,\sigma}\rangle^{*}\hat{a}_{i\sigma}$; this can even be extended to a multiband model~\cite{Larson09}, and therefore to internal states of molecules.  However, the Gutzwiller ansatz has only limited application to dynamics, and excludes spatially entangled states.  To reach the fully quantum regime, the recently developed time evolving block decimation algorithm (TEBD)~\cite{Vidal03,Vidal04}, which is equivalent to an adaptive time-dependent density matrix renormalization group method~\cite{Daley04,Schollwock05}, allows one to evolve a useful class of spatially entangled states, called matrix product states, in one spatial dimension.  This method has been adapted to molecules and is available in open source form~\cite{Opensource09,Wall09}.  Ground states of well-known condensed matter models are in many cases well represented by matrix product states~\cite{Verstraete06}; this may also be true of ultracold many body molecular systems, a point that requires investigation.  However, higher dimensional extensions of TEBD such as the projected entangled pair state algorithm (PEPS)~\cite{Verstraete04,Shi06} are still under development.  Molecules may require significant parallelization of the TEBD algorithm for realistic simulations in even one spatial dimension, let alone two and three spatial dimensions.

Many fully quantum aspects of ultracold many-body molecular systems remain inaccessible on classical computers, as can be proved in specific cases relevant to quantum computing, such as Shor's factorization algorithm.  In addition to their use for quantum information processing, ultracold polar molecules have been proposed as the fundamental elements of a \emph{quantum simulator}.  Here, rather than performing inefficient digital computations, the idea is to directly engineer a system whose Hamiltonian matches that of an interesting, fully quantum many-body system whose properties are not fully understood.  The long-range and anisotropic nature of the dipolar interaction between molecules in an optical lattice provides a novel and powerful tool for this task.  A series of key papers~\cite{Micheli06,Buchler06,Barnett06,Micheli07} have outlined specific manifestations of these ideas.  For example, Barnett {\it et al.}~\cite{Barnett06} describe how to engineer interactions equivalent to quantum magnetism, using the exchange of angular momenta between rotational states to produce long range interactions.  In a remarkable development, Micheli {\it et al.} showed how to use molecules with internal spin substructure to simulate any arbitrary Hamiltonian for interacting spins on a lattice~\cite{Micheli06}.  These are among the most intensely studied problems in hard condensed matter physics at this time.  Properly designed Hamiltonians of this type can allow, for example, a means for error-resilient qubit encoding, or for producing a topologically protected quantum memory.  Their method uses an ingenious combination of microwave-induced resonant dipolar interactions together with the internal spin-rotation coupling of the molecule.  In later work~\cite{Buchler07}, a method was devised to turn off 2-body interactions and cause 3-body interactions to dominate. Such exotic interactions can give rise to topological phases with anyonic excitations~\cite{Moore91}.  These examples provide evidence that a wide variety of condensed matter models can be simulated using polar molecules -- far more than with ultracold atoms -- due to their unique dipolar interactions.

\section{Conclusions}
\label{sec:conclusions}

This is an exciting time for the research field of cold and ultracold molecules. Fundamentally, the field represents the continued advance of our capabilities in the precise study, control, and measurement of increasingly complex quantum systems.  The range of topics covered in this review and in this special issue demonstrates that the creation and study of cold and ultracold molecules have already had a profound impact on many research areas of physics. Yet, we argue, this is just the beginning. The last three years mark the creation of ultracold molecules in the absolute ground state with high phase-space density~\cite{Ni08}, the development of technology to study collisions of tunable molecular beams with trapped molecules~\cite{Sawyer08a}, the creation of a molecular synchrotron~\cite{Heiner07}, the first experiments using cold molecules for tests of fundamental physics~\cite{Hudson06b,Tarbutt08}, etc., etc.  These experiments have advanced the research field of ultracold molecules to a new level: while most of the research in the past ten years was on the development of new methods for the production of dense ensembles of ultracold molecules, the focus of today's research is shifting to {\it applications} of cold and ultracold molecules in science and technology. This review describes many of the applications already proposed; however, it seems likely that many new applications will be discovered in the near future, as the production of dense ensembles of ultracold molecules becomes routine.

The development of this research field in the immediate future will most likely be driven by the scientific and practical goals outlined in Table 1 and in Section~\ref{sec:applications}.  Whether these goals will be achieved relies on a number of fundamental questions that remain to be answered.  We hence end this review with a list of open (and pressing) questions which need to be addressed to make progress in the study of cold and ultracold molecules. These questions include the following.\\

(1) What is needed to bridge the gap between the cold and ultracold regimes, for the wide variety of molecular species accessible only through direct cooling techniques?  As already discussed, for some types of molecules direct laser cooling may be possible.  More generally, the approach of collisional (evaporative~\cite{Ketterle96} or sympathetic~\cite{Myatt97}) cooling of trapped molecules appears very attractive in principle.  For example, sympathetic cooling of molecules by contact with laser-cooled atomic gases is being pursued by several groups, both experimentally~\cite{Schlunk07,Rieger07} and theoretically~\cite{Soldan04,Lara06,Lara07,Zuchowski08,Soldan09}.  However, two conditions are needed to make collisional cooling feasible.  First, there must be a sufficiently rapid rate of elastic collisions to ensure thermalization in a time much shorter than the trap lifetime.  This in turn requires that the product of the collision cross section and the collisional density (averaged over the trap volume) be large.  However, in the cold temperature regime cross sections are not typically enhanced by quantum effects as they are at ultralow temperatures; moreover, the low phase space densities now available with direct cooling methods make it necessary to use trap volumes very large compared to typical atomic traps.  It has been predicted that dipolar effects and/or collisional resonances could enhance the cross sections sufficiently to make collisional cooling viable~\cite{Bohn02,Kajita02,DeMille04,Bohn05,Ticknor05}.  Here the ability to cool \textit{any} strongly polar species to the ultracold regime may prove useful, if such a species can be used as an accelerant for sympathetic cooling of other species.  However, the few relevant calculations to date have focused on the \textit{total} elastic cross-section $\sigma_{\mathrm{el}} = \int{d\Omega \frac{d \sigma}{d\Omega} }$, while for collisional cooling the important quantity is instead the momentum-transfer cross-section $\sigma_{\mathrm{mt}} = \int{d\Omega \frac{d \sigma}{ d\Omega} (1-\cos{\theta})}$, which is typically much smaller.  Improved methods for trap loading, such as dissipative optical pumping methods proposed in several contexts~\cite{vandeMeerakker01,DeMille04,Narevicius09}, may allow increased density as well.

The second condition for collisional cooling is that inelastic collision rates be much smaller than the elastic rates.  This is a problem since most of the deep traps employed for directly cooled molecules are effective only for weak-field seeking states--which are always internally excited states.  It was recognized early in the development of the field that the rotational degree of freedom in molecules rather generically enhances inelastic collisions of these excited states, relative to the case of atoms.  A variety of approaches have been suggested to circumvent this problem.  For example, it may be possible to use electric and/or magnetic fields to adjust molecular energy levels in such a way as to suppress inelastic processes~\cite{Volpi02,Abrahamsson07,Tscherbul06,Tscherbul08b,Tscherbul09}.  However, any such mechanism must be viable over the large initial range of kinetic energies corresponding to merely cold molecules.   Another promising approach circumvents the issue through the use of traps that are effective on the strong-field seeking absolute ground state of the molecule.  Optical traps have this feature, but the phase space density they can capture is too low to be useful for present directly-cooled sources.  However, AC electric traps (analogous to the rf Paul trap in common use for ions) have already been demonstrated~\cite{Veldhoven05}.  Larger trap volume and depth for strong-field seekers may be possible using a proposed microwave-frequency trap, analogous to an optical trap but detuned to the red of a rotational transition~\cite{DeMille04}.  A paper in this issue argues that possible inelastic collisions between microwave-dressed molecules are unlikely to be a problem in such a trap~\cite{Avdeenkov09}.

For the indirect cooling method where the ultracold temperature regime and even the quantum degeneracy are within reach, can we go beyond bialkali systems?  What will be the most interesting (diatomic) molecular species to be explored within the current atomic laser cooling technology? Is the size of the electric dipole moment the most important deciding factor?  Are the techniques explored so far with bialkali systems universally applicable?\\

(2) Will the evaporative or sympathetic cooling of complex polyatomic molecules be feasible?
With a few exceptions, the experimental work on ultracold molecules has so
far been limited to diatomic molecules. For chemical applications, we need
more complex molecules. It has not yet been established whether cooling of
polyatomic molecules to ultralow temperatures will be possible.
Sympathetic cooling could potentially yield both translationally and internally cold
molecules but the efficiency of sympathetic cooling is based on the
magnitude of the cross sections for momentum transport in atom-molecule
collisions. Momentum transport is normally efficient in collisions of
diatomic molecules with atoms~\cite{Krems09}. Collisions of more complex molecules are,
however, characterized by many long-lived resonances which may give
rise to efficient distribution of energy between the atom-molecule and
intramolecular degrees of freedom. Incoming ultracold atoms may thus stick
to large molecules for a long period of time.  If so, sympathetic  cooling
of large polyatomic molecules may be very difficult, if not impossible on
the timescale of the experiments.\\

(3) In order to understand the dynamics of sympathetic cooling and develop new
experimental techniques for cooling molecules to ultralow temperatures
and for experimental studies of ultracold collisions, it is necessary to
understand diffusion of molecular gases in ultracold mixtures. A simple
analysis of the diffusion coefficient based on the Chapman-Enskog solution~\cite{McCourt91}
of the classical Boltzmann equation and the threshold behavior of the
collision cross sections~\cite{Wigner48} indicates that diffusion properties of molecules
with open inelastic and reactive scattering channels are dramatically
different from those of non-reactive molecules in the ground state.  The
elastic scattering cross section is independent of collision energy at low
temperatures, while the reactive scattering cross section rises to
infinity as the collision energy vanishes~\cite{Wigner48}. It should be expected therefore
that the diffusion process of molecules that change identity is
qualitatively different at ultralow temperatures from the diffusion of
non-reactive atoms and molecules. However, there are many questions that
remain open.  These include the following.  What are the quantum corrections to the
classical Boltzmann equation~\cite{Snider98} required for an adequate description of
molecular gases at temperatures below 1 Kelvin? What is the effective
cross section for momentum transport in the presence of inelastic
collisions at ultralow energies? (Does it have to include inelastic and
reactive scattering cross sections? Is it different depending on whether
the reactive scattering conserves or changes the orbital angular momentum
of the reactive complex?) How does the localization of the de Broglie wave
affect the diffusion of binary mixtures? What is the difference between
the diffusion of binary mixtures and self-diffusion?\\

(4) Can the properties of ultracold molecules be used to further improve tests of fundamental physical laws?  An intriguing possibility is to use trapped molecules to enhance coherence times and hence increase energy resolution.  However, for such experiments the perturbing influence of the trapping potential must be carefully considered~\cite{Tarbutt08}.  Here the analogue of ``magic wavelength'' traps developed for atomic clocks may be useful~\cite{Zelevinsky08,Kotochigova09}.  Sympathetic cooling, perhaps using dipolar- or resonance-enhanced collisions, could be key for preparing samples of the particular species needed for many such measurements, and also for lowering their temperature sufficiently enough so that weaker, less-perturbing trap potentials can be used.  Strategies to suppress inelastic collisions will also be important as a way to improve the signal size in such experiments.  Finally, it may be possible to use the strong dipolar interaction to engineer squeezed spin states of a molecular ensemble.  Similar ideas using blockade effects in Rydberg atoms have been formulated~\cite{Bouchole02}.  Such states can in principle dramatically increase the measurement sensitivity beyond the scaling $\propto \sqrt{N}$ in the standard quantum limit.  For optimal enhancement, different molecular systems will certainly be needed for specific measurement goals,  whether in time reversal symmetry, fundamental constants, or parity violations, so it is certain that addressing the pressing question (2) will have a huge impact to this area of research.  Ultimately, just like in ultracold atom systems, we would like to prepare molecules in single quantum states, both internally and externally, to achieve quantum-limited measurement precision accuracy.\\

(5) Can we use ultracold molecules to interface with mesoscopic quantum mechanical systems?  While dipolar interactions are weaker than Coulombic interactions, they are certainly stronger and have a much longer range than interactions in atomic systems. Can we cool a nano-mechanical structure or nano-electronic circuit~\cite{Andre06} via molecules and vice-versa?  Can we achieve strong couplings between a molecular sample (or even a single molecule) and a nano-mechanical~\cite{Singh08} or nano-electronic~\cite{Rabl06} structure, in order to prepare their coherent superpositions or map quantum states to-and-fro? Can we engineer artificial molecules that exhibit more attractive properties?\\

(6) A challenging problem is molecular detection, which is more complex and difficult than for atoms. This will become an increasingly urgent issue as we move from preparation of ultracold molecular samples to different scientific applications. Can we monitor several degrees of freedom of the molecule at the same time? Can we monitor molecules in a nondestructive manner? Can we monitor a single molecule with sufficient sensitivity (and without destroying it)? The technology development in this area will be very important to answer a number of challenging tasks ranging from detection of ultracold chemistry (where one may need to detect several internal states at high spectral resolution or perhaps even different molecular species), to monitoring many body system dynamics, to precision measurement and quantum information processing.\\

(7) Finally, what are the first many body properties that will be observed for a quantum degenerate ultracold molecular gas in a harmonic trap?  Although a lot of work has been done on ultracold dipolar gases, we know almost nothing about coupling to rotational, vibrational, and other electronic degrees of freedom in the semiclassical, or mean field regime.  How will collective modes of the dipolar gas couple to molecular degrees of freedom?  Which modes will clearly signal the onset of quantum degeneracy?  What happens at finite temperature, when a two fluid model~\cite{Griffin09} becomes necessary?  When is the system really 2D or 1D in the fully quantum description, and when is it only quasi-1D or quasi-2D, i.e. of reduced dimensionality in the mean field sense?  Is there a formal expansion~\cite{Lee57a,Lee57b} in a dilute-gas-like parameter based on the dipole, in terms of $\sqrt{\bar{n} |a_d|^3}$?  Most of the theory for dipolar gases, whether bosonic or fermionic, is for equilibrium systems and small perturbations around equilibrium.  What is the correct description of a strongly driven molecular dipolar gas, as must be the case if we are to access disparate molecular degrees of freedom?  Will a semiclassical theory be sufficient to describe dipolar gases of molecules?

Is it possible to tune the dipolar interaction is to be the \emph{largest} energy scale in the system, much larger than the difference between rotational states, for example, as might occur in a lambda-doublet system?  What kind of many body theory can describe such a system?  Do dipolar collapse dynamics change in this circumstance?  For fermions, is there a BCS-like theory for pairing of molecules in such circumstances?  Such a theory would have to incorporate rotational modes and other molecular internal degrees of freedom.  Similar to ultracold atom-based optical lattices, can we prepare the lowest energy (or lowest entropy) state for ultracold molecules in optical lattice? Can we achieve the same level of experimental control for ultracold molecules? How does the long-range interaction manifest itself in the phase transition? How does it affect the loading? What is the lowest temperature we can reach?
What about electric-dipolar domain formations and spontaneous symmetry breaking~\cite{Iskin07}?  Is there a relationship with quantum magnetism that can be exploited to address outstanding issues in naturally occurring magnetic materials?

\section{Acknowledgements}
\label{sec:acknowledgements}

L. D. Carr acknowledges support from the National Science Foundation under Grant PHY-0547845 as part of the NSF CAREER program. D. DeMille acknowledges support from NSF (DMR-0653377 and PHY-0758045), DOE, and ARO. The work of R. V. Krems is supported by NSERC of Canada.  J. Ye acknowledges support from DOE, NIST, and NSF and he thanks colleagues at JILA for their collaborations on cold and ultracold molecules over the last 9 years.

%\section{References}

%\bibliographystyle{unsrt}
%\bibliography{coldMoleculesRefs_v3_6}

\end{document}